\documentclass[useAMS,usenatbib]{mn2e}

\usepackage{txfonts}
\usepackage{natbib}
\usepackage{longtable}
\usepackage[pdftex]{graphicx}
\usepackage{multirow}
\usepackage[bookmarks,breaklinks]{hyperref}
\usepackage{aas_macro}
   \hypersetup{
     colorlinks,
     citecolor=blue,
     filecolor=magenta,
     linkcolor=blue,
     urlcolor=cyan
}

\graphicspath{{X-ray/}{Radio/}{MidIR/}}

\title[Close environment of AGN in the VIDEO survey]{Mergers as triggers for nuclear activity: A near-IR study of the close environment of AGN in the VISTA-VIDEO survey}
\author[M. Karouzos et al.]{M. Karouzos,$^{1}$\thanks{E-mail: mkarouzos@astro.snu.ac.kr}\thanks{This research has been partially supported by the EU COST Action ``Black Holes in a Violent Universe.''.}
M. J. Jarvis,$^{2,3}$
D. Bonfield,$^{4}$\\
$^{1}$Center for the Exploration of the Origin of the Universe, Seoul National University, Seoul, Korea\\
$^{2}$Astrophysics, Department of Physics, Keble Road, Oxford OX1 3RH\\
$^{3}$Department of Physics, University of the Western Cape, Private Bag X17, Bellville 7535, South Africa\\
$^{4}$Centre for Astrophysics, Science \& Technology Research Institute, University of Hertfordshire, HatÞeld, Herts AL10 9AB}

\begin{document}

\date{}

\pagerange{\pageref{firstpage}--\pageref{lastpage}} \pubyear{2013}

\maketitle

\label{firstpage}

\begin{abstract}
There is an ongoing debate concerning the driver of nuclear activity in galaxies, with AGN either being triggered by major or minor galactic mergers or, alternatively, through secular processes like cooling gas accretion and/or formation of bars.\\
We investigate the close environment of active galaxies selected in the X-ray, the radio, and the mid-IR. We utilise the first data release of the new near-IR VISTA Deep Extragalactic Observations (VIDEO) survey of the XMM-Large Scale Structure (LSS) field. \\
We use two measures of environment density, namely counts within a given aperture and a finite redshift slice (pseudo-3D density) and closest neighbour density measures $\Sigma_{2}$ and $\Sigma_{5}$. We select both AGN and control samples, matching them in redshift and apparent K$_s$-band magnitude.\\
We find that AGN are found in a range of environments, with a subset of the AGN samples residing in over-dense environments. Seyfert-like X-ray AGN and flat-spectrum radio-AGN are found to inhabit significantly over-dense environments compared to their control sample. The relation between over-densities and AGN luminosity does not however reveal any positive correlation. Given the absence of an environment density-AGN luminosity relation,we find no support for a scheme where high luminosity AGN are preferentially triggered by mergers. On the contrary, we find that AGN likely trace over dense environments at high redshift due to the fact that they inhabit the most massive galaxies, rather than being an AGN.
\end{abstract}


\section{Introduction}
\label{sec:intro}
The importance of activity in galaxies, both in terms of accretion and star-formation, is evident in models of galaxy formation and evolution. The currently accepted paradigm of a very massive black hole located at the center of each bulge-dominated galaxy (e.g., \citealt{Rees1978}, \citealt{Blandford1986}, \citealt{Richstone1998}, \citealt{Genzel2000}, \citealt{Eisenhauer2005}) turns accretion onto these central black holes and its phenomenological equivalent, active galactic nuclei (AGN), to a powerful tool for probing not only the formation and evolution of these central objects but their hosts as well. Given the currently accepted scheme of a hierarchically, bottom-up, evolving $\Lambda$CDM Universe (e.g., \citealt{Frenk1988}), mergers should play an important (if not regulating) role in such an evolution (e.g., \citealt{Toomre1972}).\\
In the last couple of decades a set of rather intriguing scaling relationships have emerged, that actually connect galaxy evolution to the aforementioned central black holes (e.g., \citealt{Merritt2001}, \citealt{Tremaine2002}, \citealt{McLure2002}, also see \citealt{Kormendy2013}) and therefore provide further support to a mechanism that leads to a co-evolution of the central object and its host galaxy. In this context, active galaxies have been investigated as possible phases of the evolutionary track of a galaxy (e.g., \citealt{Sanders1988}, \citealt{Veilleux1995}, \citealt{Canalizo2001}, \citealt{Nagar2003}, \citealt{Hopkins2006}), as a result of a merger event (e.g., \citealt{Hernquist1989}, \citealt{Kauffmann2000}, \citealt{Cattaneo2005}, \citealt{Lotz2008}).\\
One of the most straightforward ways of testing the potentially causal link between merger events and nuclear activity is to look for direct merger induced effects, such as distortions in the morphology of the AGN host galaxy. Although such studies are hampered by the nucleus usually outshining its host and thus masking or potentially even introducing spurious structures, the latest generations of telescopes working in the optical and the infrared, combined with complex  simulations, have helped make significant progress in the field (e.g., \citealt{Canalizo2001}, \citealt{Grogin2005}, \citealt{Bennert2008}, \citealt{Darg2010}, \citealt{Cisternas2011}, \citealt{Almeida2011}, \citealt{Kocevski2012}). However, the results have been contradictory, with the significance of merger imprints on AGN closely connected to the selection criteria of the AGN sample (e.g., X-ray selected, radio selected).\\
A complementary approach to the topic employs the study of the environment of active galaxies to infer the probability of such objects having taken part in a merger or galaxy interaction event. Again the results of such studies have been dependent on the wavelength of the AGN selection, the depth and wavelength of the surveys used, and the physical scales investigated. Several studies have shown AGN to reside in over-dense environments (e.g., \citealt{Best2004}, \citealt{Serber2006}, \citealt{Tasse2008}) {or that activity is enhanced in galaxies with close companions (e.g., \citealt{Ellison2011}, \citealt{Silverman2011}),} while others provide evidence for either no dependence between AGN and environment density (e.g., \citealt{Miller2003}) or even AGN inhabiting under-dense environments (e.g., \citealt{Kauffmann2004}, \citealt{Tasse2011}). These in turn have been interpreted in different ways, e.g., radio-AGN in under-dense environments being triggered by mergers, in contrast to radio-AGN in groups or clusters being fueled by cold and/or hot gas accretion from the IGM (e.g, \citealt{Tasse2008}).\\
In this paper we build upon previous studies by using the new VISTA Deep Extragalactic Observations (VIDEO; \citealt{Jarvis2013}) survey data, combined with a variety of  AGN selected at multiple wavelengths. Near-IR wavelengths are less affected by obscuration than optical surveys, enabling a statistically more complete investigation. Given the ancillary data used here, we can homogeneously compare the behavior, in terms of their environment, of active galaxies selected in different wavelengths. Moreover, we focus on the local environment, at scales relevant to small galaxy groups or even close companions/galaxy pairs, as these are the most relevant systems for possible ongoing mergers.\\
The paper is organized as follows: in Sect. \ref{sec:data} we describe the VIDEO survey data used here, as well as the multi-wavelength ancillary data at our disposal. In Sect. \ref{sec:analysis} we discuss the different tools that we employ to study the environment of AGN and in Sec. \ref{sec:results} we give our results. Section \ref{sec:discussion} follows with a discussion on the robustness of our results and the completeness of the data used here, as well as the implications of our results and comparison with other works. We conclude the paper in Sect. \ref{sec:conclusion} where a short summary and outlook is given. Throughout the paper, we assume the cosmological parameters $H_{0}=71$\,km\,s$^{-1}$\,Mpc$^{-1}$, $\Omega_{\rm M}=0.27$, and $\Omega_{\Lambda}=0.73$ (\citealt{Komatsu2011}).

\section{Data and sample selection}
\label{sec:data}
Our investigation is enabled by the availability of large multi-wavelength datasets and deep and wide surveys of the sky. In particular, for this study, we focus on the XMM Large Scale Structure (XMM-LSS; \citealt{Pierre2007}) field, which was originally observed by the X-ray telescope XMM-Newton and has since then been part of several multi-wavelength observational efforts. In particular, the field has been included in the VIDEO survey (VIDEO; \citealt{Jarvis2013}). In the following we will describe in short the VIDEO survey, the ancillary data used, as well as the selection criteria used for each wavelength regime.

\subsection{The VISTA-VIDEO survey}
\label{sec:survey}
\begin{figure}
\begin{center}
\includegraphics[width=0.5\textwidth,angle=0]{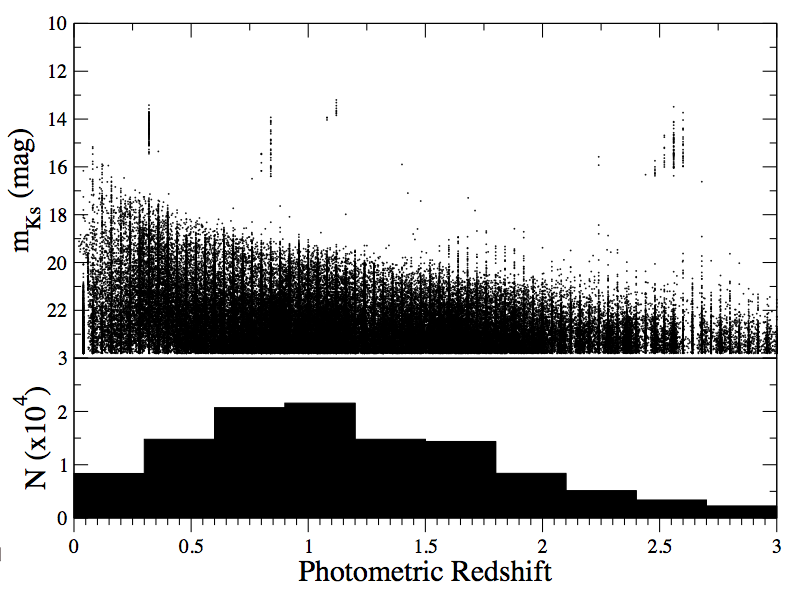}
\caption{The $m_{K_{s}}$ AB magnitude distribution as a function of photometric redshift is shown for the $5\sigma$ VIDEO sample. The absolute number distribution of the photometric redshifts of the same sample is also shown.}
\label{fig:z_histo}
\end{center}
\end{figure}

The VIDEO survey is a near-IR survey that will cover 12 square degrees of the sky, over three separate fields, in five photometric bands Z, Y, J, H, and $K_{s}$, using the ESO Visual and Infrared Survey Telescope for Astronomy (VISTA; \citealt{Emerson2010}). Once completed, the survey is planned to reach a depth of 25.7, 24.6, 24.5, 24.0, and 23.5 for Z, Y, J, H, and $K_{s}$ filters, respectively. In this paper we use data from the first data release of the VIDEO survey, covering tile 3 of the XMM-LSS field ($\sim 1.8$ deg$^{2}$), reaching a depth of 25.2, 24.6, 24.6, 24.1, and 23.8 for bands Z, Y, J, H, and K$_s$, respectively (5$\sigma$ detections; see \citealt{Jarvis2013} for full details).\\
In particular we focus on a smaller field that overlaps with the CFHTLS D1 deep optical field and covers roughly 1 deg$^{2}$, centered at [36.496,-4.494]. Image stacks have been made for each band separately using SWarp (\citealt{Bertin2002}), with stacks at seeing worse than 0.9'' FWHM being rejected. From each set of images, catalogues have been produced using SExtractor (\citealt{Bertin1996}) in double-image mode, utilizing also the CFHTLS D1 T0005 images. The final catalogues contain photometry (at several different aperture sizes) in all five near-IR bands, as well as the optical bands from CFHTLS. In addition a stellarity index is given for each source, describing its resemblance to a point source. By utilizing the multi-band photometry from both CFHTLS and VIDEO, photometric redshifts for all the VIDEO sources have been calculated. For this the Le Phare software (\citealt{Arnouts1999}, \citealt{Ilbert2006}) was used. A 2'' aperture size was used for the photo-z estimation\footnote{For consistency reasons, we also use 2'' aperture size magnitudes throughout this paper.}. The composite templates from \citet{Salvato2009} are used for the photo-z calculation of the AGN in our samples. Photometric redshifts for the rest of the galaxies are calculated as described in \citet{Jarvis2013}. In Fig. \ref{fig:z_histo}, the redshift distribution of the VIDEO sources, as well as the magnitude in the $K_{s}$ band as a function of photometric redshift are shown.\\

A number of selection criteria are used to create the main catalogue of near-IR sources used in this paper. We select our sample in the K$_{s}$ band, as this enables us to probe the old stellar population in galaxies and hence provides a proxy selection for stellar mass content. We choose to use a detection limit of 5$\sigma$, which corresponds to a limiting magnitude of $m^{lim}_{K_{s}}=23.8$. Additionally, we exclude sources with stellar optical and near-IR colors, following the color selection employed in \citet{Jarvis2013}. Finally, given the accuracy of the photometric redshifts and the study of source number density within the VIDEO field as a function of redshift, we impose a hard limit at redshift z$\sim 3$, beyond which the environments of sources can not be constrained reliably. The resulting near-IR catalogue used for our analysis (multi-wavelength sources cross-matching and control source selection) contains 113728 sources.

\subsection{Multi-wavelength datasets}
Here the multi-wavelength data used in this paper are described. All datasets are synoptically described in Table \ref{tab:crossID}.
\subsubsection{XMM-Newton}
The XMM-Large Scale Structure (XMM-LSS; \citealt{Pierre2007}) was originally designed to provide a well-defined, statistically significant X-ray galaxy cluster sample out to z=1. Aside from its cluster sample aim, the survey of the XMM-LSS field in the soft X-rays (0.5-2 keV band) has recovered 2980 non-spurious objects (3385 in total from band-merging), around half of these point-like. This presents a 90\% completeness at a sensitivity level of $4\cdot10^{-15}$ erg/s/cm$^{2}$. In the overlapping region of VIDEO and XMM-LSS a total of 1049 X-ray sources are detected, for which fluxes are available in both soft (0.5-2 keV) and hard (2-10 keV) X-ray bands.\\
Using the hard X-ray band (2-10 keV) fluxes we can divide XMM detected sources in QSO-like or Seyfert-like, according to whether their spectrum is dominated by their optical (infrared) or high-energy (X-ray) emission. We quantify this through the ratio of K$_{s}$ to [2-10 keV] flux (following \citealt{Bradshaw2011}). We assume a limiting value between QSO-like and Seyfert-like of K/X=2. {In addition, we set a luminosity limit at the [2-10 keV] band at $10^{42}$ erg/s, below which we expect severe contamination from local starbursting galaxies. Taking these two criteria in account, we find 153 XMM-detected sources as QSO-like (K/X$<2$) and 178 as Seyfert-like (K/X$>2$). Only sources detected in both $K_{s}$ and with L$_{2-10keV}\geqslant10^{42}$ erg/s are considered for this comparison.}

\subsubsection{VLA-GMRT}
The XMM-LSS field was observed at radio wavelengths with the Very Large Array (VLA) at 1.4GHz (\citealt{Bondi2003}; also see \citealt{McAlpine2012})  as part of a multi-wavelength effort complementing the spectroscopic VIRMOS VLT Deep Survey (\citealt{Lefevre2002}). Using the VLA in B configuration a resolution of 6 arcsec was achieved together with a detection limit of 60$\mu$Jy. The final radio catalogue of the field contains 1054 individual sources, some of which consist of multiple components. For the purpose of this paper multi-component sources have been substituted by the assumed core of the source (also provided in the catalogue) in order to avoid multiple cross-identifications. In total 19 such multi-component sources were substituted.

In addition the field has been observed using the Giant Meter-Wave Radio Telescope (GMRT) at 0.6GHz (\citealt{Bondi2007}). The resolution of the array at this frequency is 6 arcsec, matching that of the VLA observations.  The 3$\sigma$ sensitivity of the GMRT survey is 150$\mu$Jy. The GMRT catalogue of the XMM-LSS field contains 514 individual sources (5$\sigma$ detections), with 17 of these showing multi-component structure. Given the dual-band observations in the field, a radio spectral index $\alpha_{0.61}^{1.4}$ has also been calculated. Including also 3$\sigma$ detections, radio spectral indices of 741 sources, for which counterparts are found at both frequencies, were calculated. For sources without detections in both bands, upper limits were used to calculate their spectral indices.

We employ two different AGN selection methods in the radio regime using (1) the radio luminosity at 1.4 GHz and (2) the spectral index of the sources. Following \citet{Condon1992} we can define a limiting non-thermal radio-luminosity above which a source can be classified as an AGN:
\begin{equation}
\left (\frac{L_{N}}{W\cdot Hz^{-1}}\right) \sim 5.3\cdot 10^{21}\left (\frac{\nu}{GHz}\right )^{-\alpha}\left [ \frac{SFR(M\geqslant5M_{\odot})}{M_{\odot}yr^{-1}}\right ],
\end{equation}
where $L_{N}$ is the non-thermal luminosity produced from ongoing star-formation with a star formation rate SFR, at an observing frequency $\nu$ and with a radio spectral index $\alpha=0.8$. We set a threshold of 100 solar masses per year, which gives a critical luminosity of $L_{lim}=7.24\cdot 10^{30}$erg/s/Hz. This corresponds to a radio power of $\simeq10^{24}$W/Hz and roughly matches the turn-over power of the local luminosity function of radio sources, above which radio-loud AGN dominate the radio-source population (\citealt{Mauch2007}).  After K-correcting the 1.4GHz luminosities of the VLA sources, those objects more luminous than the above limit are assumed to be AGN. Finally we also make a division using the spectral index information, choosing $\alpha=-0.5$\footnote{Here we use the convention $F\propto\nu^{\alpha}$.} as the limit between flat and steep-spectrum radio sources. In total we classify 321 AGN through their flat radio spectral index, while 497 AGN are classified through their 1.4GHz luminosity.

\subsubsection{Spitzer Space Telescope}
The Spitzer Wide-area InfraRed Extragalactic (SWIRE; \citealt{Lonsdale2003}) survey is one of the Legacy Surveys of the Spitzer Space Telescope using both IRAC (3.6-8$\mu$m) and MIPS (24-160$\mu$m) instruments. The SWIRE survey achieved a sensitivity of 450 $\mu$Jy (5$\sigma$ detections at 24$\mu$m), with the master catalogue of the survey (merged bands) containing 250733 individual sources. For our study, we require that sources are detected at 24$\mu$m, where emission should be dominated by the dusty AGN torus (although contribution from star-formation is also expected). We find 2329 in the common VIDEO-SWIRE area.

\begin{figure}
\begin{center}
\includegraphics[width=0.5\textwidth,angle=0]{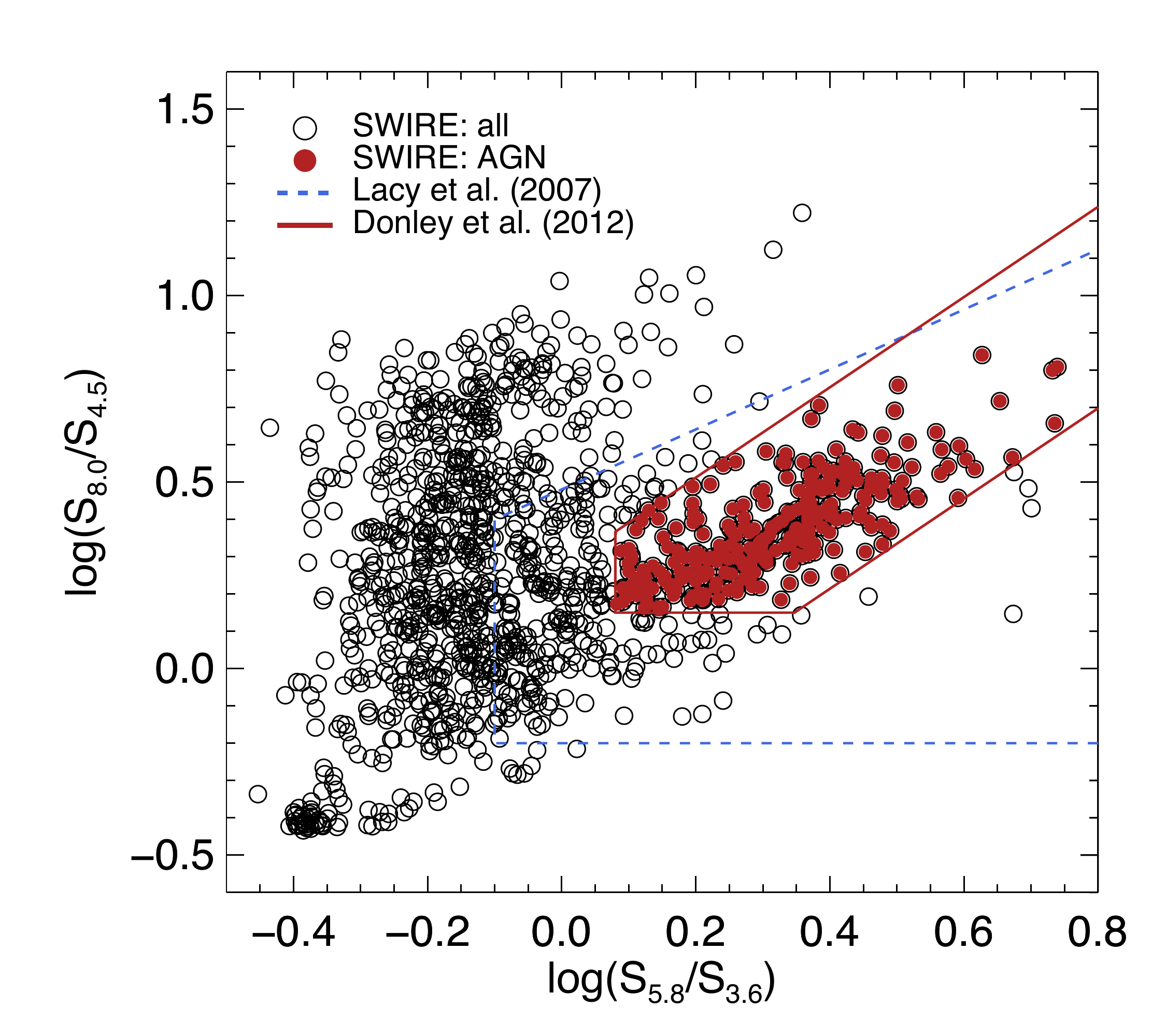}
\caption{{IRAC colour-colour diagram for the parent sample of 24$\mu$m-selected sources (empty black circles) and the selected dust-obscured AGN (filled red circles). The colour cuts from \citealp{Lacy2004,Lacy2007} (dashed blue) and \citet{Donley2012} (solid red) are shown as well.}}
\label{fig:IRAC_color}
\end{center}
\end{figure}

{Using mid-IR colors of the selected sources, we can make further selections. \citet{Donley2012} updated the empirical mid-IR colour cuts, originally derived by\citep{Lacy2004,Lacy2007}, to select dust-obscured AGN. We use these cuts, selecting 269 dust-obscured AGN candidates. We show the IRAC colour-colour distribution of the parent sample and the selected AGN in Fig. \ref{fig:IRAC_color}.} 

\section{Tools and analysis}
\label{sec:analysis}
We use a number of selection criteria at different wavelengths in an effort to create samples of active galaxies representative of different intrinsic properties and potentially of different evolutionary stages. The first step in doing this is the cross-matching of the multi-wavelength catalogues available for the XMM-LSS field with the catalogue of the VIDEO survey. To do this we follow the method of \citet{Downes1986}, taking into account both the angular distance to the source and the near-IR magnitude of the counterpart. In short, around each multi-wavelength source (i.e., X-ray, radio, or mid-IR)  we calculate the Poisson probability for each near-IR source to be within a circle of radius $r_{c}$. This Poisson probability is defined as:
\begin{equation}
P^{*}=1-e^{-\pi r^{2}N_{m}},
\end{equation}
where r is the distance of the candidate counterpart from the multi-wavelength source, and $N_{m}$ is the surface number density within a radius r and limiting near-IR magnitude m. $r_{c}$ is defined through the positional uncertainties of the VIDEO survey (assumed here $\sigma_{VIDEO}$=0.2 arcsec) and the survey of each respective wavelength (X-ray, $\sigma_{XMM}$=2 arcsec; radio, $\sigma_{VLA}$=2 arcsec deg; mid-IR, $\sigma_{Spitzer}$=2 arcsec\footnote{This is the positional uncertainty at 3.6 $\mu$ and should reflect the accuracy of the SWIRE band-merged catalogue.}). The expected number of events (i.e., near-IR sources) with $P\leqslant P^{*}$ can then be approximated (for a finite search radius $r_{c}$) as 
\begin{equation}E=P_{c}=\pi r_{c}N_{T},
\end{equation} 
for $P^{*}\geqslant P_{c}$, and 
\begin{equation}E=P^{*}(1+\ln{P_{c}/P^{*}}),
\end{equation}
for $P^{*}<P_{c}$. $P_{c}$ is a critical Poisson probability, defined by the surface number density, $N_{T}$, at the limiting magnitude of the survey.
Finally, the probability of a chance cross-identification of the source can be calculated as $1-e^{-E}$. The near-IR candidate with the lowest such probability is chosen to be the true counterpart\footnote{For the case of XMM-Newton sources, 20 pairs of X-ray sources are identified with the same VIDEO source. For these cases, for each pair of XMM-Newton sources assigned to the same VIDEO source, we keep the identification with the lowest $1-e^{-E}$ value, while for the other one we adopt the identification with the second lowest such value. For one pairs this was not possible and thus only one of the sources was retained. Similarly, this procedure was followed for 2 pairs of SWIRE sources.}. In Table \ref{tab:crossID} we give the number of cross-identified sources for each of the multi-wavelength catalogues considered here.
\begin{table*}
\caption{Information concerning the multi-wavelength data used in this paper. The wavelength (Col. 1), telescope (Col. 2), flux limit (Col. 3), and total (Col.4) and cross-identified (Col. 5) number of sources in each catalogue are given.}
\begin{center}
\begin{tabular}{c c c c c}
\hline
Regime	&Instrument	&$f_{lim}$		&Number		&Cross-ID\\
\hline
0.5-2.0 keV	&XMM-Newton	&$4\cdot10^{-15}$ erg/s/cm$^{2}$	&1049	&709\\
\hline
\multicolumn{1}{c}{} &\multicolumn{2}{|c|}{QSO-like (K/X$<$2)}	&\multicolumn{2}{|c|}{153}\\
\multicolumn{1}{c}{} &\multicolumn{2}{|c|}{Seyfert-like (K/X$>$2)}	&\multicolumn{2}{|c|}{178}\\
\hline
1.4 GHz		&VLA		&60 $\mu$Jy	&1054		&1041\\
\hline
\multicolumn{1}{c}{} &\multicolumn{2}{|c|}{AGN ($L_{1.4GHz}>L_{N}$)}	&\multicolumn{2}{|c|}{497}\\
\multicolumn{1}{c}{} &\multicolumn{2}{|c|}{AGN ($\alpha>-0.5$)}			&\multicolumn{2}{|c|}{321}\\
\hline
3.6-24 $\mu$m	&Spitzer		&280 $\mu$Jy	&2329		&2261\\
\hline
\multicolumn{1}{c}{} &\multicolumn{2}{|c|}{Obscured AGN}		&\multicolumn{2}{|c|}{580}\\
\end{tabular}
\end{center}
\label{tab:crossID}
\end{table*}%
\subsection{Environment density parameters}
We use a pseudo-3D number density parameter (being one of the simplest and most straightforward measures of density, widely used in the literature, e.g., \citealt{Cooper2005}, \citealt{Strand2008}, \citealt{Lee2010}) to study the environment of AGN in the near-IR. To do this, we employ the photometric redshifts available for the VIDEO sources. For each VIDEO source that is cross-identified with a source in another wavelength, we calculate the surface number density $\rho$ (measured in kpc$^{-2}$) within cylindric annuli with depth $\Delta z=0.2\cdot (1+z)$ and fixed surface area of 12000$\pi$ kpc$^{2}$\footnote{Although the choice of the area of the annulus is somewhat arbitrary, it is optimized with two factors in mind. We wanted the first annulus to probe the close environment of the AGN, i.e., $\sim200$ kpc. At the same time, given the statistical fluctuations and the finite number of sources per redshift slice, we wanted to have enough sources within each annulus, so that a robust estimation of the density can be achieved.},
\begin{equation}
\rho =\frac{N_{r}^{\Delta z}}{\pi r^{2}}.
\end{equation}
The widths of the annuli ($r_{out}-r_{in}$) range from 18 to 194 kpc, and distances to the central source ranging from 0.1 to $\sim1$ Mpc.

For the redshift range of the VIDEO sample this corresponds to angular distances out to $\sim 2 $arcmin. In this way, we study the close environment of AGN (checking for companion galaxies), while simultaneously probing the medium-scale environment. It is known (e.g., \citealt{Cooper2005}) that edge-effects influence environment density parameters. For each radius $r_{i}$ of the cylinder within which the density is measured, we correct for this effect by excluding all sources that are found at distances $\leqslant r_{i}$ from the edges of the VIDEO field.\\
We also employ the projected distance to the \textit{n}th closest neighbor, again within a redshift interval of $\Delta z=0.2\cdot (1+z)$, in order to calculate the surface density parameter $\Sigma_{n}$ (originally described by \citealt{Dressler1980}),
\begin{equation}
\Sigma_{n}=\frac{n}{\pi d_{n}^{2}}.
\end{equation}
We choose to investigate the surface density using both n=2 and n=5, as we are interested in the close environment of the sources. Following \citet{Cooper2005}, to minimize the contamination from edge effects, we exclude sources at a distance of 2 arcmins from the field edges. For the calculation of both density parameters and projected distance to the \textit{n}th closest neighbor we consider the total, band-merged, VIDEO sample, i.e., all sources detected in at least one VIDEO band, excluding only sources with stellar like colours.
 
\subsection{Control samples}
\subsubsection{Stellar mass matching}
We carefully select control samples for each of the multi-wavelength samples studied here. In particular, we want to account for possible dependence of the environment on the luminosity, stellar mass, and redshift of an object. We choose the following procedure to select our control sources: for each multi-wavelength source, we randomly select 20 VIDEO-sources that are matched in both redshift ($\Delta z=0.2\cdot(1+z)$, accounting for the increasing photometric redshift uncertainties with redshift) and $K_{s}$ magnitude ($\Delta m_{K}=0.2$). By means of these two criteria, we select sources of similar $K_{s}$ luminosity, as well as similar host galaxy stellar mass (e.g., \citealt{Gavazzi1996}). Moreover, we require that the control sources are at projected distances from the multi-wavelength source larger than 60 arcsecs. This ensures that the control sources are unaffected by any local over-densities around the AGN source. Control sources are selected from the same sample that is used for the cross-identification of the multi-wavelength sources and which was described in Sect. \ref{sec:survey} above, after excluding the NIR sources that have been cross-matched with any of our multi-wavelength samples.

Because of the $K_{s}$ magnitude-based selection of both our main sample and means of choosing the control samples, we do not consider stellar-like sources from the VIDEO catalogue. As was discussed previously, this is done by means of optical and near-IR colors. We also exclude known optical quasars in the VIDEO field. The latter have near-IR emission that is dominated by the luminous AGN at their centers. Consequently, when using the $K_{s}$ magnitude as a proxy for stellar mass, we would be biased toward selecting more massive sources to build the control sample. Given a dependency of the environment density on the host stellar mass (e.g., \citealt{Deng2011}, \citealt{Li2006}, \citealt{Kauffmann2004}), these sources would appear to inhabit under-dense environments due to this selection effect. Finally, we check that our control sources are not found within the vicinity of the handful of very bright stars in the VIDEO field, as these areas suffer from lower S/N and as a result might show artificially low source densities.

For each of the control sources, the same environment density parameters are calculated as for the multi-wavelength sources. These parameters are then averaged over all control sources for each multi-wavelength source, resulting in a control value for each environment density parameter for each AGN. We can then define a density difference or ratio between the AGN-candidates and their control values. In this way, for ratios larger than 1 and differences larger than zero, the AGN sample is in denser environments than its control sample sources. 

\subsubsection{Random control}
In addition to the K$_s$-band magnitude and redshift-matched control sample, we also select a random control sample. For each multi-wavelength source we select 40 random positions within the VIDEO-CFHTLS field. We do not require the coincidence of a VIDEO source. Therefore, the random sample is decoupled from the presence of a source. For each random position and for the respective redshift slice of the multi-wavelength source, density parameters are calculated similarly to the process described previously. The average over all the random position reflects then the average density of the field and provides a further means of comparison to the AGN samples.

\section{Results}
\label{sec:results}
\begin{figure}
\begin{center}
\includegraphics[width=0.5\textwidth,angle=0]{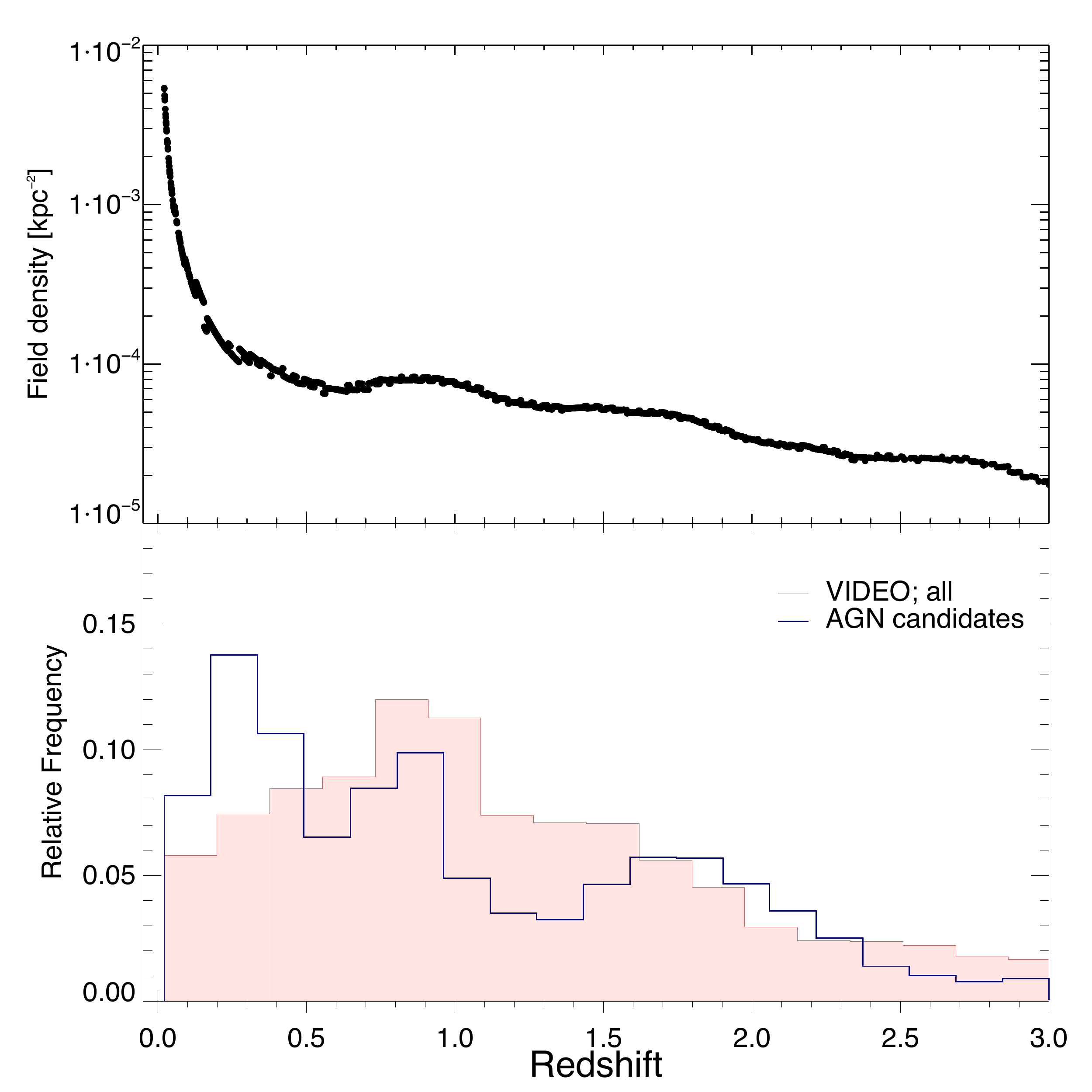}
\caption{({\it top}) VIDEO-CFHTLS field density as a function of redshift.  ({\it bottom}) Normalized histograms of the photometric redshift distribution for the total VIDEO sample (filled red) and of the AGN candidate sample (open blue).}
\label{fig:field}
\end{center}
\end{figure}

We first turn our attention to the field density of the VIDEO-CFHTLS as described in the previous section. In Fig. \ref{fig:field} the field number density is plotted as a function of redshift. In addition the photometric redshift distributions of both the AGN candidate sample (here defined as the combination of all three multi-wavelength samples) and that of the total VIDEO sample (which is used for the calculation of densities) out to $z=3$. From the upper panel of Fig. \ref{fig:field} we can see that the field density is smoothly decreasing with redshift as expected. The lower panel shows that most of our AGN sources are found at $z<3$ and mainly found at $z<1$, with a secondary concentration at $z\sim 1.8$. Compared to the photometric redshift distribution of the total sample we can see that even at $z\sim 2$, there are enough sources for the density of the small-scale environment to be constrained.

\subsection{X-ray AGN environment}

\begin{figure}
\begin{center}
\includegraphics[width=0.5\textwidth,angle=0]{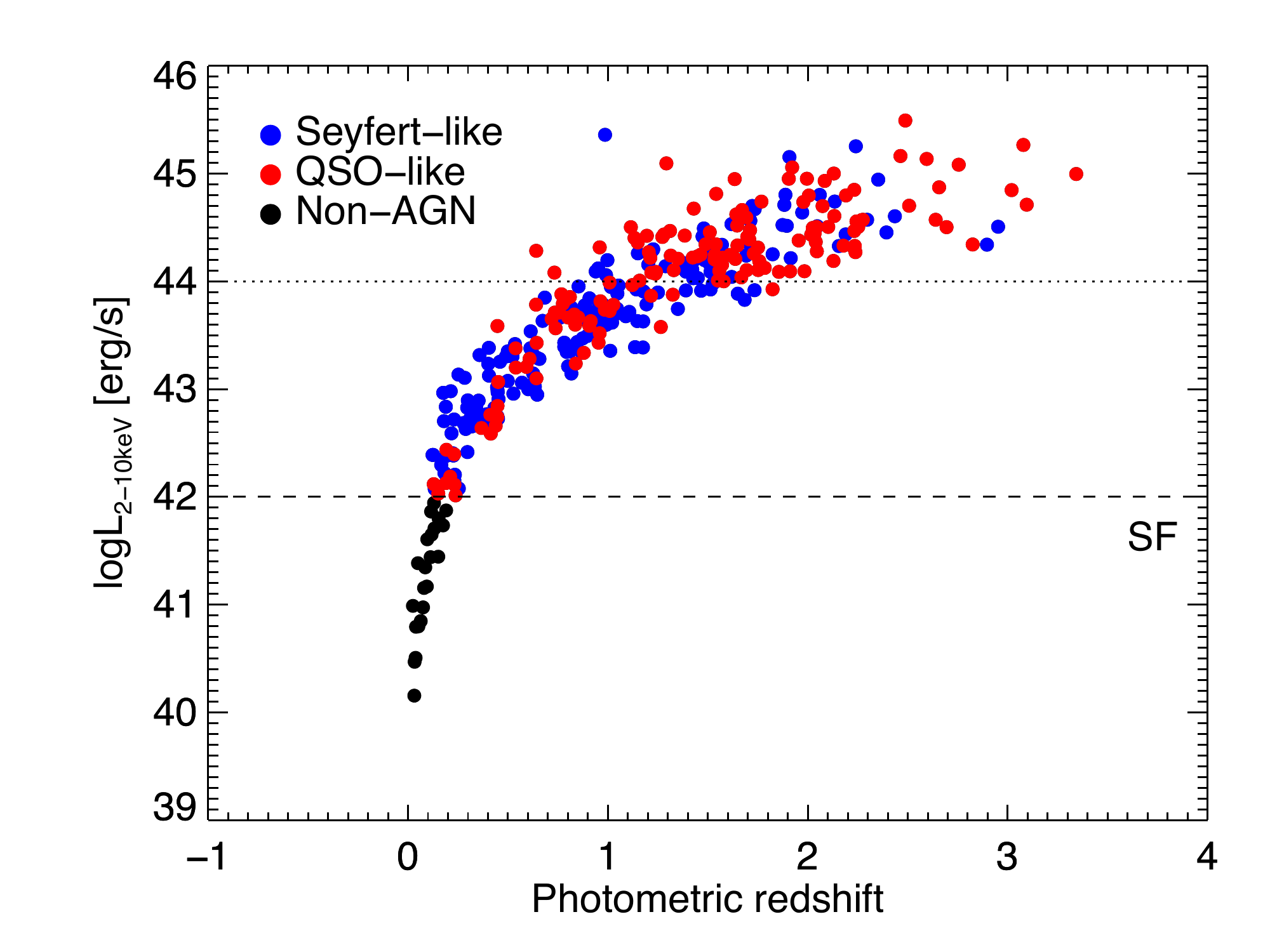}
\caption{{Hard (2-10keV) X-ray luminosity distribution over our photometric redshift range for all hard X-ray detected sources. Seyfert-like (blue), QSO-like (red), and non-AGN (black) sources are shown separately. The dotted line shows the limit above which traditionally X-ray bright QSOs are found. The dashed line shows the lower limit required in our selection for a source to be considered as an AGN.}}
\label{fig:X_zlum}
\end{center}
\end{figure}

{In Fig. \ref{fig:X_zlum} we present the 2-10 keV hard X-ray luminosity distribution for the X-ray cross-matched sample. As expected for a flux-limited survey, the luminosity is tightly correlated with the redshift and we show QSO-like and Seyfert-like sources separately (red and blue symbols, respectively). The bulk of Seyfert-like sources are within an X-ray luminosity of $10^{42}<L_{2-10keV}<10^{44}$ erg/s, although a few sources with luminosities typical of powerful QSOs are classified as Seyfert-like according to our K/X criterion. Conversely. QSO-like sources spill over the $10^{44}$ erg/s limit, covering the whole range of hard X-ray luminosities considered here. In total, 22 sources have luminosities $<10^{42}$ erg/s and are not considered in our AGN selection. Most of them are around or below $z=0.1$.}

\begin{figure}
\begin{center}
\includegraphics[width=0.5\textwidth,angle=0]{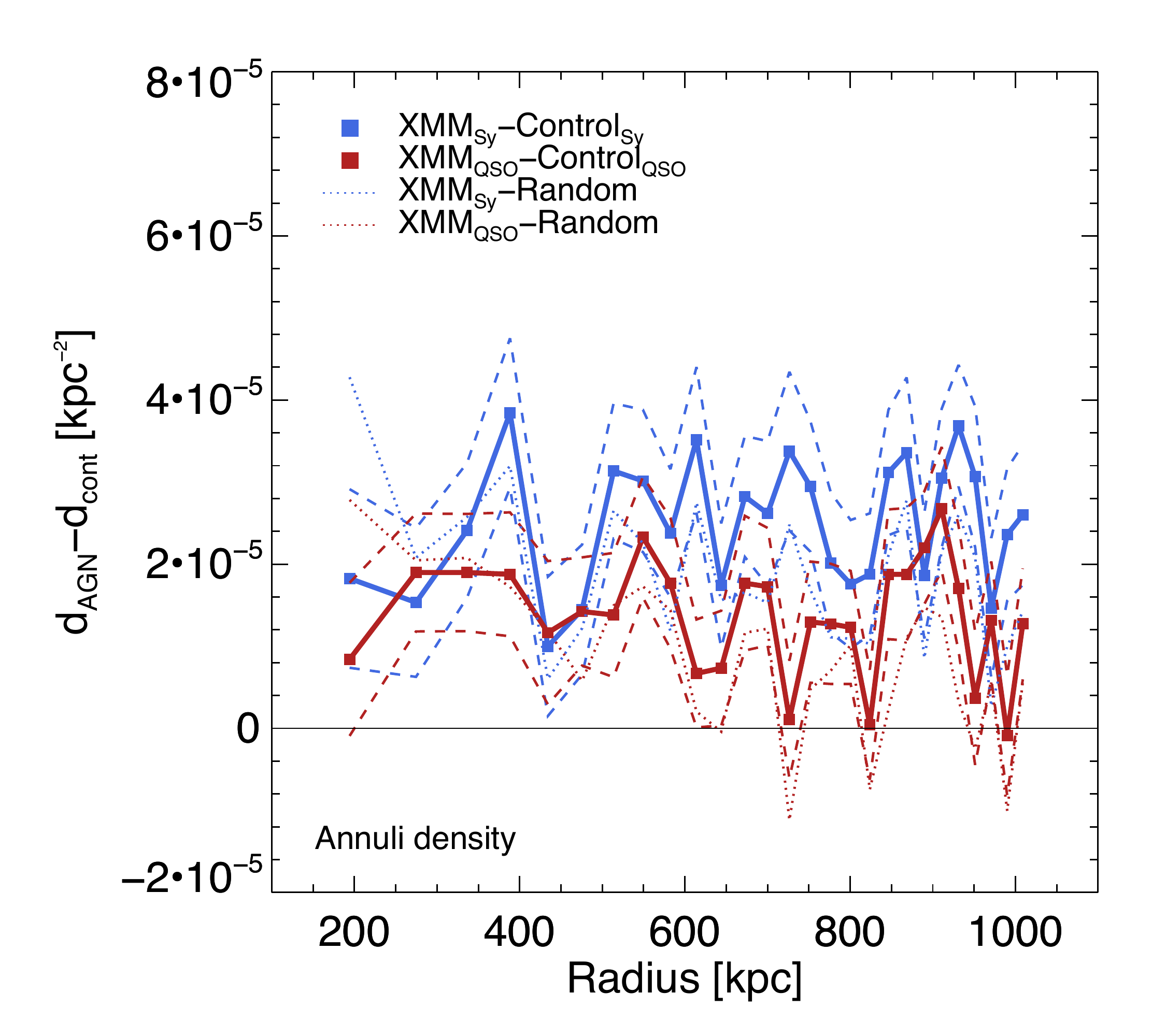}
\caption{Averaged surface number density difference, in annuli, of QSO-like sources (solid red line), and Seyfert-like sources (solid blue line) with their control samples, as a function of  the outer radius of the annulus within which the density has been calculated. In addition the density difference between the respective X-ray samples and the random sample is also shown with dotted lines. The line for a density difference of zero is also shown. Uncertainties (1$\sigma$) are shown as loci in dashed lines.}
\label{fig:X_surfacedens}
\end{center}
\end{figure}

In Fig. \ref{fig:X_surfacedens} we show the average surface density difference, calculated in annuli, between the X-ray samples and their respective control samples, as a function of the distance from a source. We also show the density difference of the X-ray samples and the random sample. Given the relatively small surface area of the annuli, we observe strong fluctuations of the density difference as a function of radius. {However, it is clear that Seyfert-like sources (shown in blue) reside in denser environments  compared to their control sample, an average of 30 more sources  per Mpc$^{2}$. QSO-like sources show a moderate over-density out to a radius of 500 kpc, with an average of 10 sources more per Mpc$^{2}$ compared to their control sample. Beyond a radius of 600 kpc, QSO-like sources inhabit environments which are consistent with those of their control sample.}

The dotted lines also show the respective density differences between the AGN samples and the random sample. We can see that for the first few bins, these are found to lie significantly above the solid lines for both QSO-like and Seyfert-like AGN, meaning that compared to the average field density, X-ray AGN show more over-dense environments compared to the field than compared to their control sources. This is expected in the sense that environment density around a source should be higher than around an empty field, hence a stronger over-density compared to the random positions than the control sources. These lines also demonstrate that, in absolute sense, Seyfert-like AGN reside in similar environments to the QSO-like sample (i.e. the comparison with random positions are similar for both QSO-like and Seyfert-like).

\begin{figure}
\begin{center}
\includegraphics[width=0.5\textwidth,angle=0]{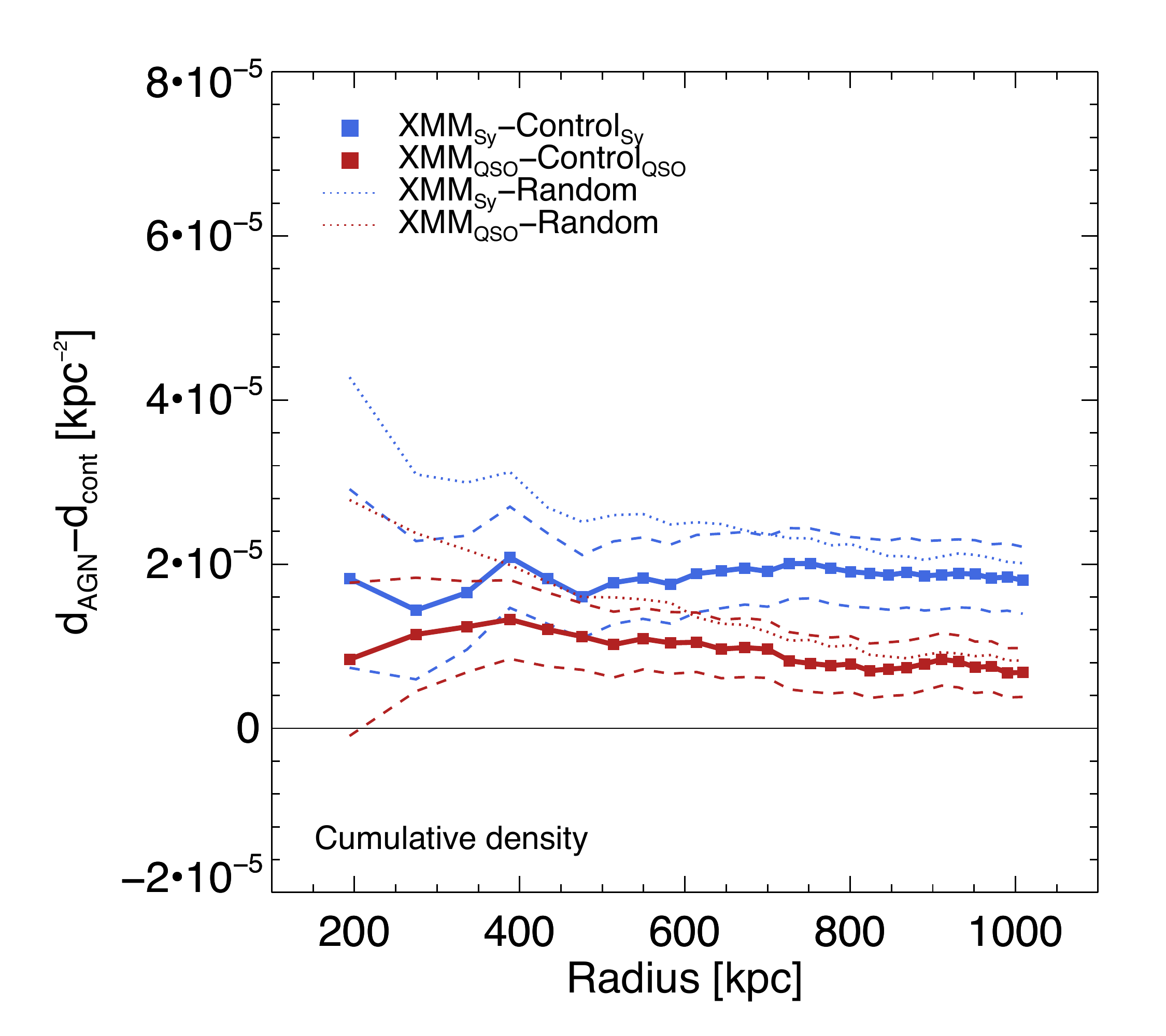}
\caption{Averaged surface number density difference, in circles, of QSO-like sources (solid red line), and Seyfert-like sources (solid blue line) with their control samples, as a function of  the outer radius of the circle within which the density has been calculated. In addition the density difference between the respective X-ray samples and the random sample is also shown with dotted lines. The line for a density difference of zero is also shown. The left plot shows the calculated density differences for each bin, while the right plot uses a radial binning of 20 kpc to smooth the curves. Uncertainties (1$\sigma$) are shown as loci in dashed lines.}
\label{fig:X_surfacecumdens}
\end{center}
\end{figure}

In Fig. \ref{fig:X_surfacecumdens} we  again show the surface number density differences for the different X-ray samples, but this time with densities calculated within circles rather than annuli. We can directly compare our surface density measures with, for example, the density parameters calculated through the closest neighbours projected distances. {As seen in Fig. \ref{fig:X_surfacedens}, Seyfert-like sources show significant over-densities, with  now smaller uncertainties given the increasing  surface area with increasing radius. QSO-like sources show similar over-densities (within the uncertainties) with Seyfert-like sources, out to a radius of 600 kpc. Their behaviour appears to diverge beyond that radius, with QSO-like sources converging towards the density of their control sample. }The dotted curves show the density difference between the AGN samples and the random sample. These appear consistently above the solid lines, as expected. The over-density persists for Seyfert-like sources out to distance of 1~Mpc from the central AGN. As a consistency check, we have also looked at the density difference out to distances of 3 Mpc. Both QSO-like and Seyfert-like samples converge to a density difference of zero as expected.

\begin{figure*}
\begin{center}
\includegraphics[width=0.33\textwidth,angle=0]{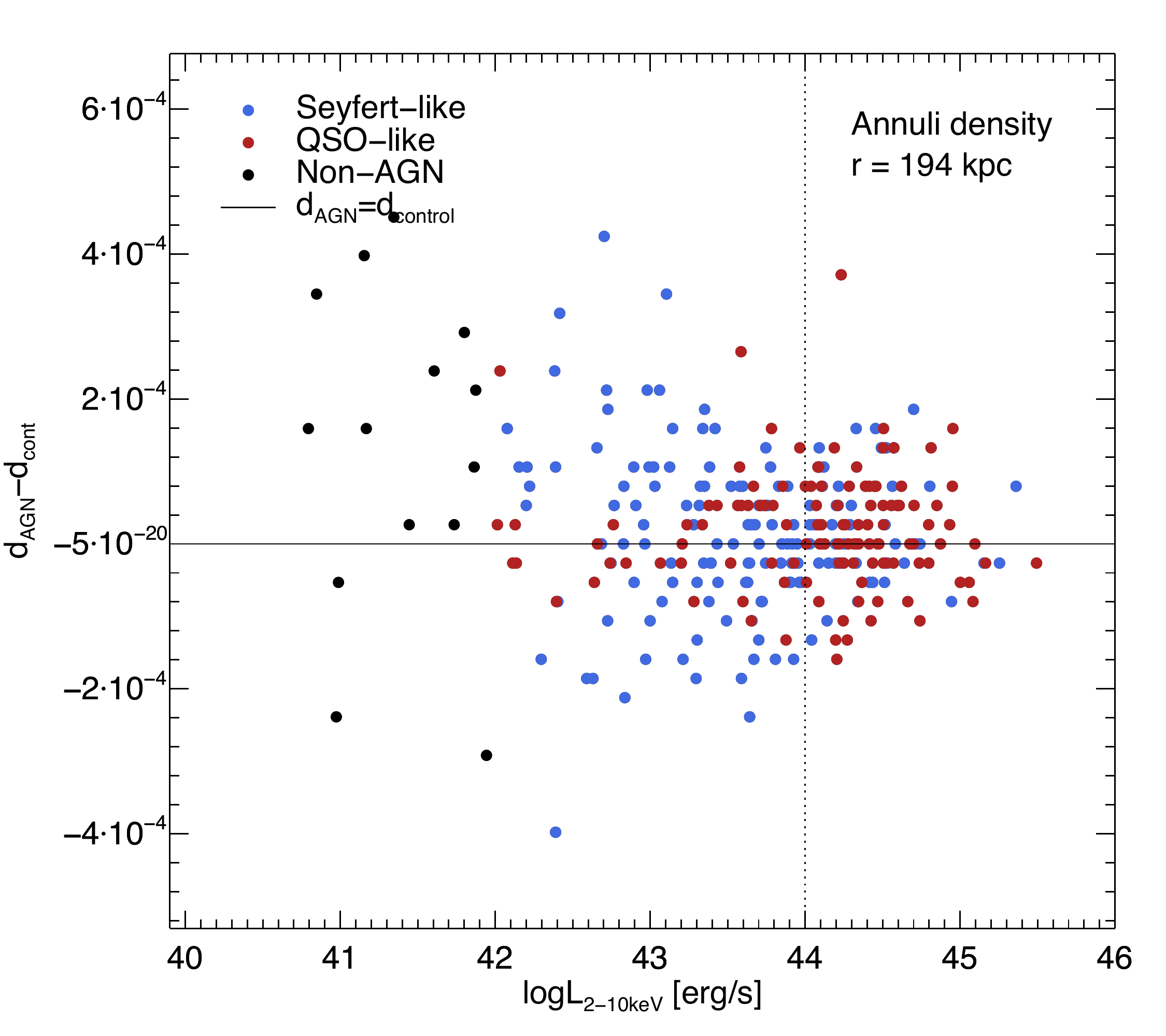}
\includegraphics[width=0.33\textwidth,angle=0]{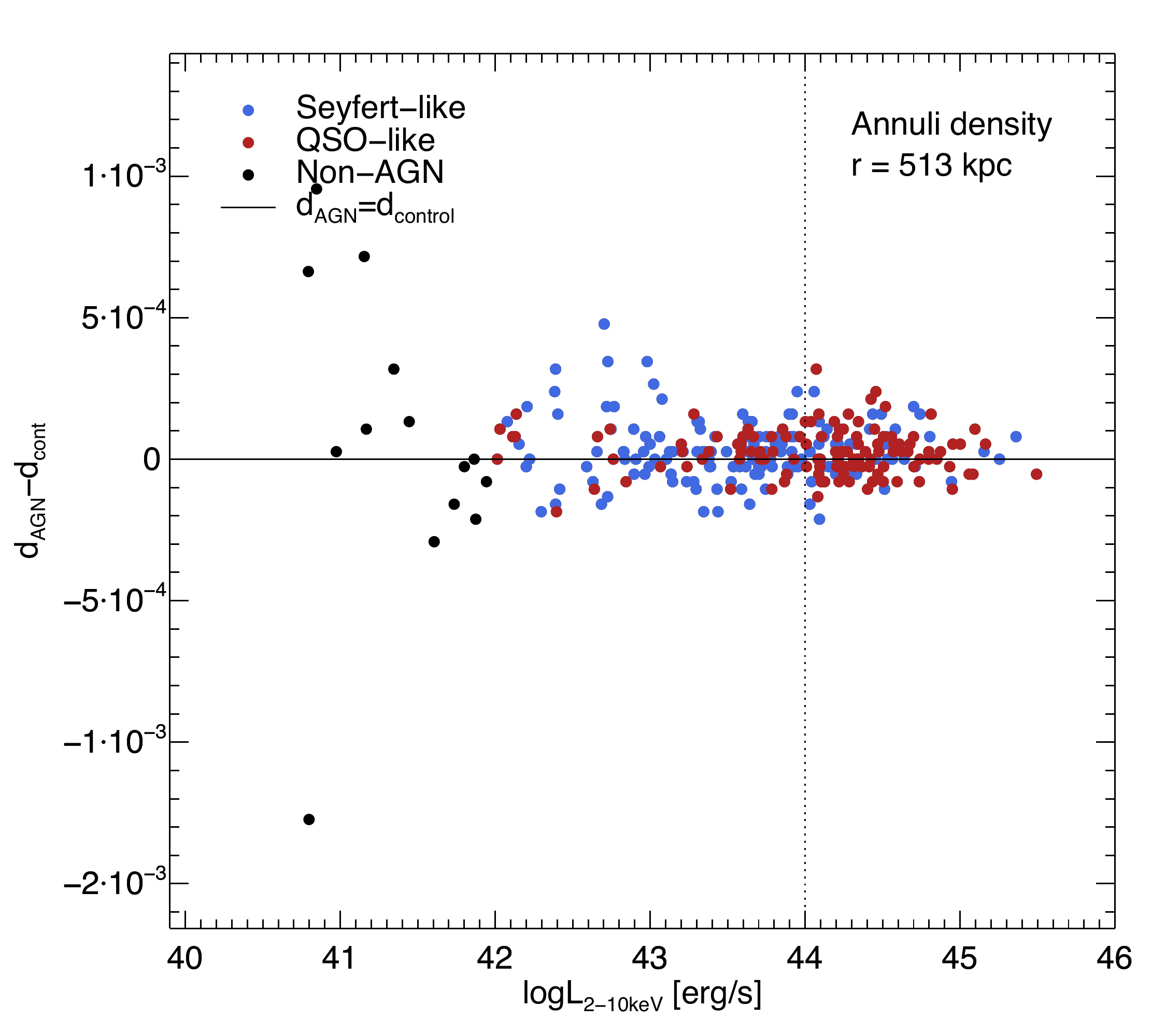}
\includegraphics[width=0.33\textwidth,angle=0]{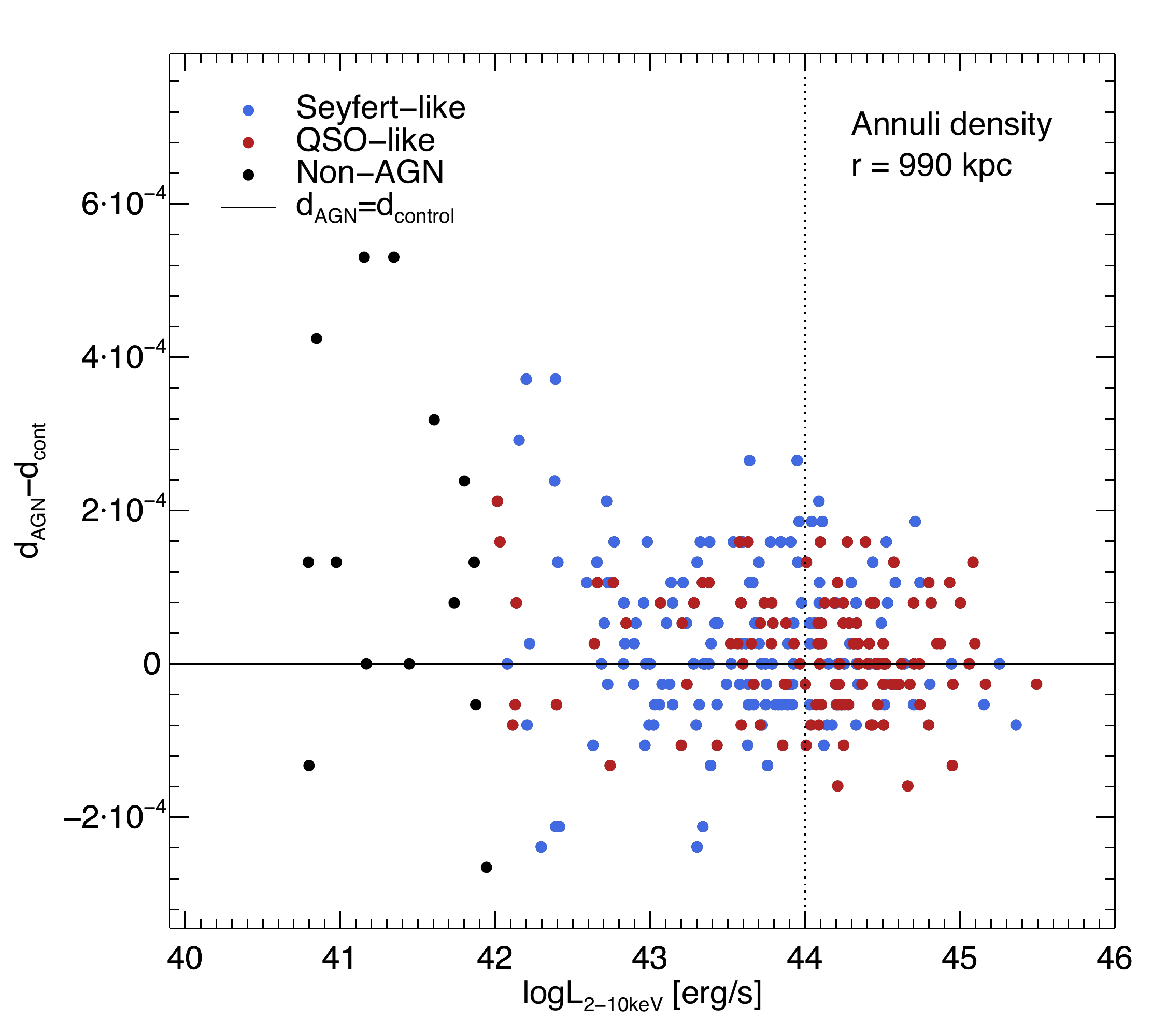}
\caption{Surface number density difference, in annuli, of all hard X-ray sources, divided in Seyfert-like (blue) and QSO-like X-ray AGN (red), and non-AGN sources ($L_{2-10keV}<10^{42}$; black), with their respective control samples, as a function of their 2-10 keV hard X-ray luminosity. This is shown for three different radial distances from the central source, $\sim200$ kpc (left), $\sim500$ kpc (middle), and $\sim1000$ kpc (right). The vertical dashed line denotes the luminosity limit above which typically powerful quasars are found.}
\label{fig:X_dif_lum}
\end{center}
\end{figure*}

In Fig. \ref{fig:X_dif_lum} we plot the density difference between X-ray AGN and their control samples as a function of their hard X-ray (2-10 keV) luminosity. Independent of the scale at which this is probed, there appears to be no correlation between the two properties, with points clustered homogeneously slightly above zero independent of X-ray luminosity. Moreover, different sub-samples show the same behaviour, with Seyfert-like sources being on average at higher positive values (as was already shown in Fig. \ref{fig:X_surfacedens}). We also see no particular trend for X-ray sources above the 10$^{44}$ erg/s limit, usually considered a boundary for the most luminous quasars. For the lower luminosity sources (L$_{2-10keV}<10^{42}$) we see a higher scatter of values. {Conversely, sources below $10^{42}$ erg/s (shown in black) appear to show the highest over-densities, especially at intermediate and large radii. These sources might represent starburst whose intense SF is triggered by ongoing mergers and hence are found in the densest environments.}

\begin{figure}
\begin{center}
\includegraphics[width=0.5\textwidth,angle=0]{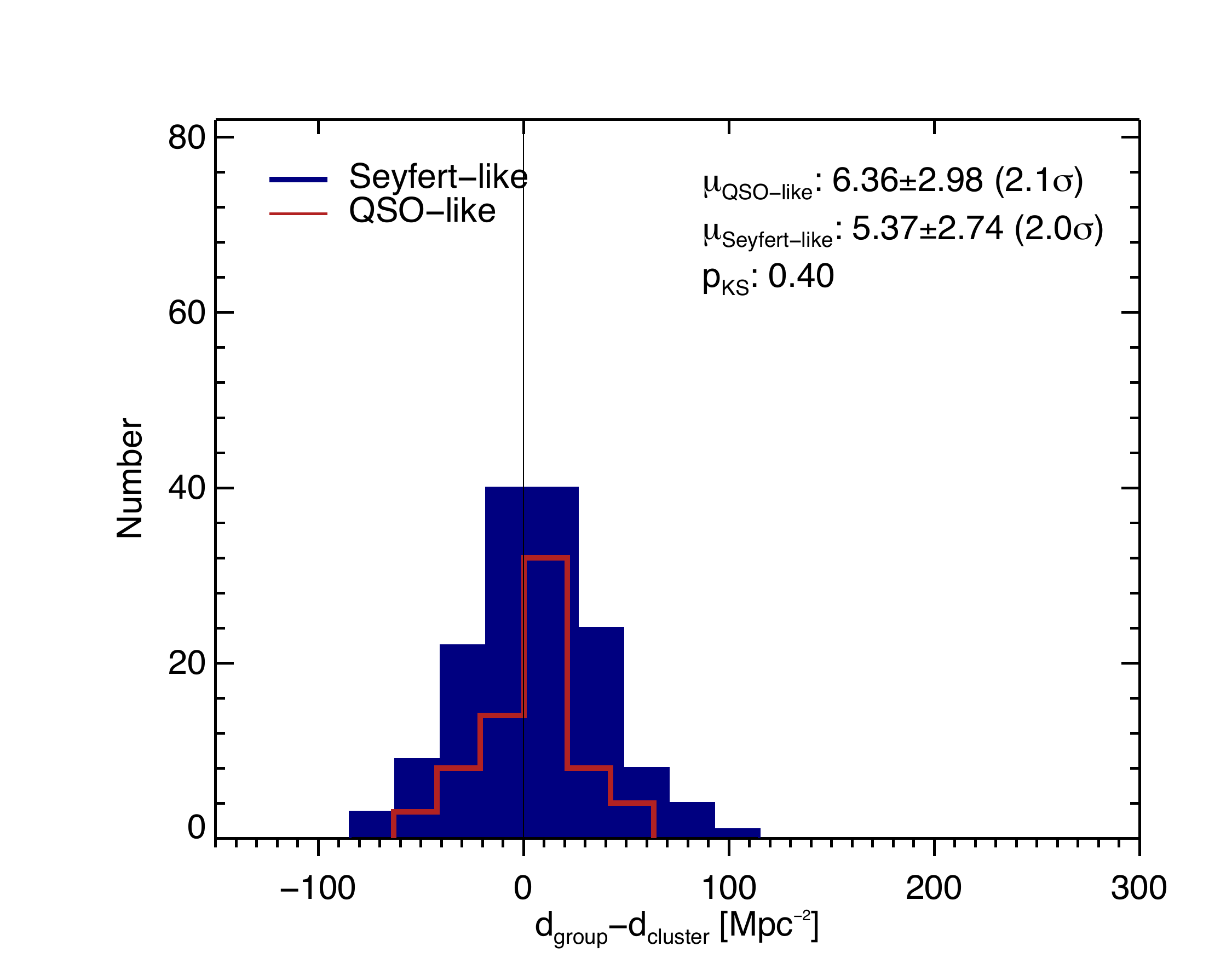}
\caption{Histogram of the distributions for QSO-like (red empty histogram) and Seyfert-like (blue filled histogram) X-ray AGN for the difference between densities calculated at scales of 200 and 800~kpc. This difference reflects and relative importance between small- and large-scale environments. Average values with standard errors for the respective samples are shown. The p-value of a two-sample K-S test for the two samples is also given.}
\label{fig:X_group_cluster}
\end{center}
\end{figure}

We also investigate the relation between the environment at small scales (usually associated with small structures like galaxy groups) and at larger scales (usually associated with large structures like galaxy clusters). This might give us some insight about the processes more relevant to the triggering of an AGN. In particular, it is known that the high velocities encountered in the central regions of clusters are prohibitive to gravitational interactions and mergers, while small groups provide the ideal circumstances for mergers to happen. As such, in Fig. \ref{fig:X_group_cluster} we show the distribution of the difference between the density at 200 and 800 kpc for the X-ray samples. We see that both distributions peak at a value of around zero, implying that neither sample shows any evidence of an enhancement in the source density on small scales. When considering average values of the difference, these are positive from which we can infer that the density is larger towards the AGN when compared to the larger scales, as one would expect if the AGN were indeed tracers of proto-clusters at high redshift, but at a low significance ($\sim2\sigma$).

\begin{figure*}
\begin{center}
\includegraphics[width=0.4\textwidth,angle=0]{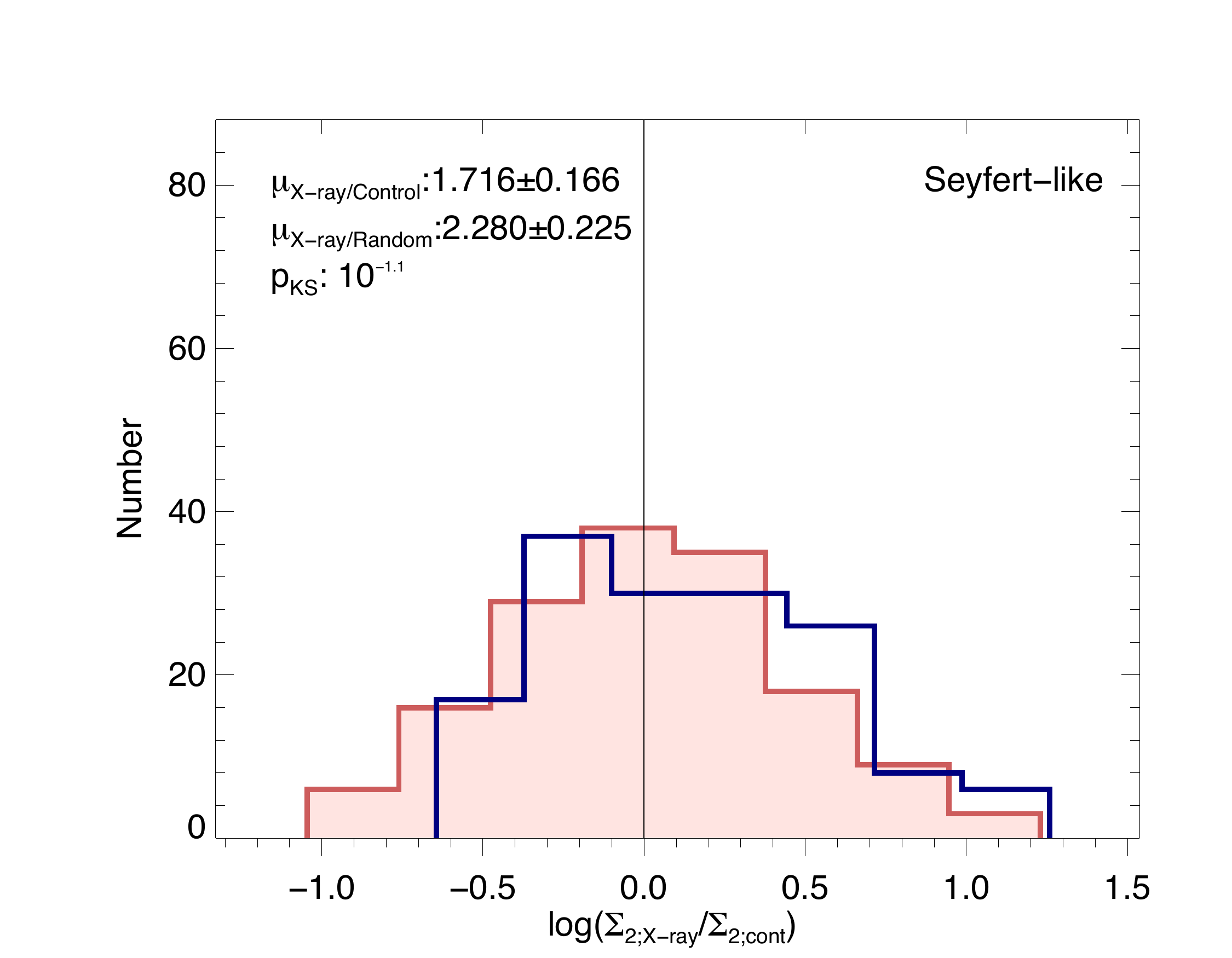}
\includegraphics[width=0.4\textwidth,angle=0]{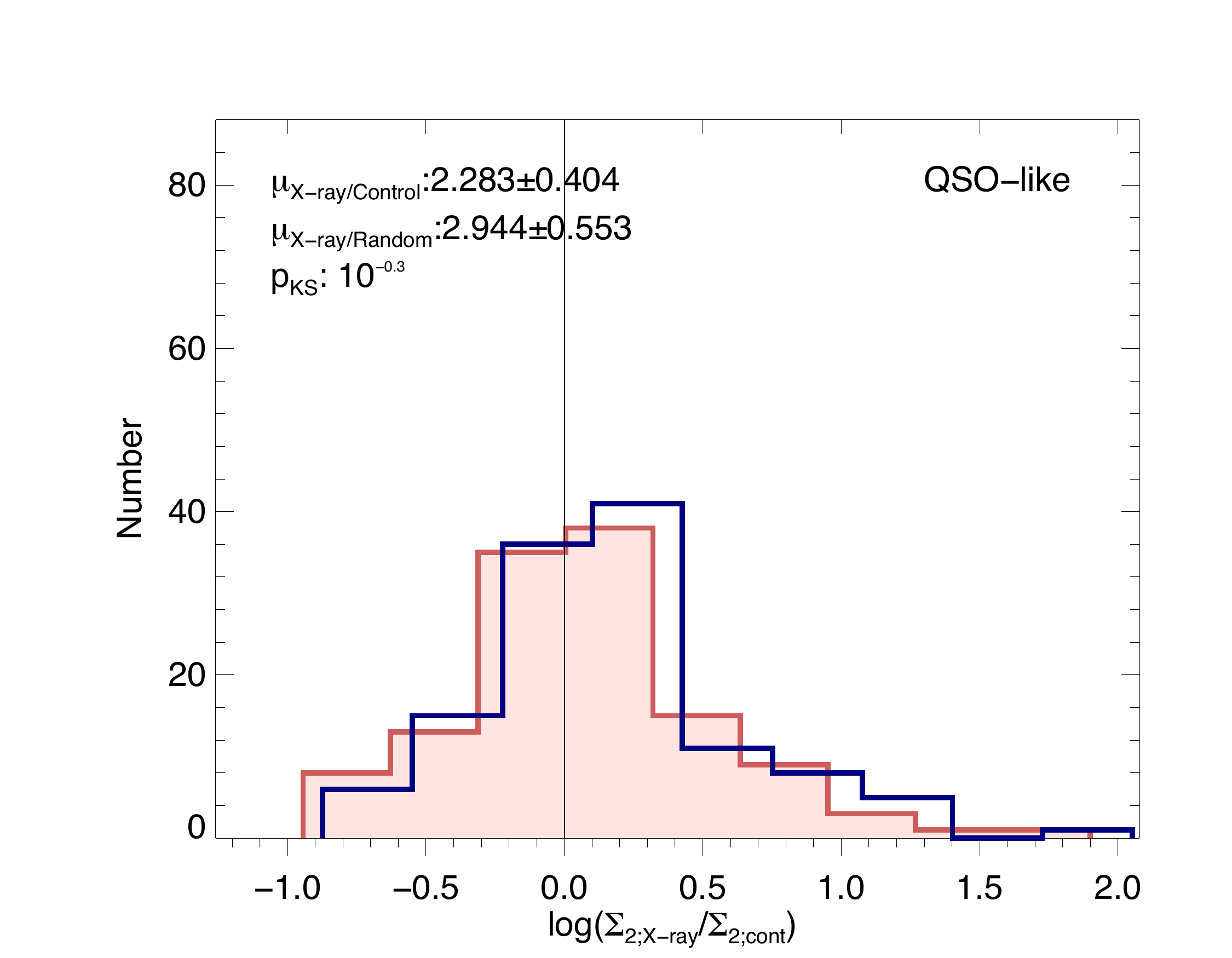}
\includegraphics[width=0.4\textwidth,angle=0]{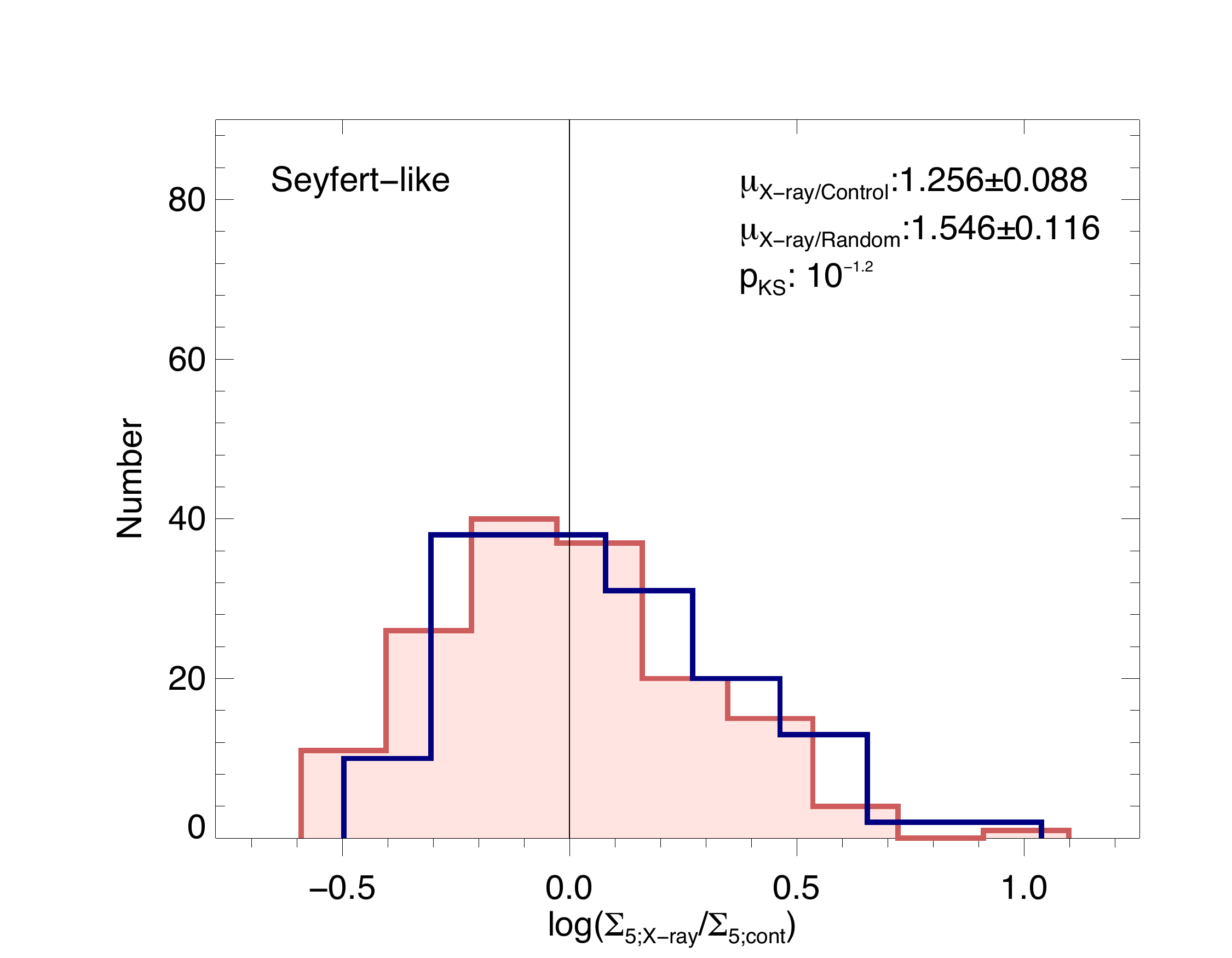}
\includegraphics[width=0.4\textwidth,angle=0]{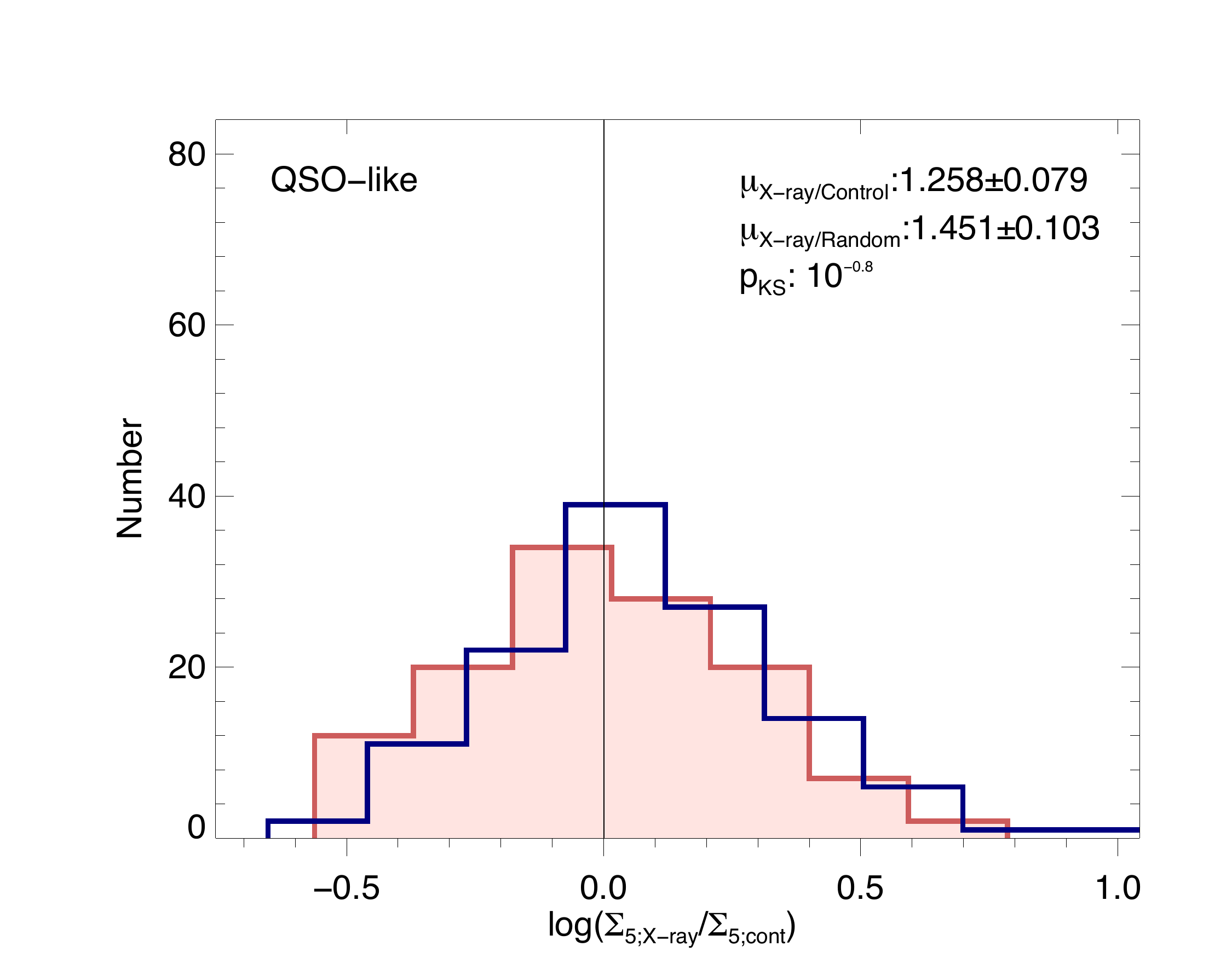}
\caption{The distributions of the logarithm of the ratio between the $\Sigma$ parameters of the  X-ray sources and that of their control sample (filled red) and the random sample (open blue). In addition, the  average values with their respective standard error are given. The probability parameter p from a K-S comparison of the AGN/control and AGN/random density ratio distributions is also given. For reference a vertical line at a ratio value of one is drawn as well. Left panels are for Seyfert-like sources, while the right panels are for the QSO-like ones.}
\label{fig:X_histo_S2S5}
\end{center}
\end{figure*}

We now turn to the second density measure we introduced in the previous section, namely the $\Sigma_{2}$ and $\Sigma_{5}$ density parameters. These are defined in terms of the second and fifth closest neighbour to a source. By definition they are length-scale independent and can only have non-zero values.  In Fig. \ref{fig:X_histo_S2S5} the distributions for the $\Sigma_{main}$ to $\Sigma_{control}$ density ratios are shown (where both the magnitude matched control sample, red histogram, and the random positions control sample, blue histogram, are considered). We show distributions for QSO-like and Seyfert-like X-ray sources separately.  All distributions appear to peak at around a ratio of one, implying that the bulk of the sources in our samples are not in significantly different environments than either their control sources or the field. However, when considering the average values of each sample, we find a significant over-density for both samples and for the $\Sigma_2$. For $\Sigma_5$ the samples still show over-densities, but at a lower significance levels (3$\sigma$ and 5$\sigma$ for the control and random samples, respectively). This in turn indicates the presence of an extended tail of high density sources for both our X-ray AGN samples. 

\begin{figure}
\begin{center}
\includegraphics[width=0.5\textwidth,angle=0]{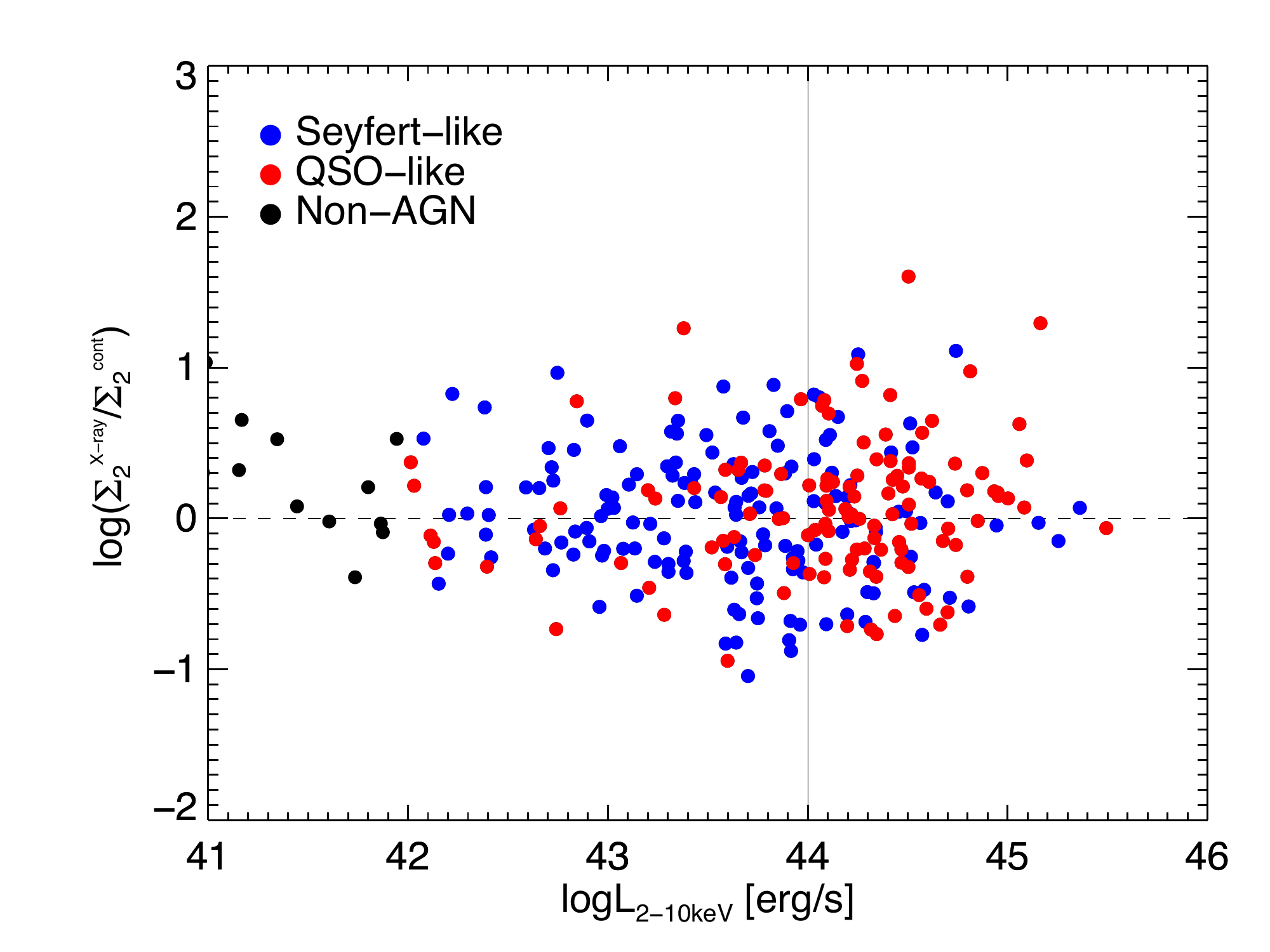}
\caption{The logarithm of the ratio of $\Sigma_{2}$ density parameter between the QSO-like (open red) and Seyfert-like (filled blue) and non-AGN (black) sources and their control samples as a function of their 2-10 keV X-ray luminosity. The zero line denotes the boundary between a source being in an over- or under-dense environment compared to its control sources ($\Sigma^{X-ray}/\Sigma^{control}=1$).}
\label{fig:X_S2_lum}
\end{center}
\end{figure}

\begin{figure}
\begin{center}
\includegraphics[width=0.5\textwidth,angle=0]{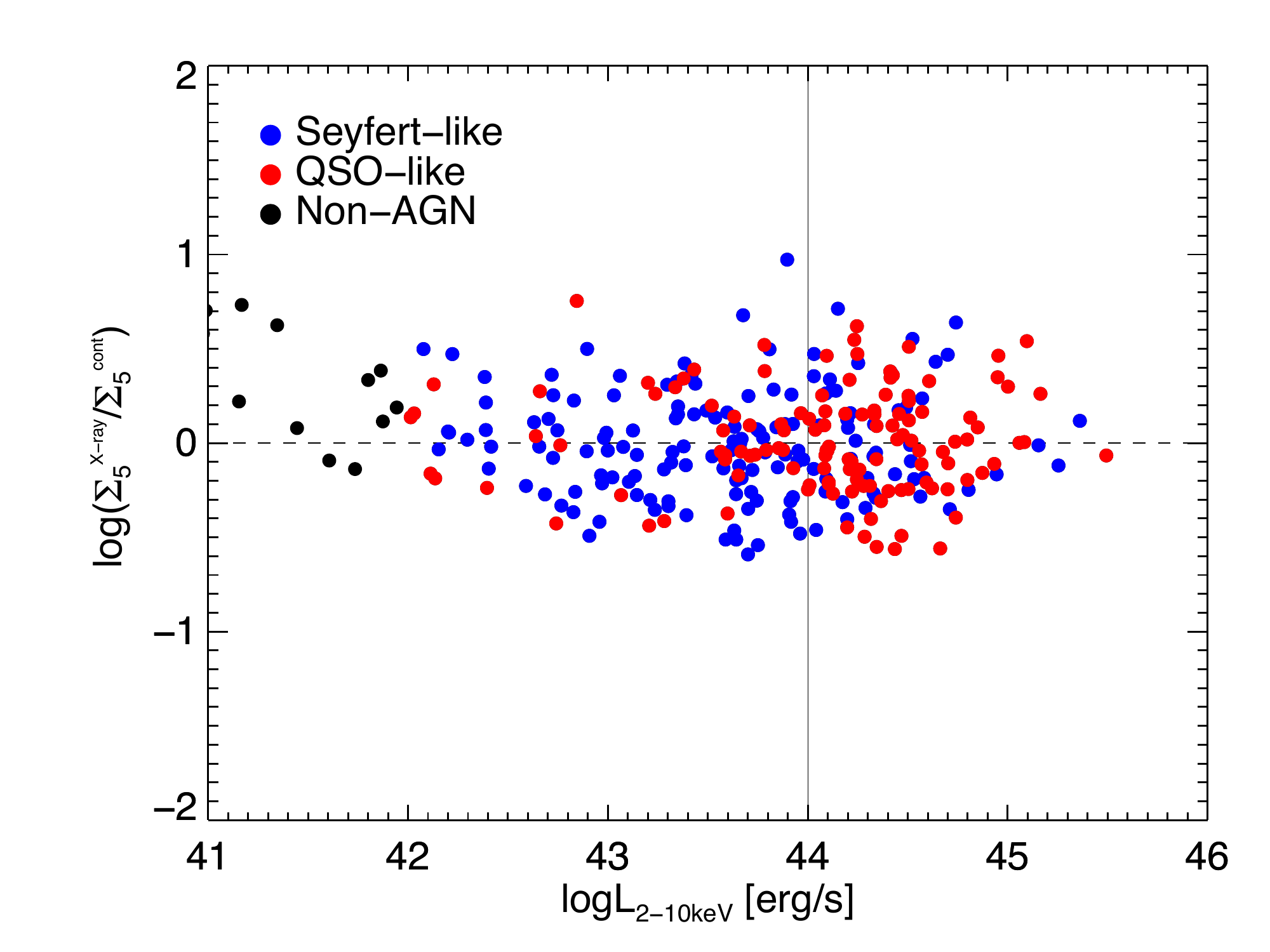}
\caption{As in Fig. \ref{fig:X_S2_lum} but for $\Sigma_{5}$ density parameter.}
\label{fig:X_S5_lum}
\end{center}
\end{figure}

In Figs.~\ref{fig:X_S2_lum} and \ref{fig:X_S5_lum} we investigate  the relationship between the $\Sigma_{main}$ to $\Sigma_{control}$ ratio and the hard X-ray luminosity. For the $\Sigma_{5}$ ratio, most objects have values around one, with Seyfert-like sources showing hints for an anti-correlation. In the low-luminosity regime, there appears to be a surplus of sources in over-dense environments, as was already seen in Fig. \ref{fig:X_dif_lum}. However, small numbers at the low-luminosity end preclude any definite statements. No specific change in behaviour is seen at $L_{2-10keV}\gtrsim10^{44}$~erg/s, a limit above which generally powerful X-ray bright AGN lie (e.g., QSO-mode). 

It should be noted that densities based on closest neighbours are coupled to the distance of the neighbours rather than a fixed distance scale (as in the case of the surface number density). This in effect translates to a mixing of different distance scales within a single density parameter. As a guideline we can say that for our X-ray samples, distances to the second closest neighbour lie between 10 and 700 kpc (averaging at $\simeq160\pm10$ kpc), while distances to the fifth closest neighbour probe scales between 20 and 970 kpc (averaging at $\simeq240\pm14$ kpc). If an underlying relation between a certain population of AGN and environment density exists, then this would lead to an increased scatter in the $\Sigma$ density parameters.\\

In summary, we find evidence for X-ray AGN residing in over-dense environments compared to both a magnitude matched control sample and a sample of random position within the field. Seyfert-like AGN in our sample show the highest degree of over-density, while QSO-like sources appear to inhabit environments consistent with those of their control sources. We find no correlation between AGN luminosity in the hard X-rays and the environment of the AGN. 

\subsection{Radio AGN environment}

\begin{figure}
\begin{center}
\includegraphics[width=0.5\textwidth,angle=0]{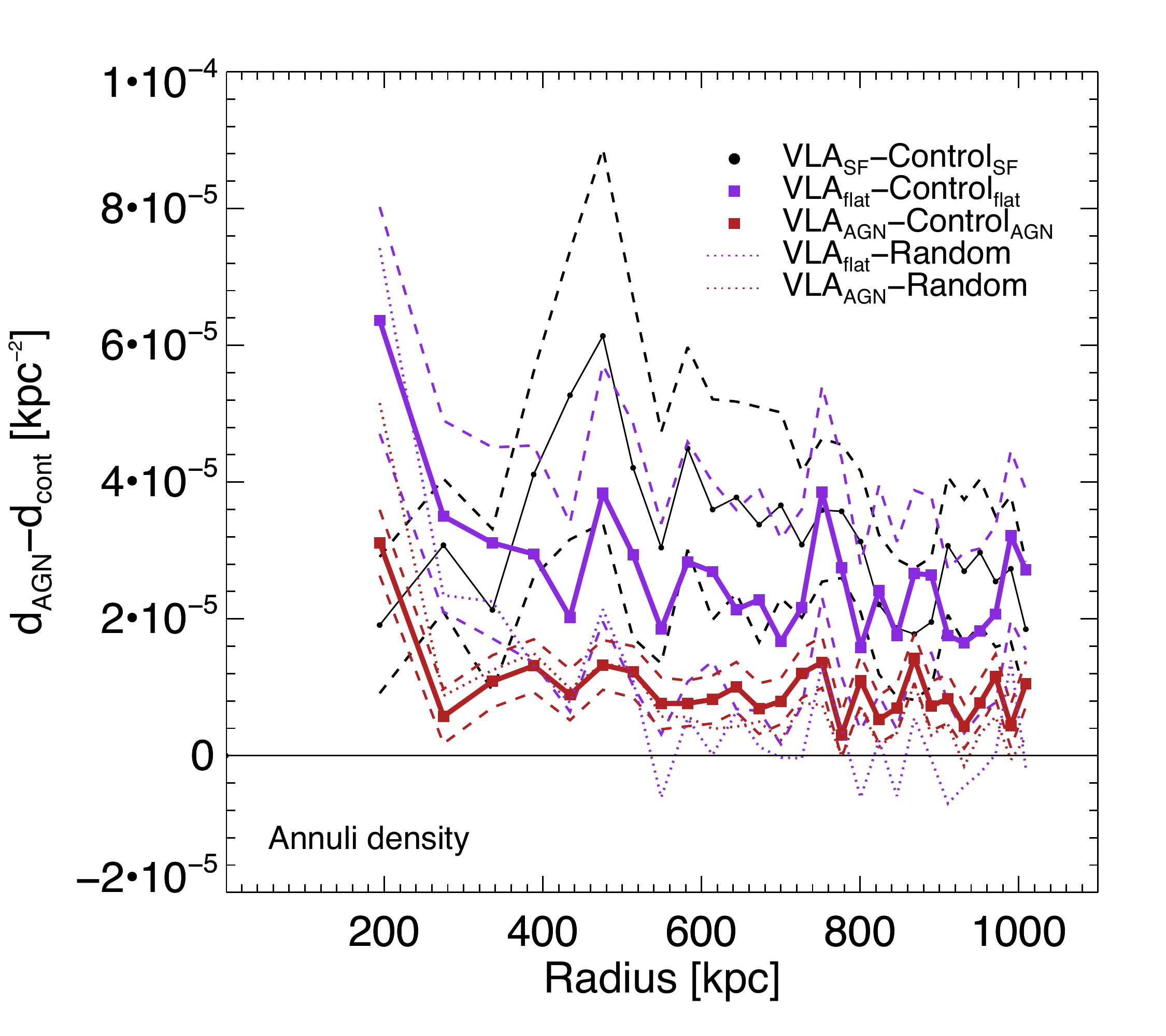}
\caption{Averaged surface number density difference, in annuli, of non-AGN radio sources (solid black line), luminosity-selected radio-AGN (solid red line), and flat-spectrum AGN (solid purple line) with their control samples, as a function of  the outer radius of the annulus within which the density has been calculated. In addition the density difference between the respective radio samples and the random sample is also shown with dotted lines. The line for a density difference of zero is also shown. Uncertainties (1$\sigma$) are shown as loci in dashed lines.}
\label{fig:R_surfacedens}
\end{center}
\end{figure}

We follow the same analysis for our radio-selected samples. In Fig.~\ref{fig:R_surfacedens} we show the average surface density difference between the radio samples and their respective control samples, as a function of the distance from a source. Differences for all 1.4-GHz radio sources, luminosity-selected AGN, and flat-spectrum radio sources are shown. In addition we show the density difference  between the radio samples and the random-position control sample. All samples appear in over-dense environments compared to both their respective control samples as well as the average field density. Of the three, flat-spectrum sources appear to be found in denser environments than their control sample but also show large scatter. In contrast, luminosity-selected radio-AGN show a smaller degree of over-density but with smaller scatter. We also find that those sources that occupy the supposed "non-AGN" dominated part of our radio sample ($L_{\rm 1.4GHz}<7.24\cdot10^{30}$ erg/s/Hz) to have the highest degree of over-density.

\begin{figure}
\begin{center}
\includegraphics[width=0.5\textwidth,angle=0]{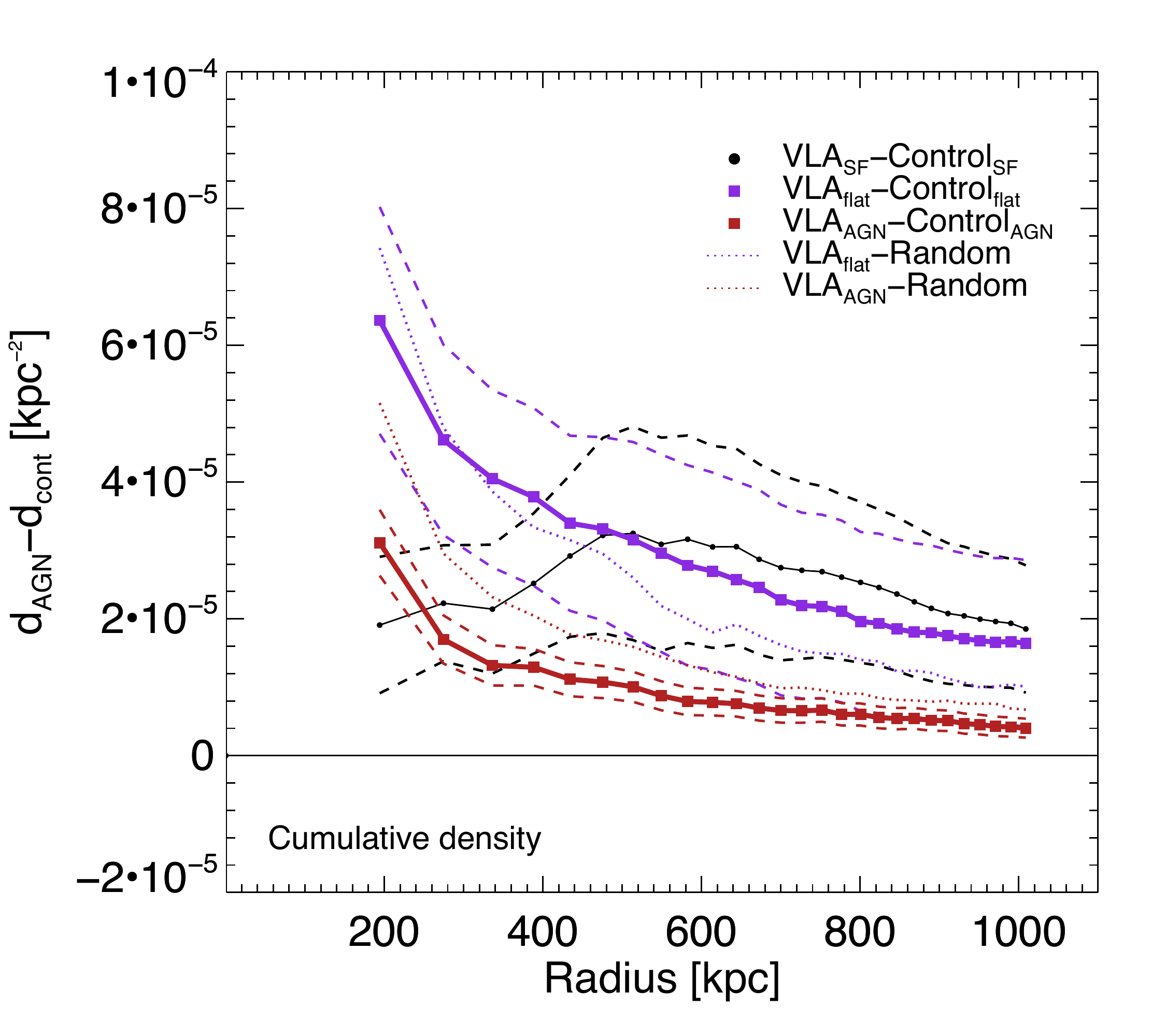}
\caption{Average surface number density difference, in circles, of non-AGN radio sources (solid black line), luminosity-selected radio-AGN (solid red line), and flat-spectrum AGN (solid purple line) with their control samples, as a function of  the outer radius of the circle within which the density has been calculated. In addition the density difference between the respective radio samples and the random sample is also shown with dotted lines. The line for a density difference of zero is also shown. Uncertainties (1$\sigma$) are shown as loci in dashed lines.}
\label{fig:R_surfacecumdens}
\end{center}
\end{figure}

As can also be seen in Fig.~\ref{fig:R_surfacecumdens},where we plot the density difference in expanding circles, rather than annuli, the higher over-densities are observed for the first bins, at scales $<200$~kpc for both radio-AGN samples. Conversely, in an absolute sense, the flat-spectrum radio-AGN sample shows a higher degree of over-density compared also to the X-ray AGN samples.

We now investigate whether there is any  correlation between radio luminosity and environmental over-density, as might be expected if the radio emission was boosted in luminosity due to the increased density of the working surface. Similar to Fig.~\ref{fig:X_dif_lum} we do not see any particular trend with radio luminosity, with sources showing average values above a zero density, homogeneously with respect to their radio luminosity. Again we observe a larger scatter for low-luminosity sources.  For the highest radio luminosity sources ($L_{\rm1.4GHz}>10^{33}$ erg/s/Hz) we note a trend for sources residing exclusively in over-dense environments, but given that only a handful of sources are found in that luminosity range, we can not draw any solid  conclusions.

\begin{figure*}
\begin{center}
\includegraphics[width=0.33\textwidth,angle=0]{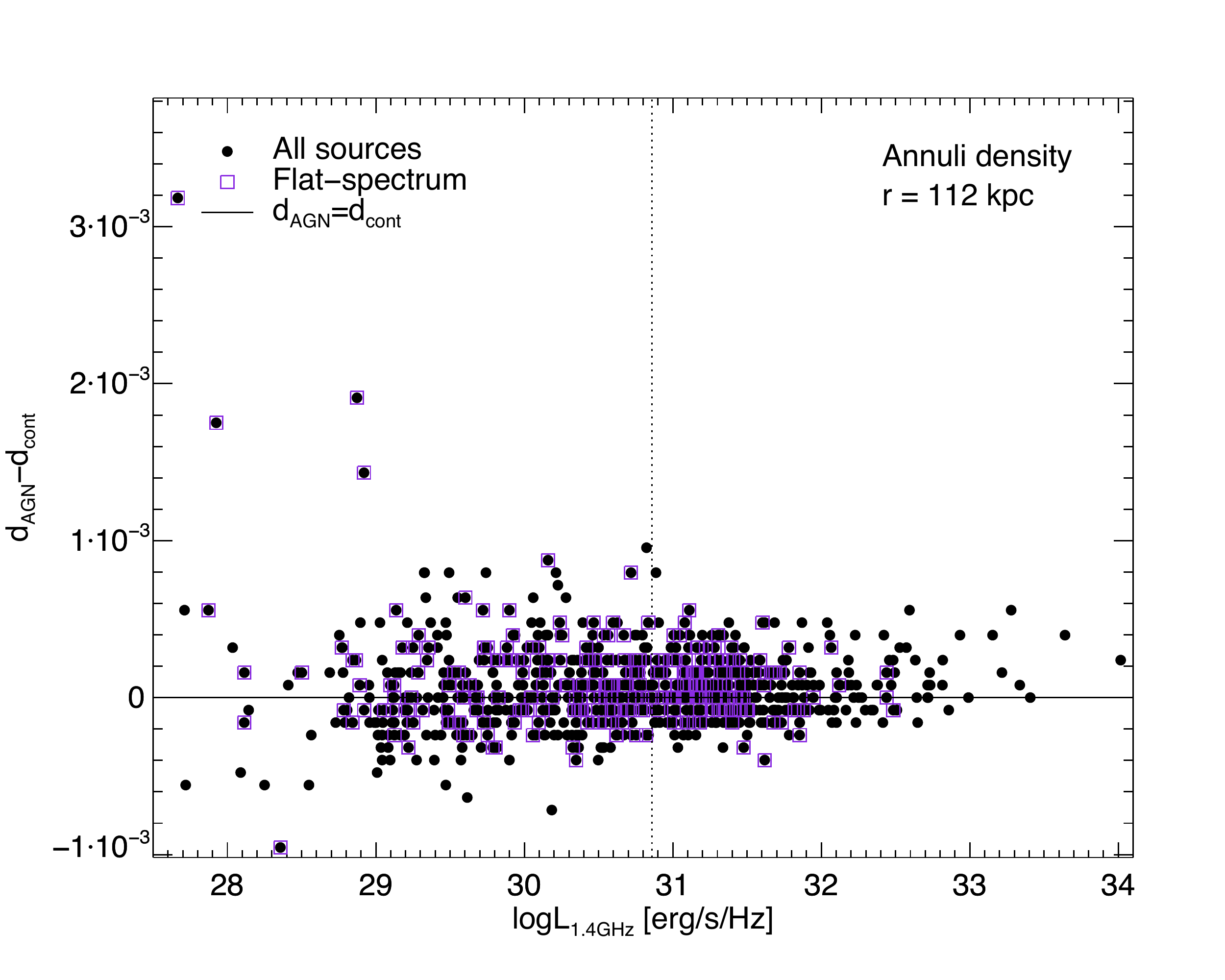}
\includegraphics[width=0.33\textwidth,angle=0]{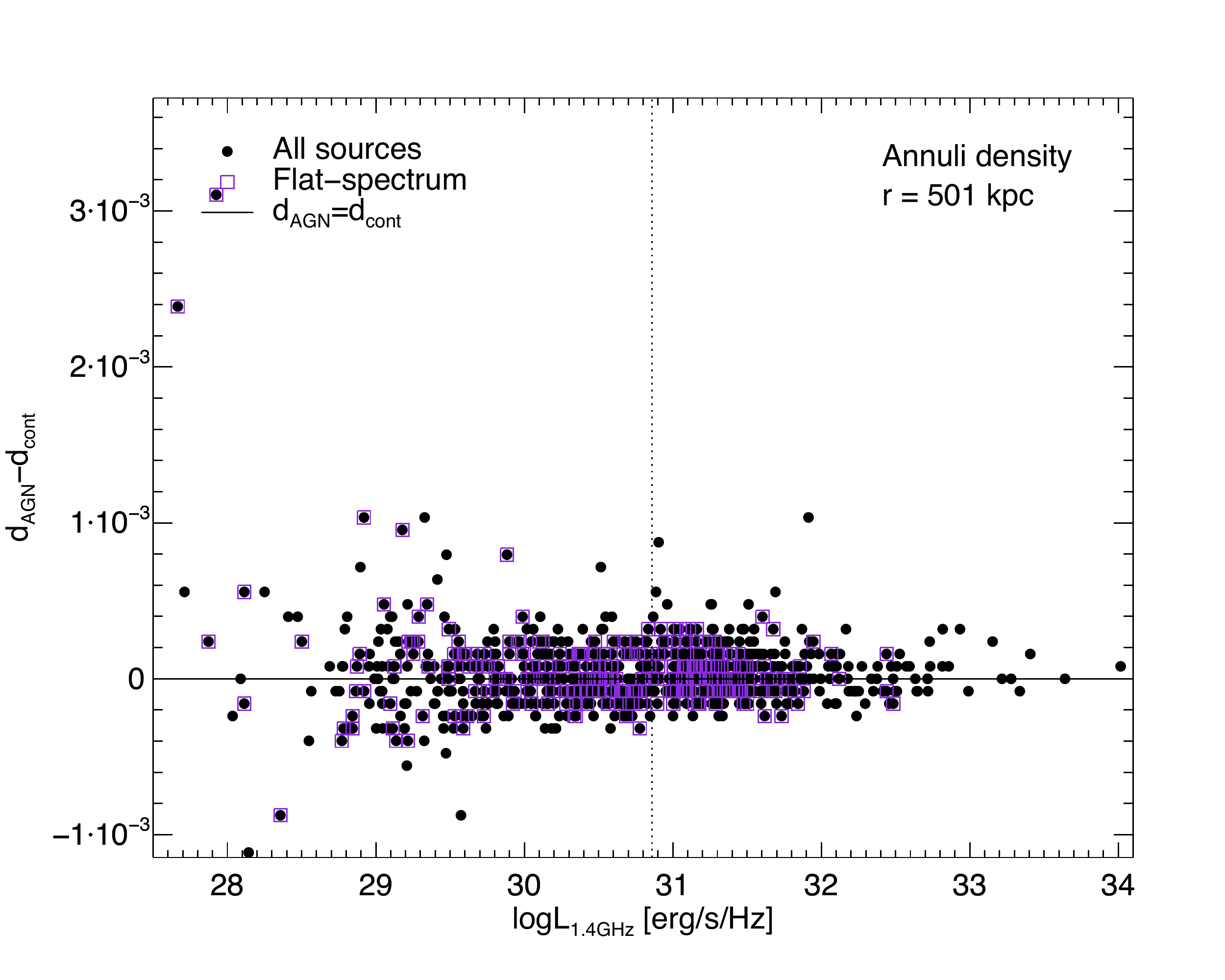}
\includegraphics[width=0.33\textwidth,angle=0]{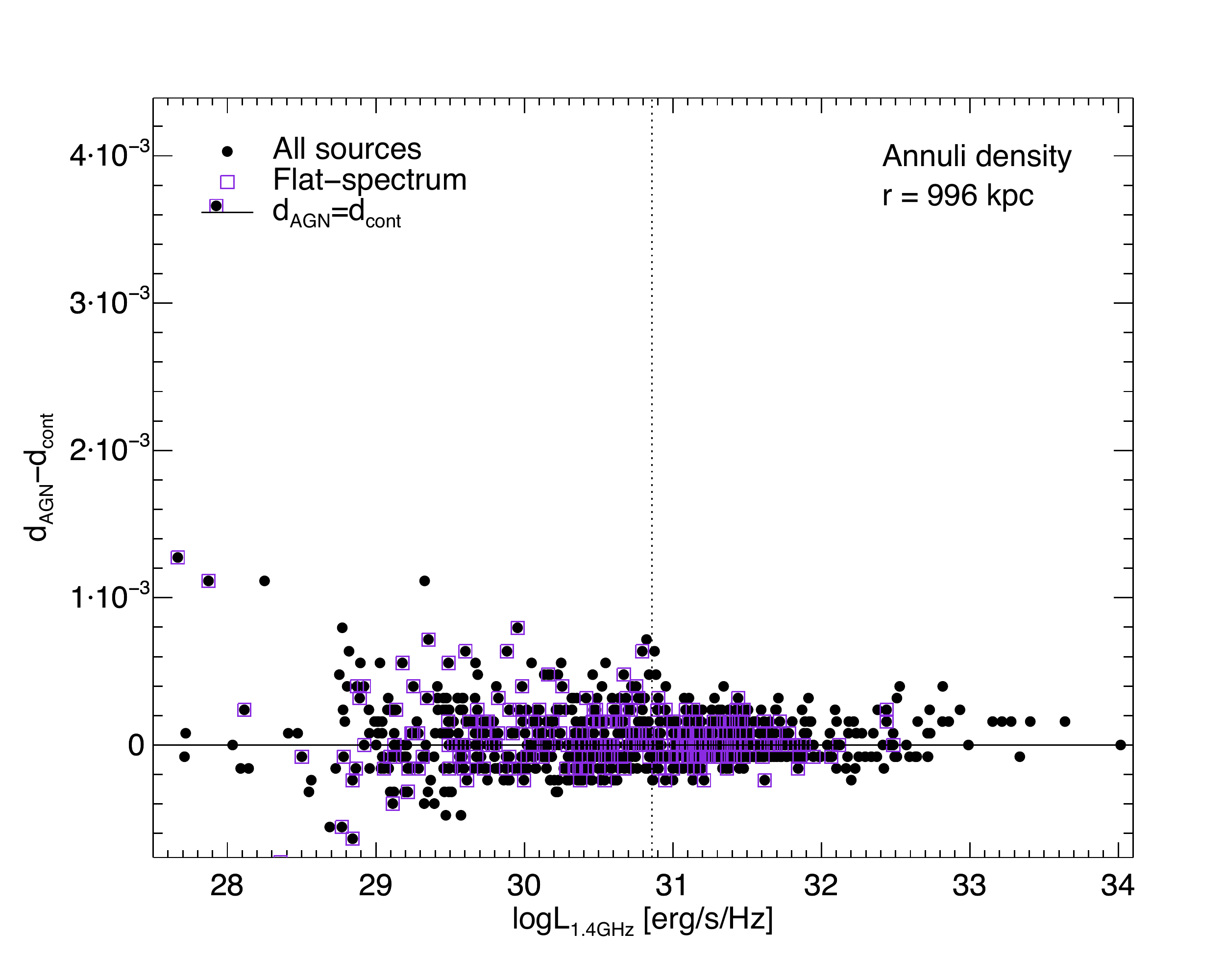}
\caption{Surface number density difference, in annuli, of all radio sources (black circles) and flat-spectrum radio-AGN (purple squares) with their respective control samples, as a function of their K-corrected 1.4 GHz luminosity. This is shown for three different radial distances from the central source, $\sim100$ kpc (left), $\sim500$ kpc (middle), and $\sim1000$ kpc (right). The vertical dashed line denotes the luminosity limit above which radio-AGN are defined.}
\label{fig:R_dif_lum}
\end{center}
\end{figure*}

\begin{figure}
\begin{center}
\includegraphics[width=0.5\textwidth,angle=0]{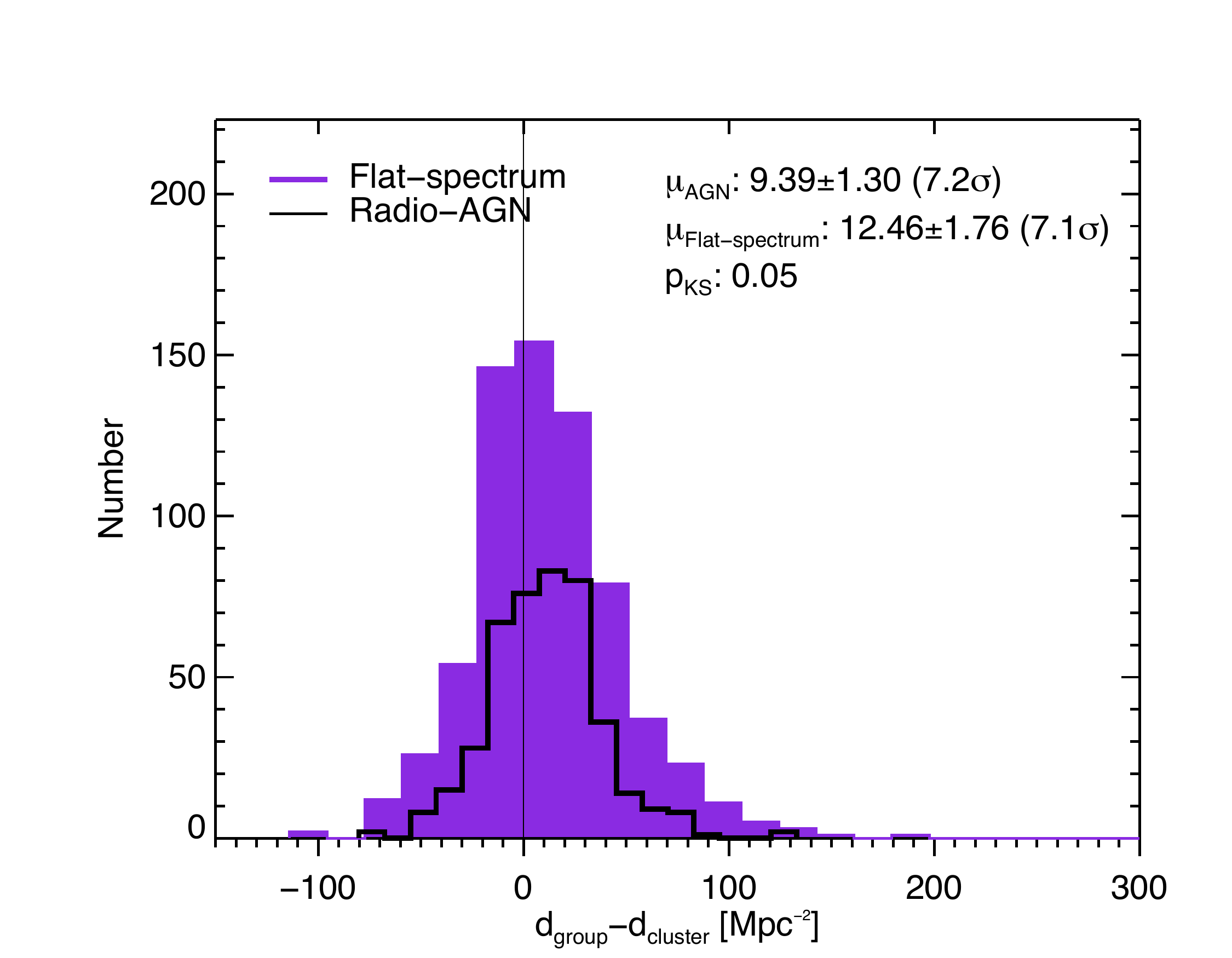}
\caption{Histogram of the distributions for radio-luminosity selected AGN (black empty histogram) and flat-spectrum radio-AGN (purple filled histogram) of the difference between densities calculated at scales of 200 and 800 kpc. Average values with standard errors for the respective samples are shown. The p-value of a two-sample K-S test for the two samples is also given.}
\label{fig:R_group_cluster}
\end{center}
\end{figure}

Similarly to Fig. \ref{fig:X_group_cluster}, in Fig. \ref{fig:R_group_cluster} we show the distribution of the difference of densities at 200 and 800 kpc for the radio samples. We see that similar to the X-ray sample, we find that radio sources show the peak of their distribution at 0 (for the flat-spectrum radio sample), while the luminosity selected AGN show the peak of their distribution at slightly higher value. However, when considering  average values of the density difference for the two samples, they are significantly above zero (at $\sim7\sigma$). This implies that there is a sizeable fraction of radio-AGN that have a large number of relatively nearby neighbours. This may be indicative of the importance of mergers for these sources, or that to generate powerful radio emission a dense working surface is required to enhance the radio emission so that it enters radio-flux limited samples.

\begin{figure*}
\begin{center}
\includegraphics[width=0.4\textwidth,angle=0]{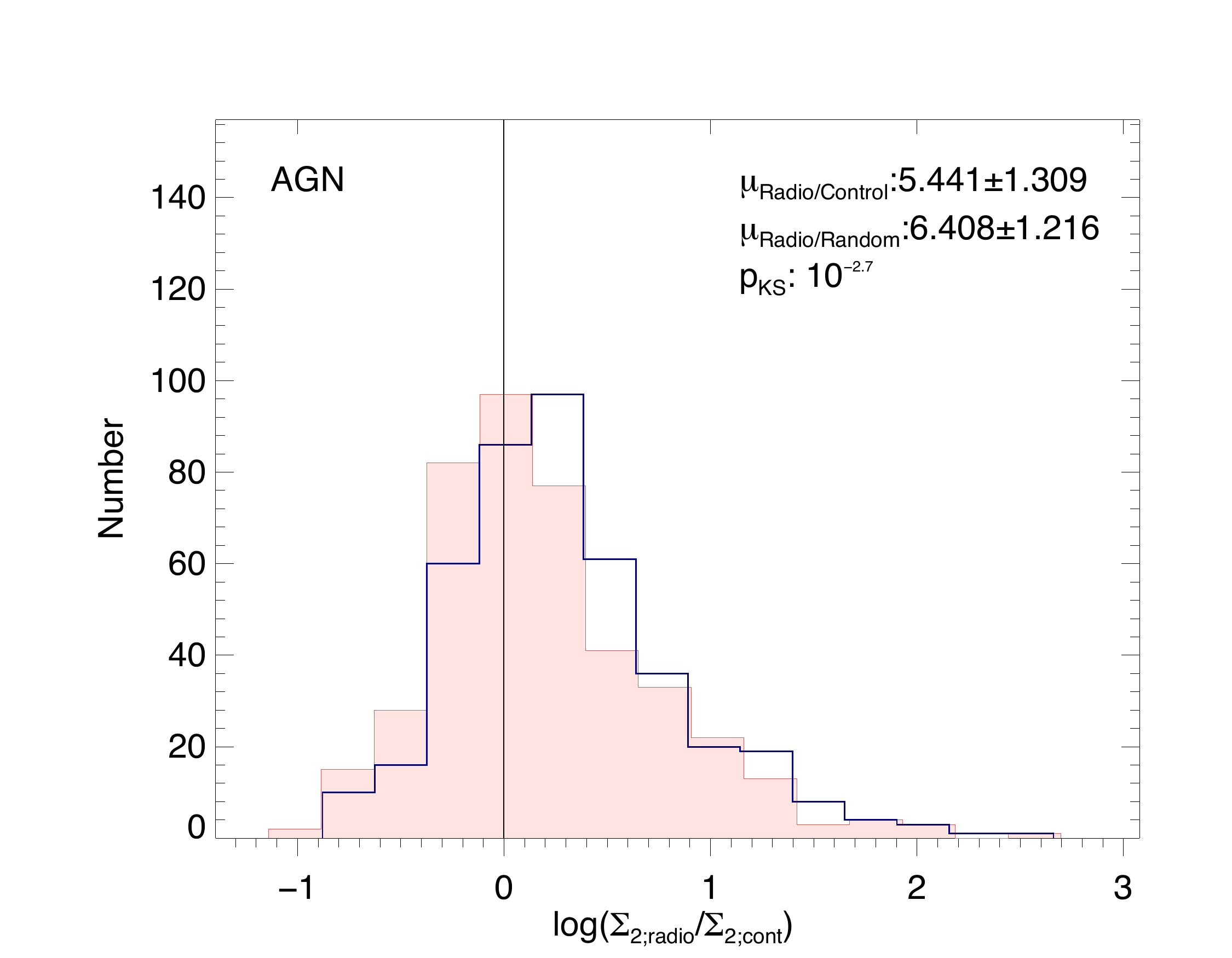}
\includegraphics[width=0.4\textwidth,angle=0]{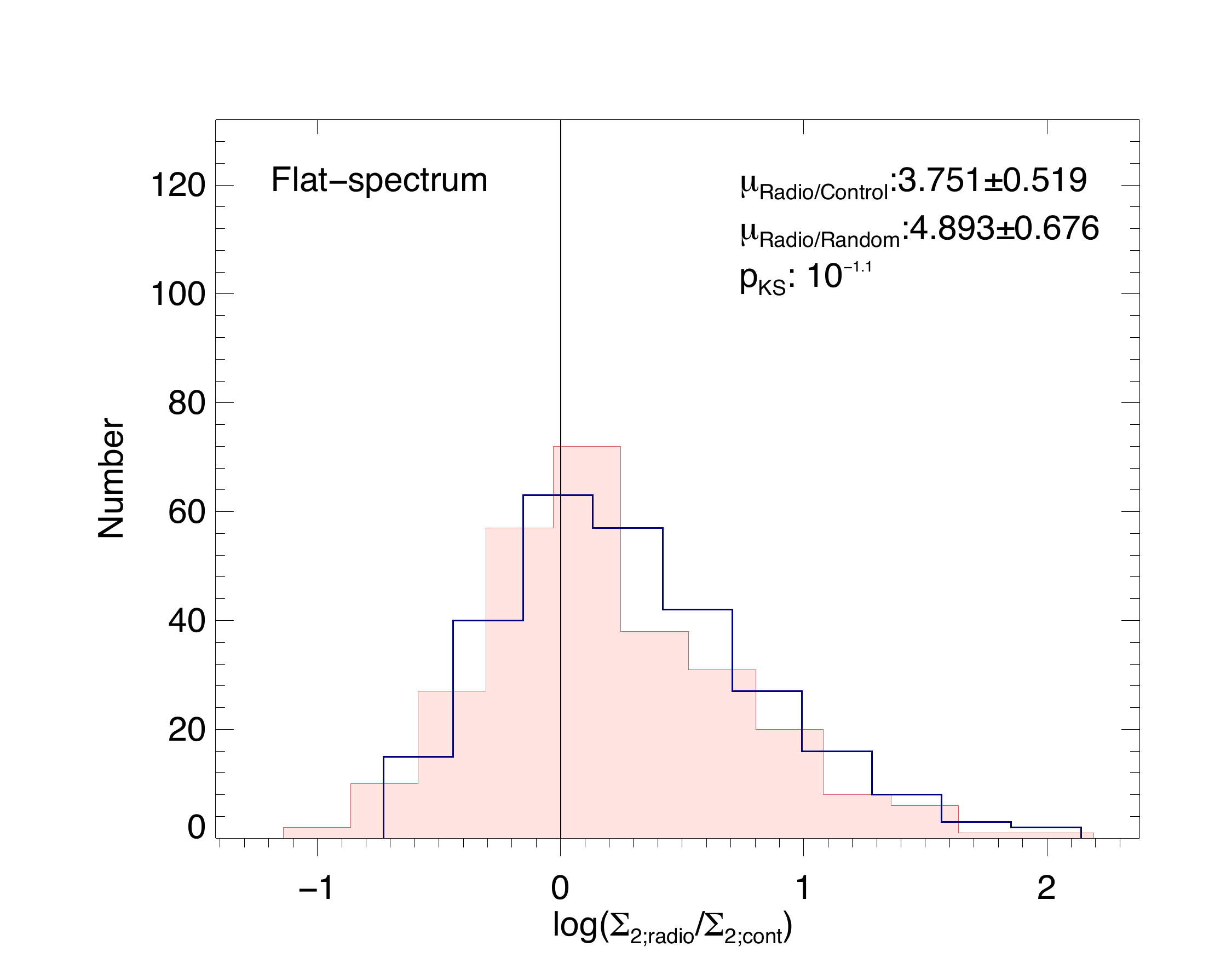}
\includegraphics[width=0.4\textwidth,angle=0]{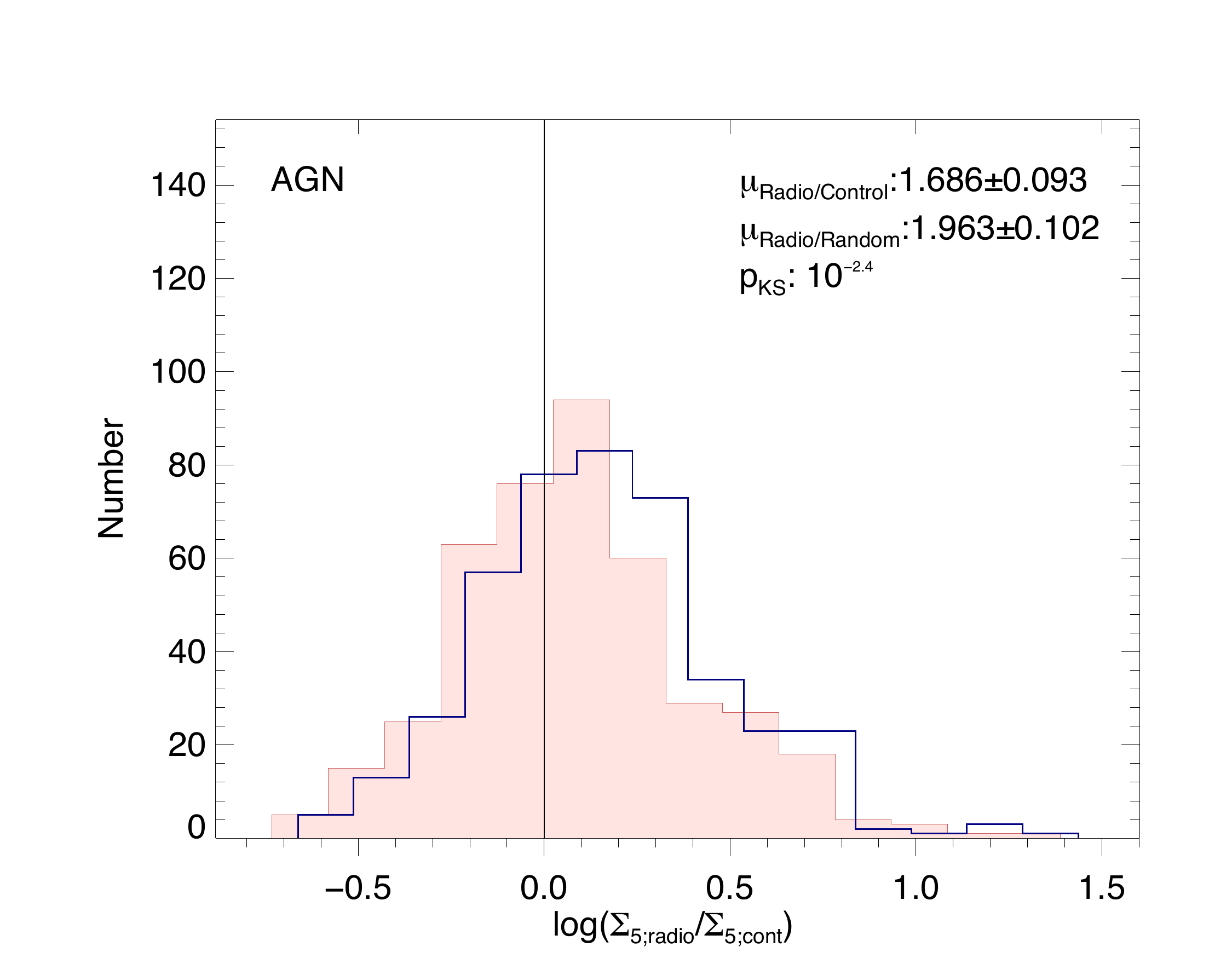}
\includegraphics[width=0.4\textwidth,angle=0]{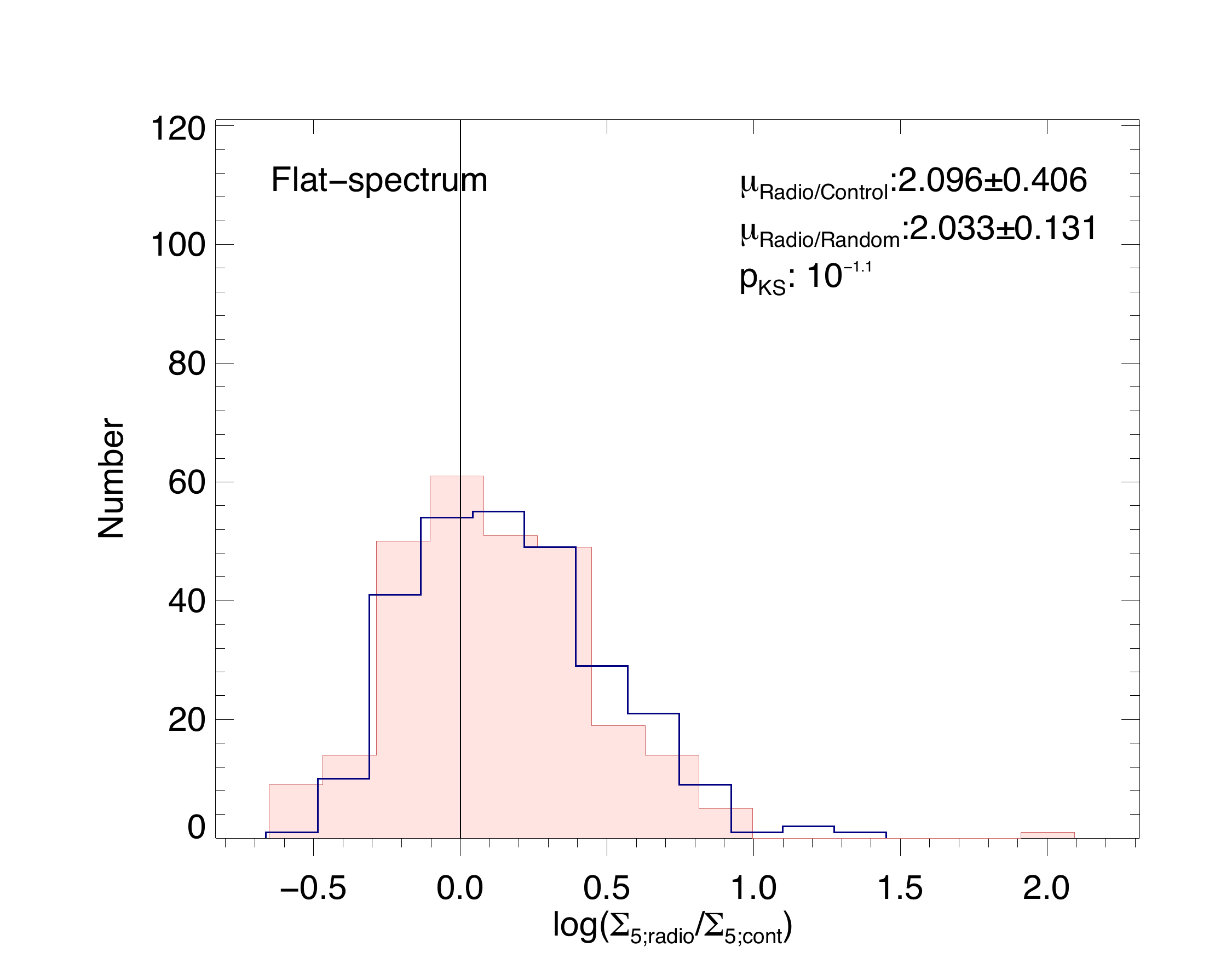}
\caption{The distributions of the logarithm of the ratio between the $\Sigma$ parameters of the radio sources and that of their control sample (filled red) and the random sample (open blue). Central values and $\sigma$ values for the distributions are given. The probability parameter p from a K-S comparison of AGN/control and AGN/random density ratio distributions is also given. For reference a vertical line at a ratio value of one is drawn as well. Left panels show the distributions of luminosity-selected radio-AGN and right ones those of flat-spectrum radio-sources.}
\label{fig:r_histo_S2S5}
\end{center}
\end{figure*}

In Fig. \ref{fig:r_histo_S2S5} we show the histograms for the density ratio $\Sigma_{main}/\Sigma_{control}$ both for the magnitude matched control sample and for the random positions. Again for both the luminosity-selected radio-AGN and flat-spectrum radio-AGN, there appears to exist a tail of high density objects. As a result, the average value for each sub-sample are significantly higher than one. This is most evident for the $\Sigma_2$ distribution, while the differences become more mooted for the $\Sigma_5$ density parameter. This is in agreement with the  previous results (e.g., Figs. \ref{fig:R_surfacedens} and \ref{fig:R_dif_lum}). We also note that albeit with a large scatter, the flat-spectrum radio-AGN sub-sample shows persistently stronger over-densities even when considering the $\Sigma_5$ density parameter.

\begin{figure*}
\begin{center}
\includegraphics[width=0.45\textwidth,angle=0]{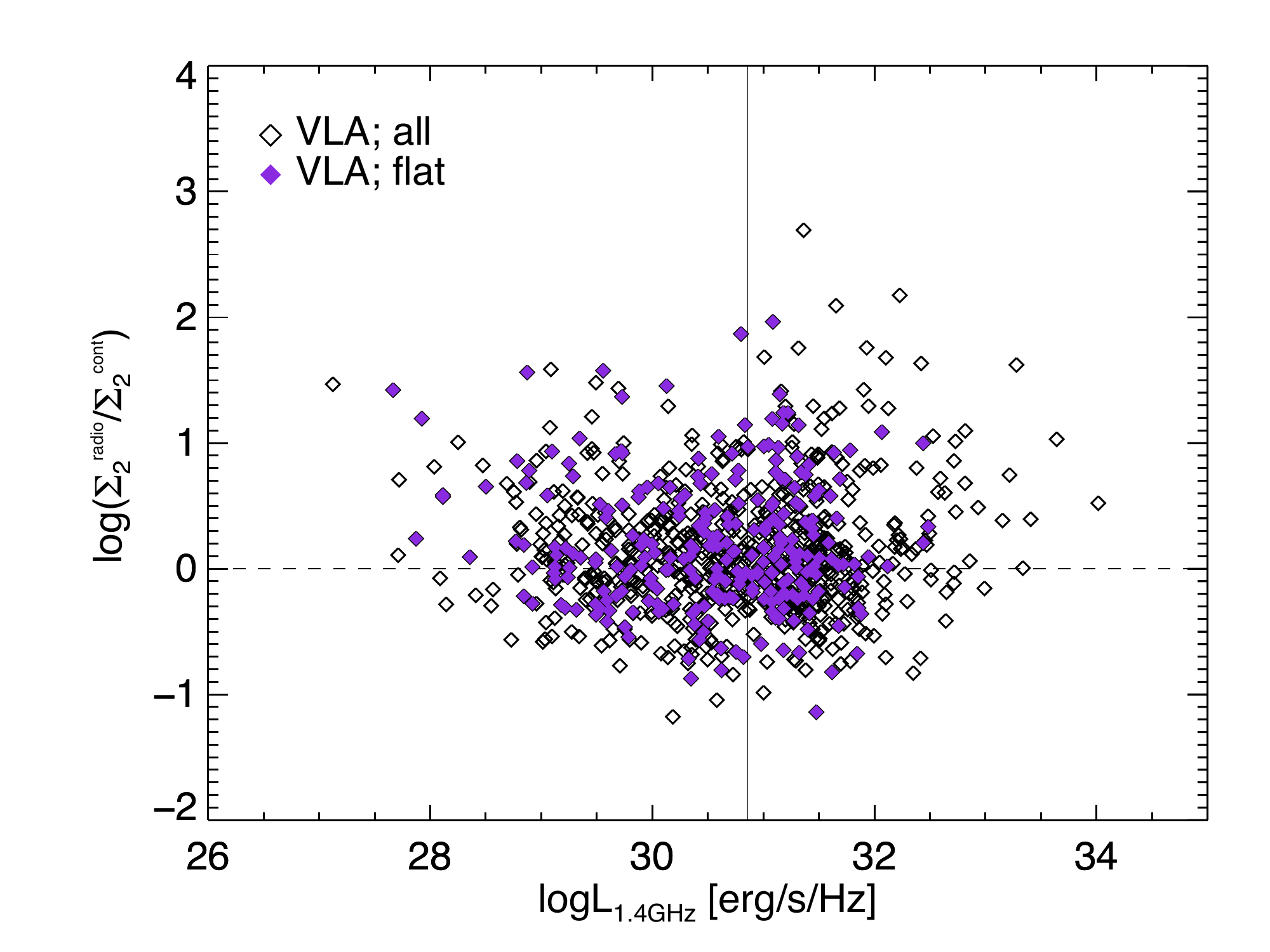}
\includegraphics[width=0.45\textwidth,angle=0]{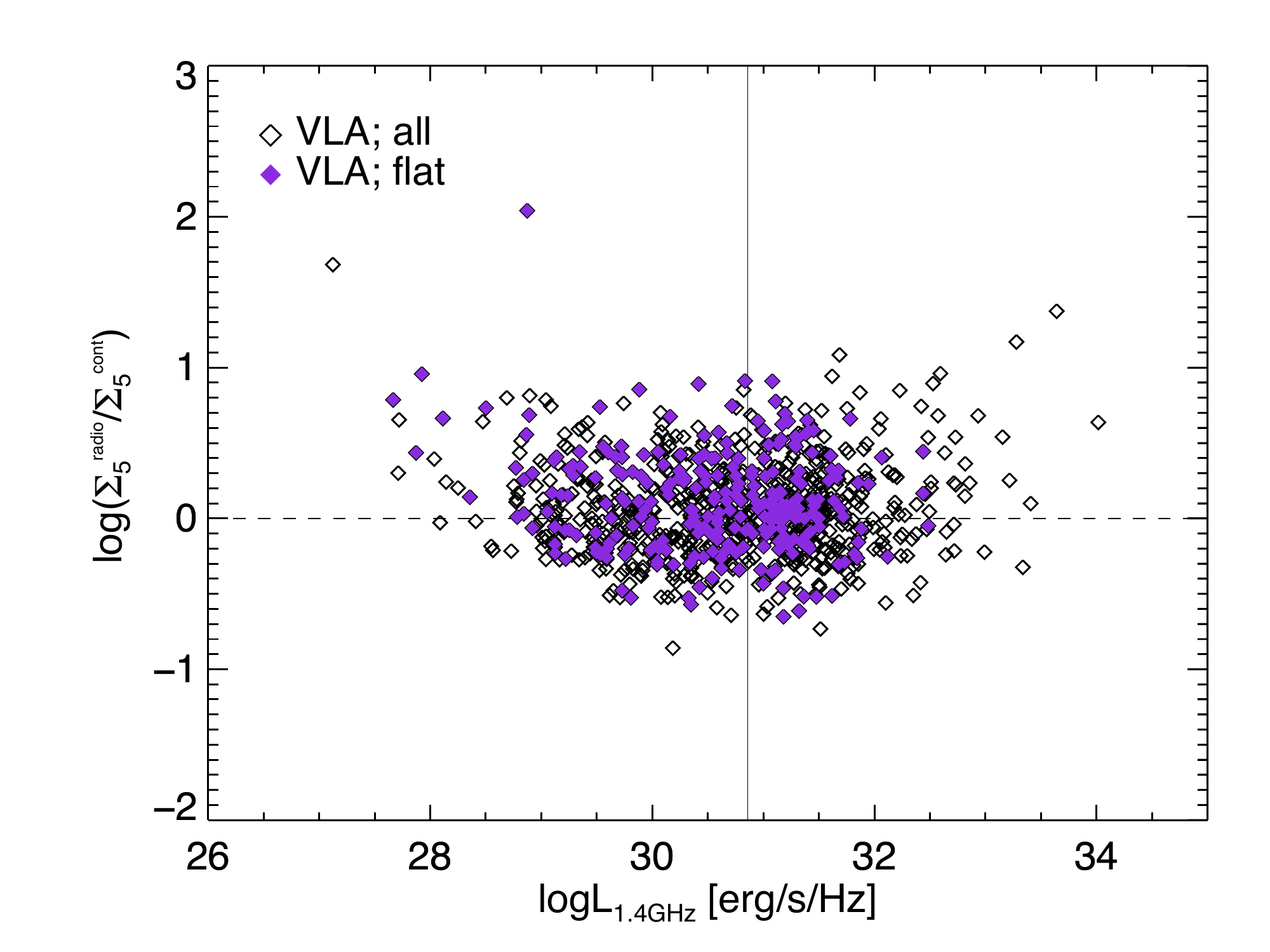}
\caption{The logarithm of the ratio of $\Sigma_{2}$ (left) and $\Sigma_{5}$ (right) density parameters between all (open black) and flat-spectrum (filled purple) radio-sources and their control samples as a function of their 1.4 GHz radio luminosity. The zero line denotes the boundary between a source being in an over- or under-dense environment compared to its control sources ($\Sigma^{radio}/\Sigma^{control}=1$).}
\label{fig:r_S2S5_lum}
\end{center}
\end{figure*}

In a similar manner to Fig. \ref{fig:X_S2_lum} we check the $\Sigma_{2}$ and $\Sigma_{5}$ density parameter values ratios for all radio-sources and for flat-spectrum radio-sources as a function of their 1.4 GHz radio luminosity, $L_{1.4GHz}$ (Fig. \ref{fig:r_S2S5_lum}). No strong trend is seen. Especially for $\Sigma_{5}$, most sources have ratios equal to one. Flat-spectrum radio sources do not show any appreciably different behavior than their parent sample. When considering our luminosity-based AGN selection, no significantly different environments are seen above that luminosity limit. There is however a hint for the most luminous radio-sources being in over-dense environments, while the most radio-faint are found below the zero line.\\

\subsection{Mid-IR AGN environment}

\begin{figure}
\begin{center}
\includegraphics[width=0.5\textwidth,angle=0]{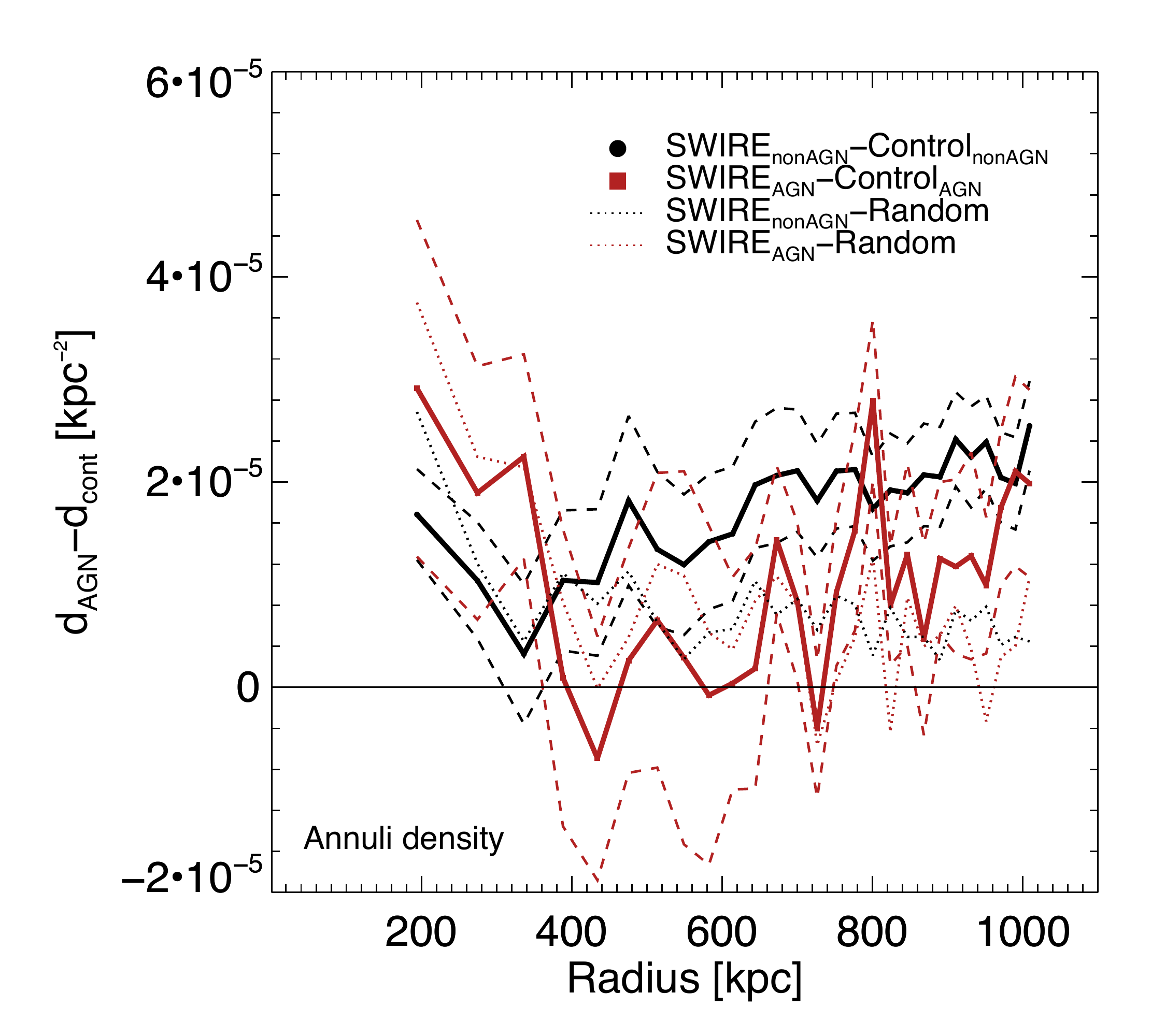}
\caption{Average surface number density difference, in annuli, of all mid-IR sources (solid black line), and colour-selected obscured AGN (solid red line) with their control samples, as a function of  the outer radius of the annulus within which the density has been calculated. In addition the density difference between the respective mid-IR samples and the random sample is also shown with dotted lines. The line for a density difference of zero is also shown. Uncertainties (1$\sigma$ are shown as loci in dashed lines.}
\label{fig:M_surfacedens}
\end{center}
\end{figure}

\begin{figure}
\begin{center}
\includegraphics[width=0.5\textwidth,angle=0]{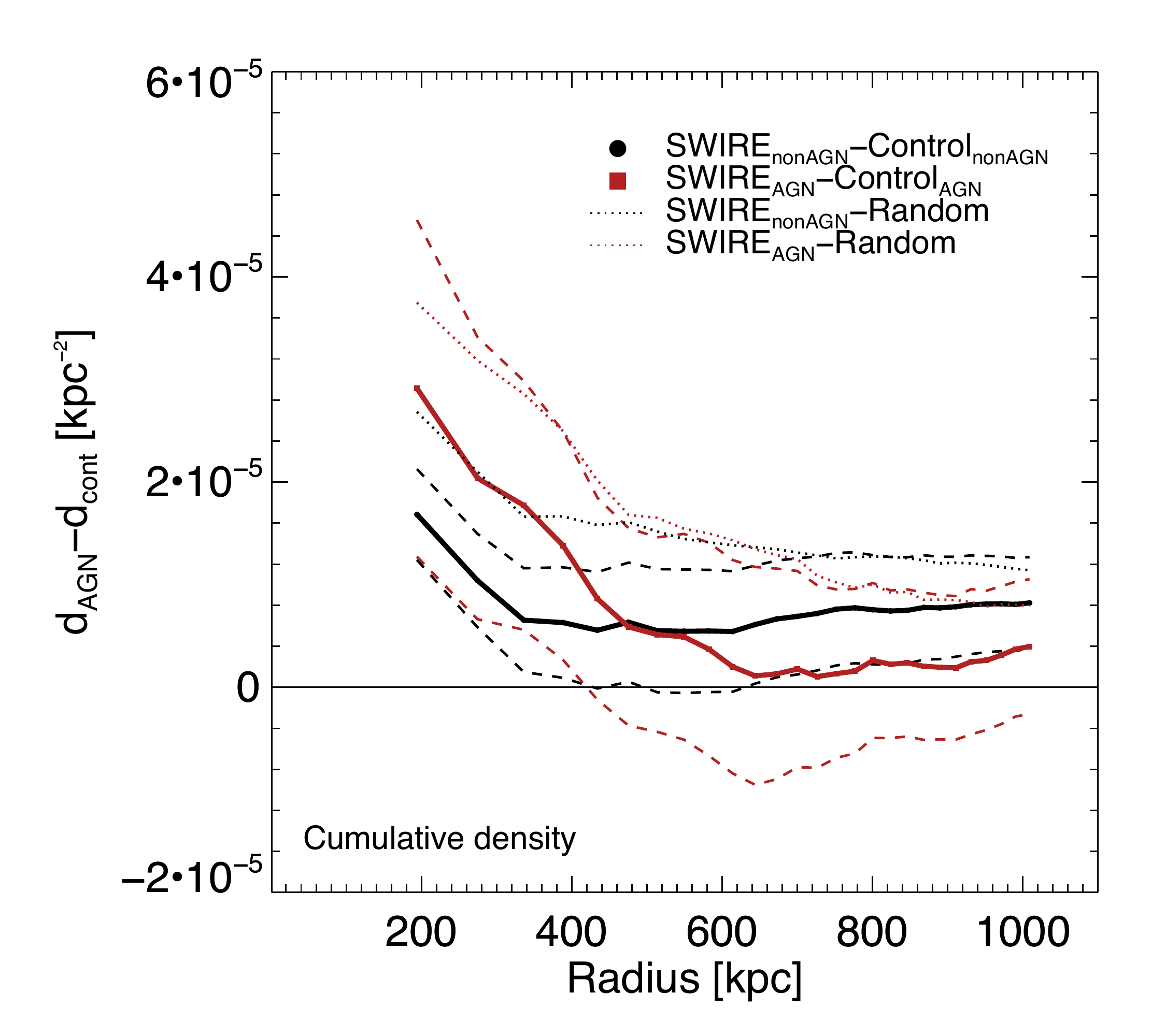}
\caption{Average surface number density difference, in circles, of all mid-IR sources (solid black line), and colour-selected obscured AGN (solid red line) with their control samples, as a function of  the outer radius of the circle within which the density has been calculated. In addition the density difference between the respective mid-IR samples and the random sample is also shown with dotted lines. The line for a density difference of zero is also shown. Uncertainties (1$\sigma$) are shown as loci in dashed lines.}
\label{fig:M_surfacecumdens}
\end{center}
\end{figure}

Finally we turn to our 24$\mu$m-selected samples. In Fig.~\ref{fig:M_surfacedens} we show the average surface density difference, in annuli, between the mid-IR samples and their respective control samples. Density differences for all 24 $\mu$m mid-IR sources and colour-selected obscured candidate AGN sources are shown. The average density difference between the mid-IR sources and the random control sample is also shown. We find that, on average, mid-IR sources reside in  environments with around 10 excess sources Mpc$^{-2}$ compared to the control sample{, but with a large scatter}. Mid-IR selected non-AGN (here defined as all sources except the colour-selected obscured AGN) show somewhat more over-dense environments with respect to their control sample, particularly above 600~kpc, and also in an absolute measurement (compared to the random positions). {However, both samples show considerable uncertainties, and for most radii their environments appear consistent with those of their control samples.}
In Fig. \ref{fig:M_surfacecumdens} we show the cumulative density calculated within circles as a function of the distance to the AGN. The difference between the main sample and the random positions sample stands out for the colour-selected obscured AGN, showing strongly over-dense environments at scales $<400$ kpc. This implies that when compared to the average field density, they indeed inhabit over-dense environments. {At similar scales ($r<400$ kpc), the comparison with the magnitude-selected control sample shows obscured AGN to reside in over-dense environments (albeit at the 1-2$\sigma$ level).}

\begin{figure}
\begin{center}
\includegraphics[width=0.5\textwidth,angle=0]{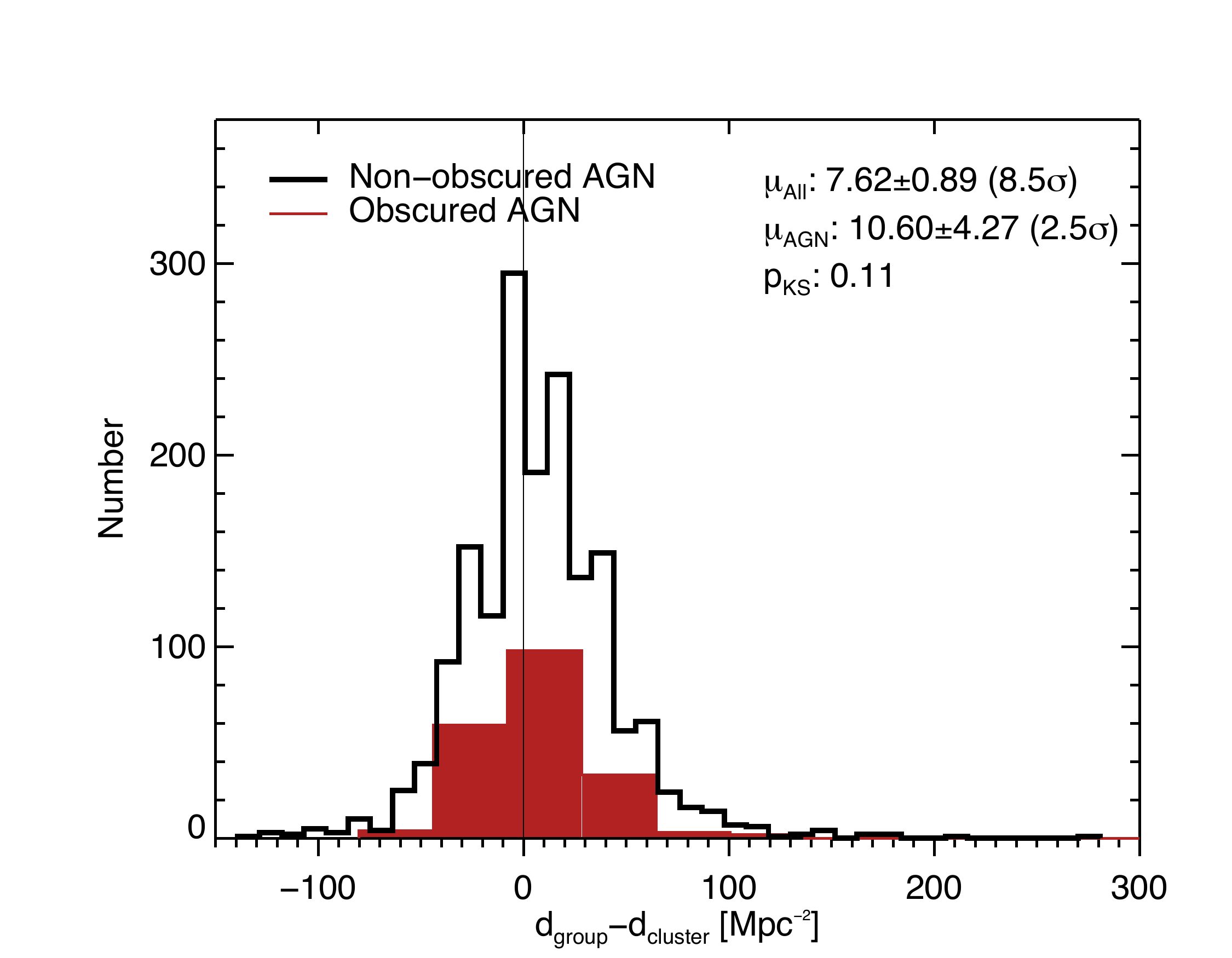}
\caption{Histogram of the distributions for color-selected obscured AGN (red filled histogram) and all mid-IR selected sources (black empty histogram) of the difference between densities calculated at scales of 200 and 800 kpc. This difference reflects and relative importance between group and cluster environments. Average values with standard errors for the respective samples are shown. The p-value of a two-sample K-S test for the two samples is also given.}
\label{fig:M_group_cluster}
\end{center}
\end{figure}

In Fig. \ref{fig:M_group_cluster} the density difference distribution between scales of 200 and 800 kpc is plotted. For both samples the distribution peaks at around 0, but a significant tail extends to strongly positive values. The average values are positive, significantly above zero for the whole mid-IR sample, but not so for the color-selected obscured AGN sub-sample. Given an expected strong contamination of star-forming galaxies for the total mid-IR sample, we conclude that mid-IR selected AGN do not show a significant preference for dense close environments over cluster environments.

\begin{figure*}
\begin{center}
\includegraphics[width=0.33\textwidth,angle=0]{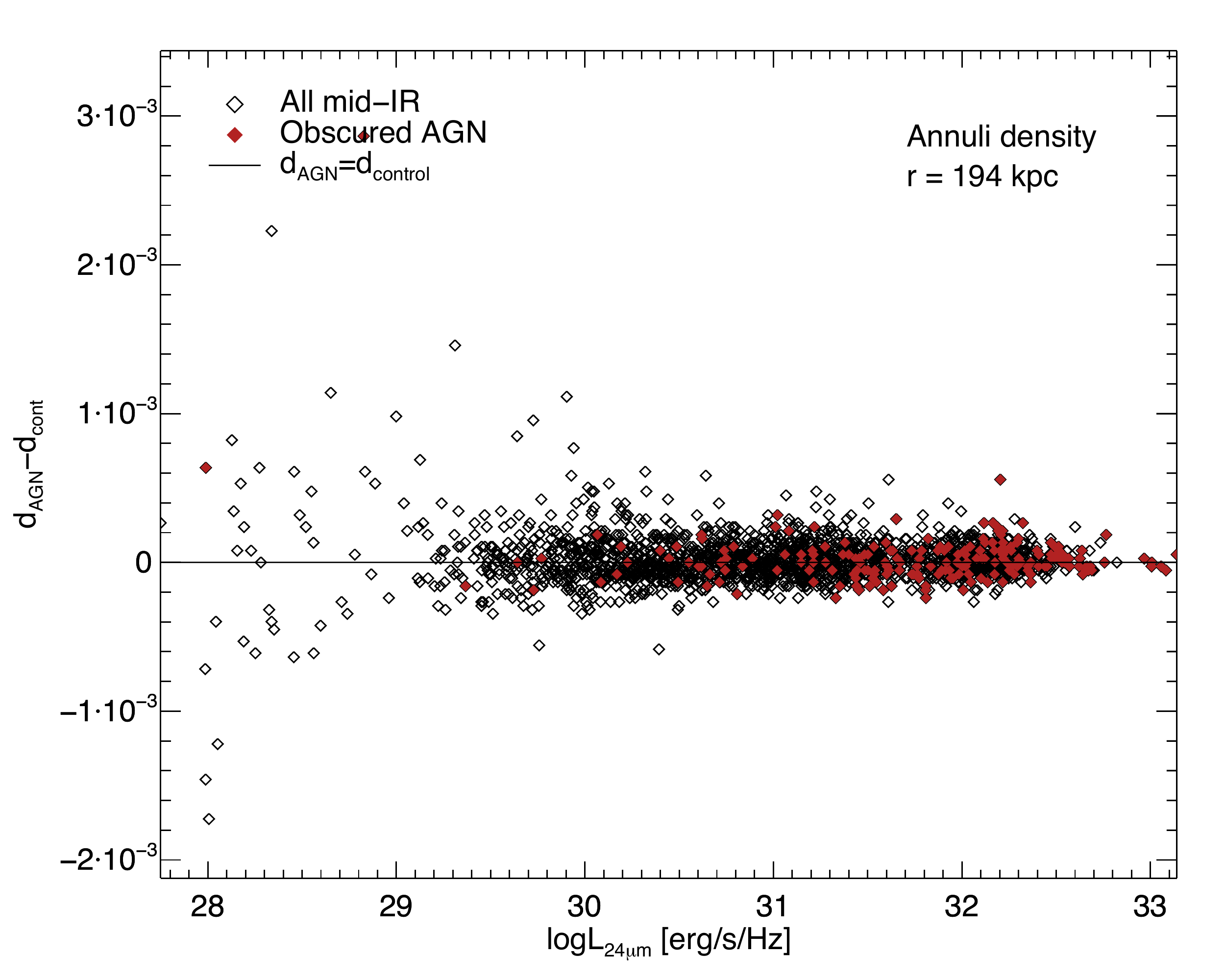}
\includegraphics[width=0.33\textwidth,angle=0]{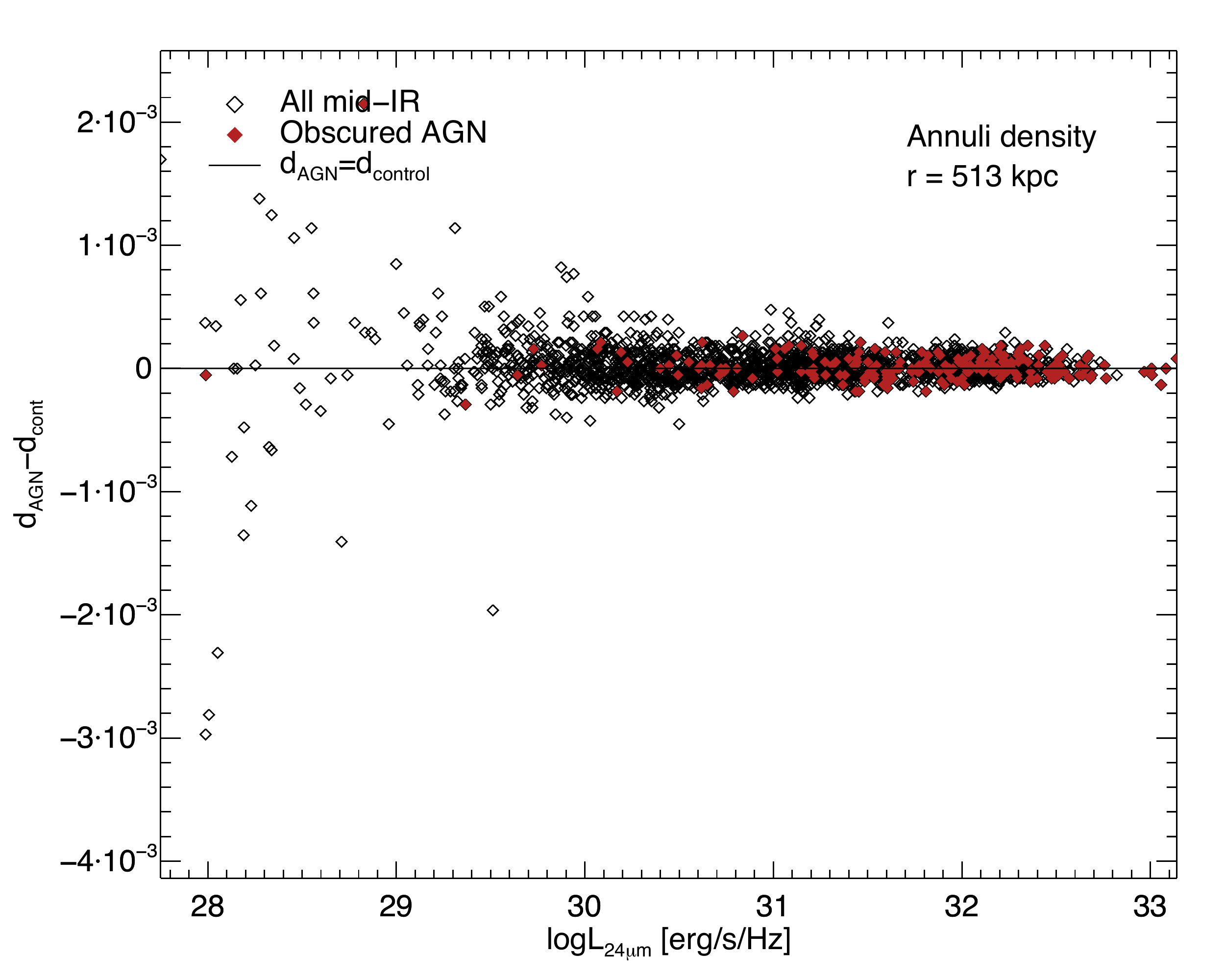}
\includegraphics[width=0.33\textwidth,angle=0]{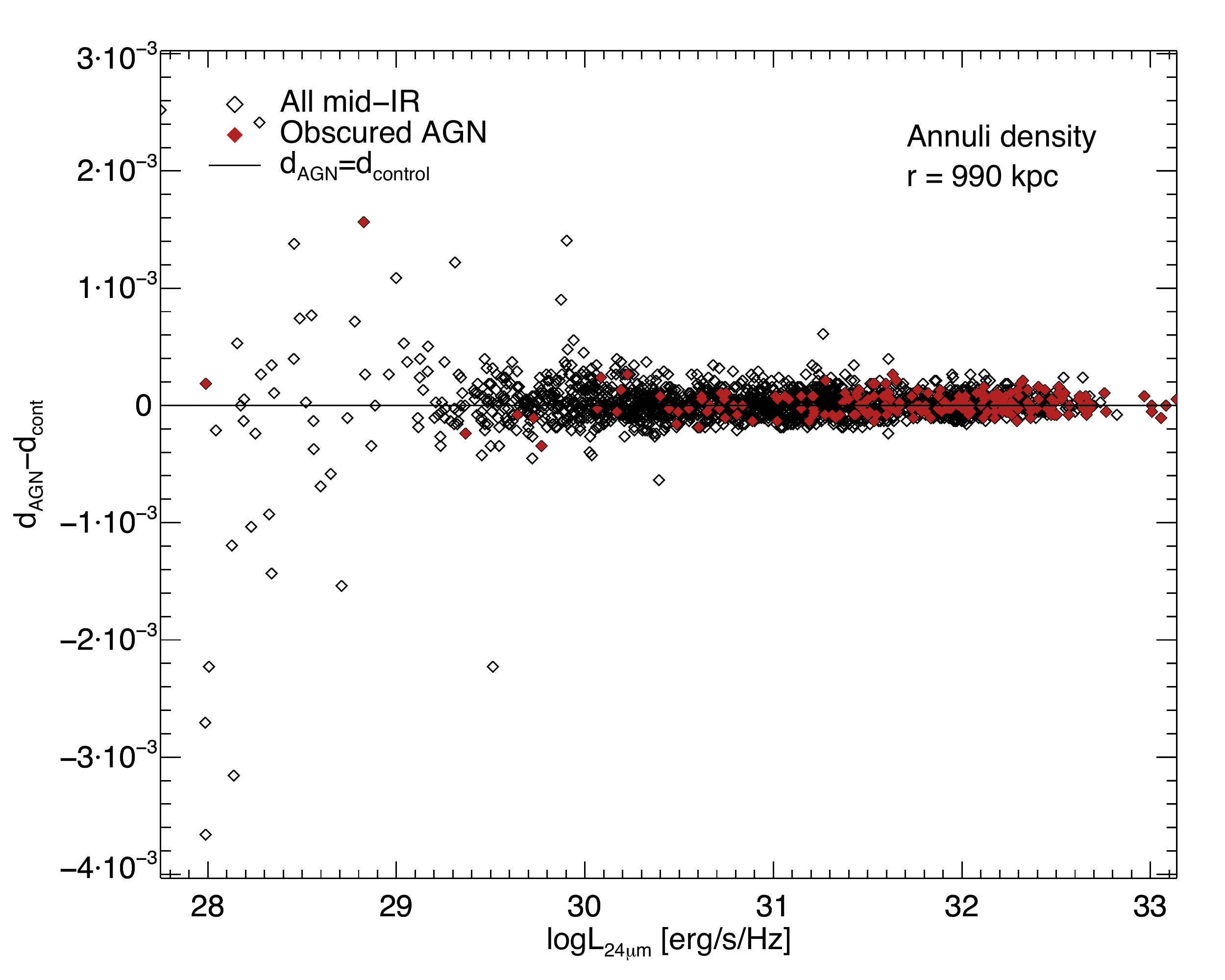}
\caption{Surface number density difference, in annuli, of all mid-IR sources (black circles) and color-selected obscured AGN (red squares) with their respective control samples, as a function of their 24 $\mu$m luminosity. This is shown for three different radial distances from the central source, $\sim100$ kpc (left), $\sim500$ kpc (middle), and $\sim1000$ kpc (right).}
\label{fig:M_dif_lum}
\end{center}
\end{figure*}

In Fig. \ref{fig:M_dif_lum} we show the dependence of the density difference on the mid-IR luminosity. Similarly to the other wavelengths, we do not find any strong correlation between the two. In particular, all of the colour-selected obscured AGN sample lie very close to a density difference of zero, exhibiting smaller scatter than the total mid-IR sample. This might reflect the more heterogeneous nature of sources included in the general mid-IR sample, compared to the obscured AGN sub-sample.

\begin{figure*}
\begin{center}
\includegraphics[width=0.4\textwidth,angle=0]{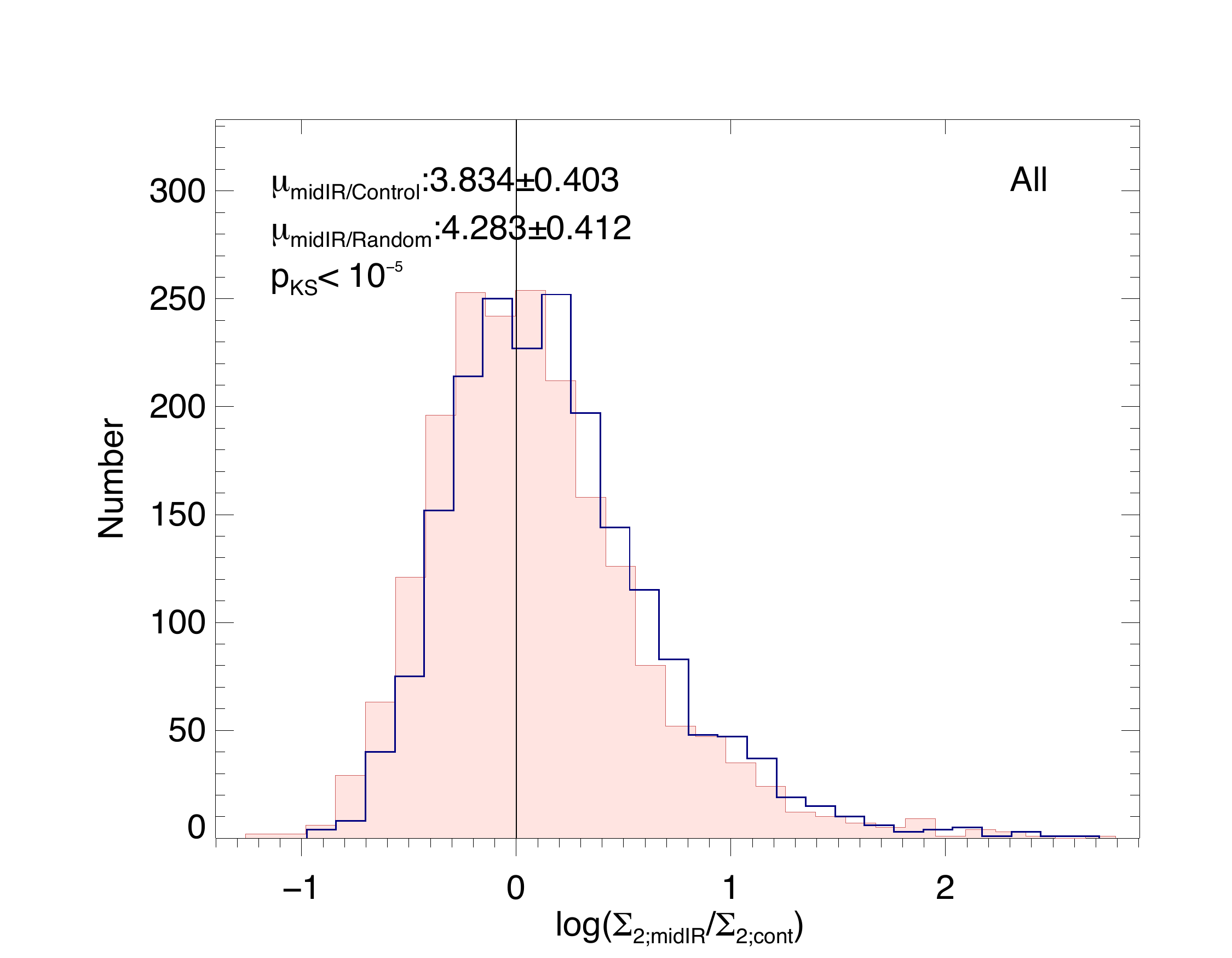}
\includegraphics[width=0.4\textwidth,angle=0]{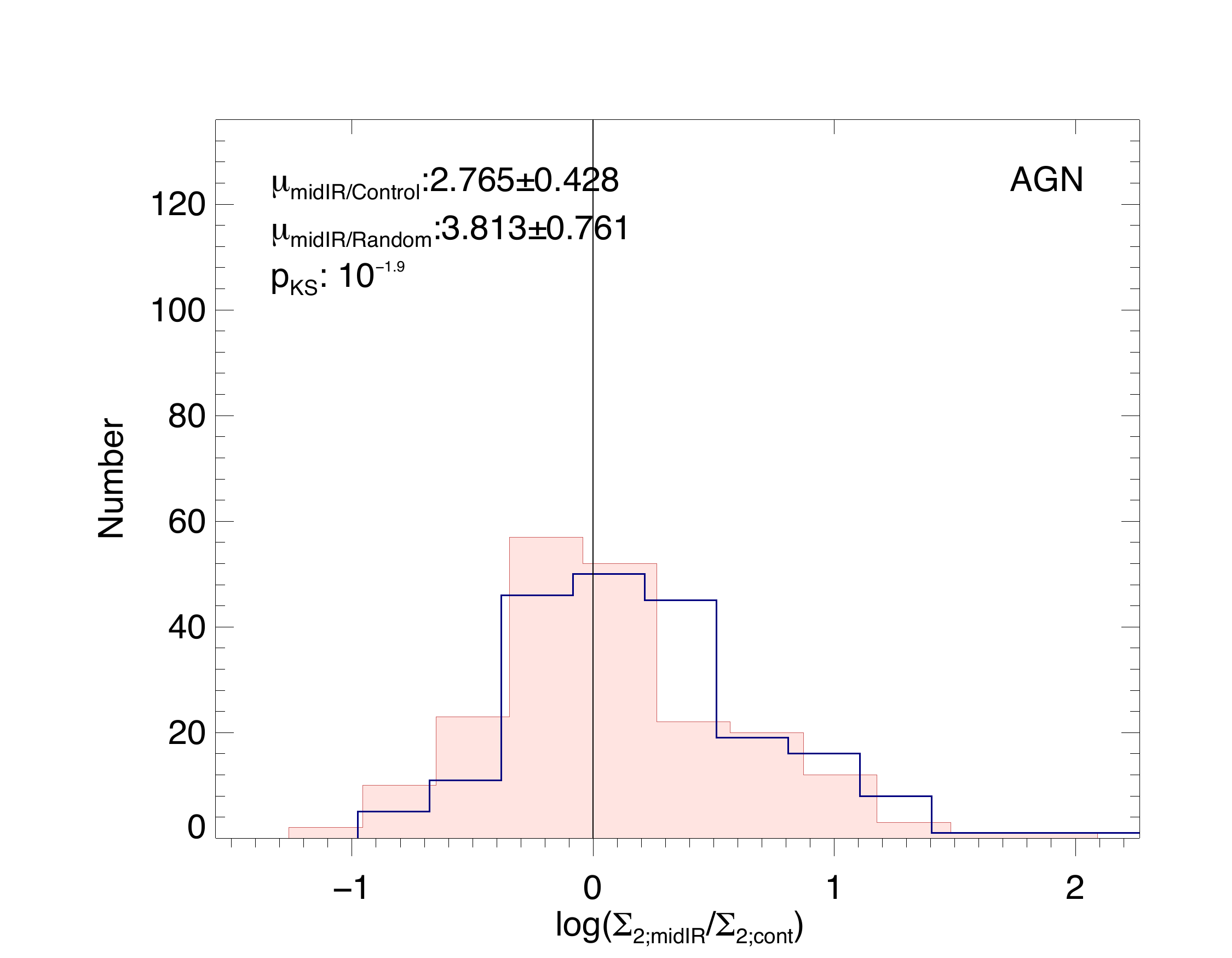}
\includegraphics[width=0.4\textwidth,angle=0]{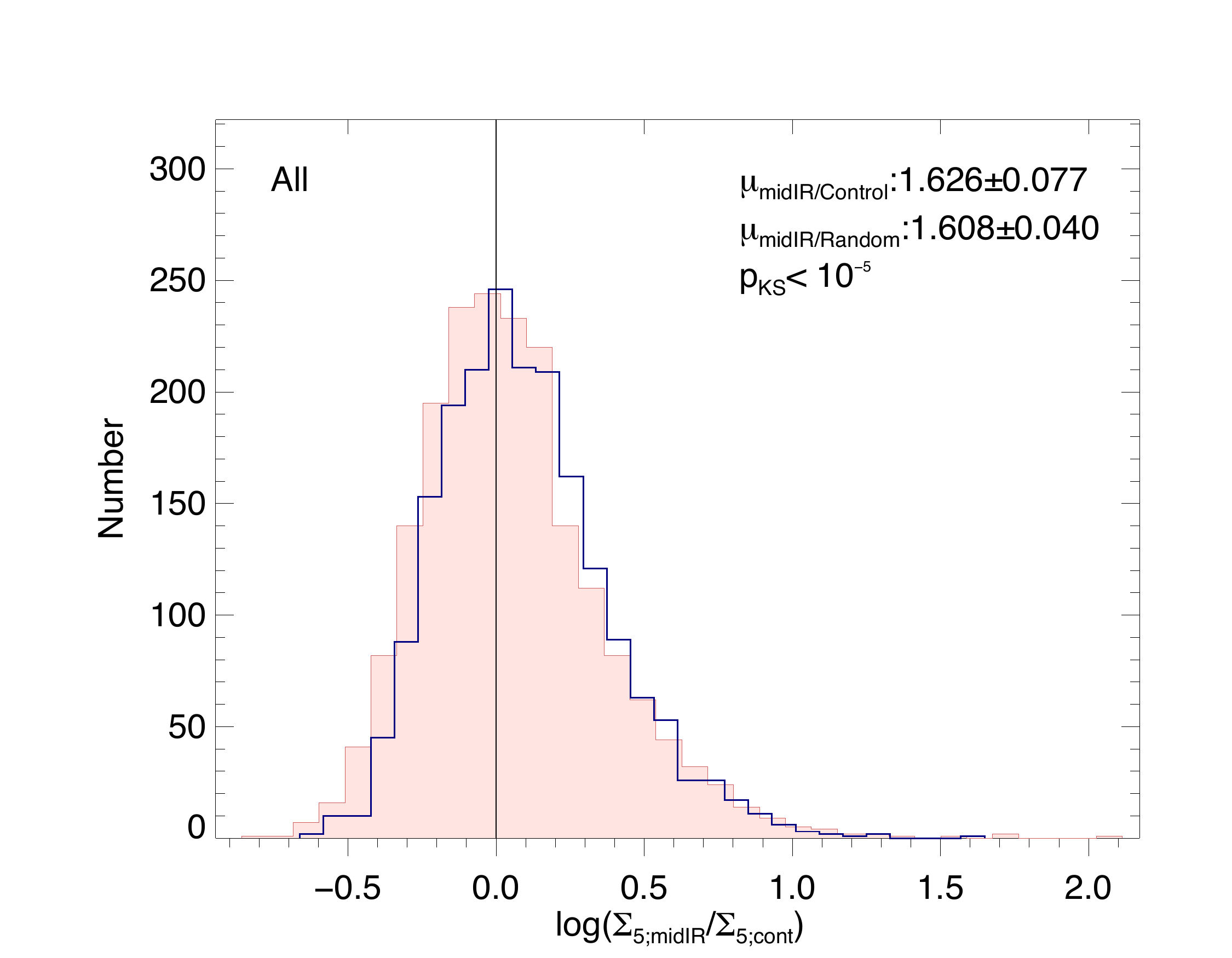}
\includegraphics[width=0.4\textwidth,angle=0]{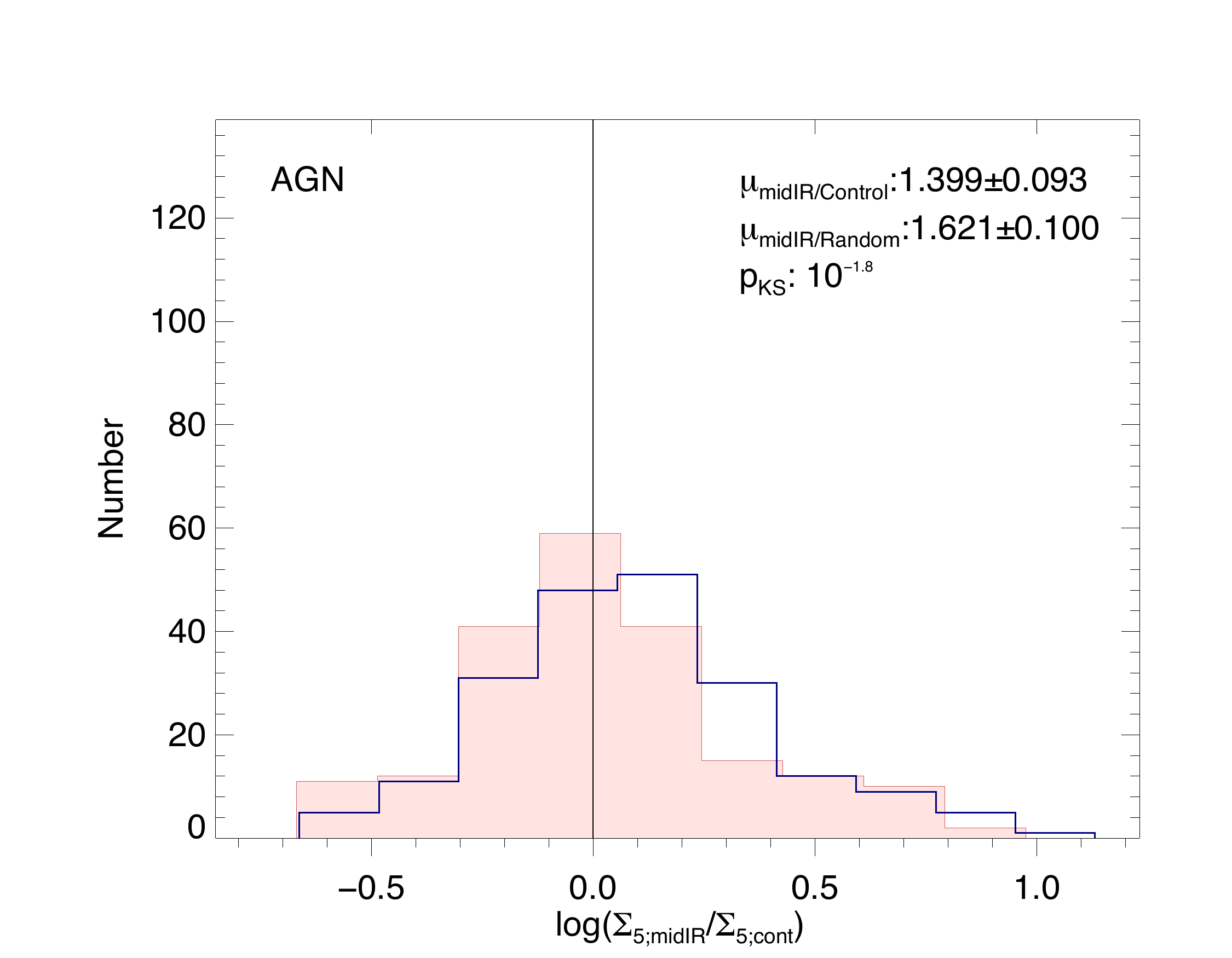}
\caption{The distributions of the logarithm of the ratio between the $\Sigma$ parameters of the mid-IR sources and that of their control sample (filled red) and the random sample (open blue) are plotted. In addition, the distributions are fitted with single Gaussians, which are shown as well. Central values and $\sigma$ values for the Gaussians are given in the each panel. The probability parameter p from a K-S comparison of the AGN/control and AGN/random density ratio distributions is also given. For reference a vertical line at a ratio value of one is drawn as well. Left panels show the distributions of all mid-IR sources while right ones show those of color-selected obscured AGN.}
\label{fig:m_histo_S2S5}
\end{center}
\end{figure*}

\begin{figure*}
\begin{center}
\includegraphics[width=0.45\textwidth,angle=0]{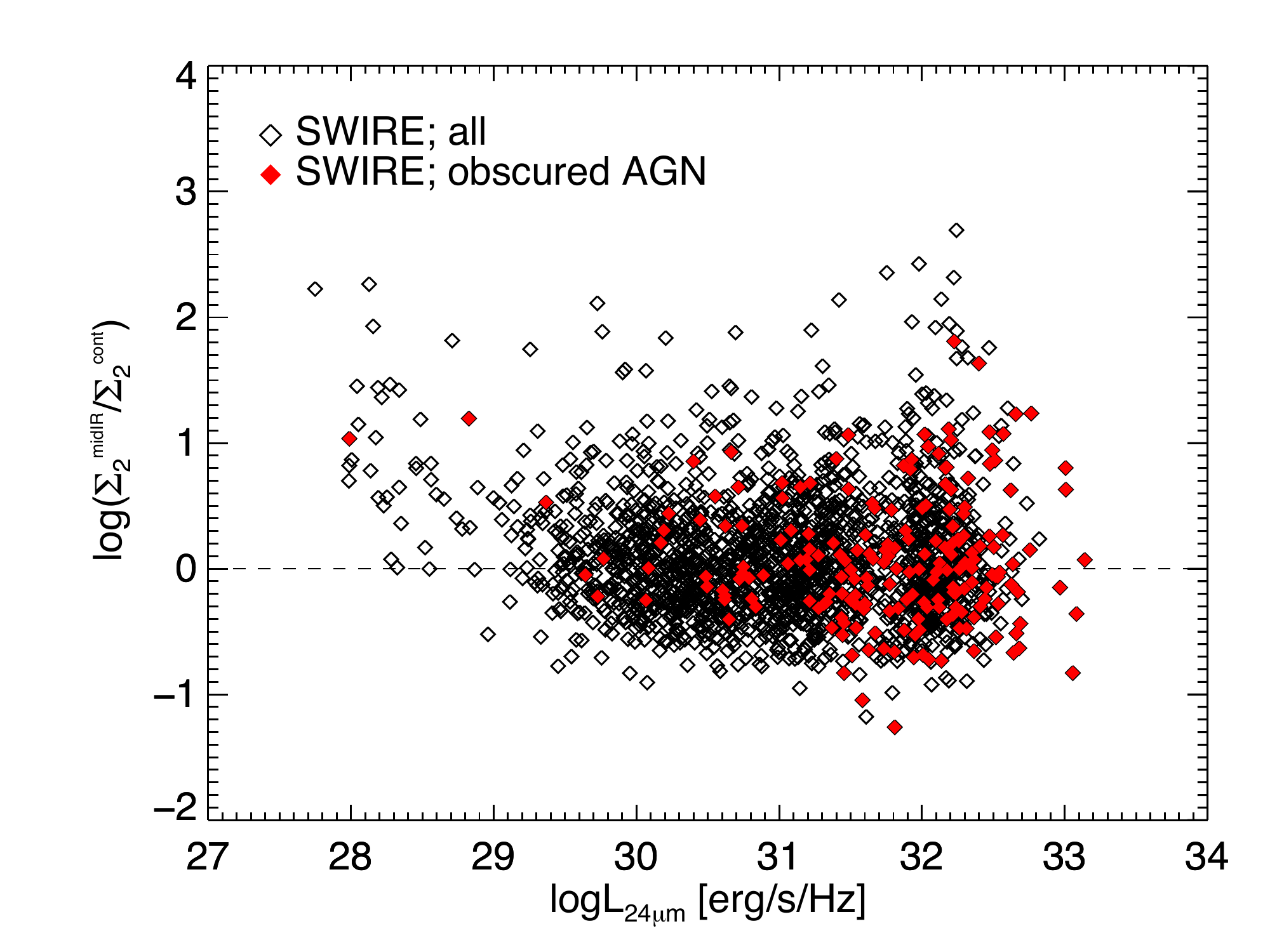}
\includegraphics[width=0.45\textwidth,angle=0]{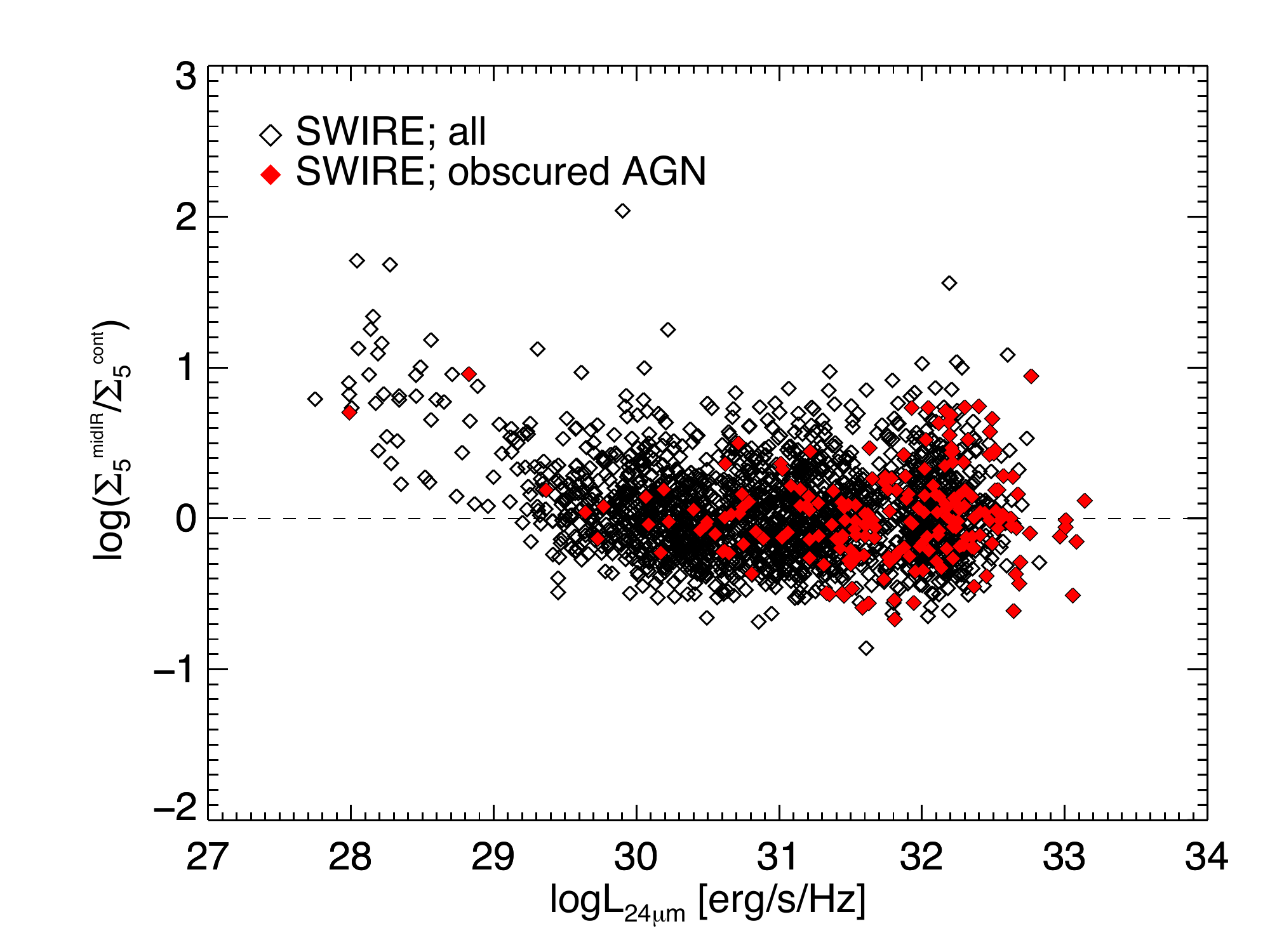}
\caption{The logarithm of the ratio of $\Sigma_{2}$ (left) and $\Sigma_{5}$ (right) density parameters between all mid-IR sources (open black) and color-selected obscured AGN (filled red) and their control samples as a function of their $24\mu m$ mid-IR luminosity. The zero line denotes the boundary between a source being in an over- or under-dense environment compared to its control sources ($\Sigma^{mid-IR}/\Sigma^{control}=1$).}
\label{fig:m_S2S5_lum}
\end{center}
\end{figure*}

The $\Sigma$ density parameters, in particular the $\Sigma_{midIR}/\Sigma_{control}$ ratio are shown in Fig. \ref{fig:m_histo_S2S5}, and Fig. \ref{fig:m_S2S5_lum}. Unsurprisingly we see the same behaviour as in our previous plots, namely that distributions peak around a ratio value of 1 (zero in logarithmic scale in the plots). However both all mid-IR and the colour-selected obscured AGN samples show an extended non-Gaussian tail reaching to high values of the density ratio. This is reflected in the average values for the $\Sigma_{2}$ distributions which are significantly higher than one. These plots imply that there is a fraction of these AGN that reside in very over-dense environments, even though on average, most of these samples reside in environments very similar to their control sources. 

In a similar manner to Fig. \ref{fig:X_S2_lum}, we check the $\Sigma_{2}$ and $\Sigma_{5}$ density parameter values for all mid-IR and for the colour-selected obscured AGN candidate sources as a function of their 24 $\mu$m mid-IR luminosity, $L_{24\mu m}$ (plotted in Fig. \ref{fig:m_S2S5_lum}). No strong trend is seen, with sources having density ratio values lying tightly around one in both $\Sigma_{2}$ and $\Sigma_{5}$ cases. Obscured AGN candidate sources do not show any different behaviour to their parent sample. Similar to the X-ray sample, here we note the low-luminosity regime, where there appears to be less scatter in density ratio values, with both parent sample and AGN sub-samples having density ratio values close to one. Similarly to the X-ray samples, the high-luminosity regime shows much larger scatter of density ratios but is still centred around one.\\

\section{Discussion}
\label{sec:discussion}
In the previous section we have presented our results concerning the close environment of AGN selected at different wavelengths. Here we discuss our results in the context of currently accepted galaxy evolution models and results from similar or adjacent studies.

\subsection{Photometric redshifts}
{Before  discussing  our results, in this section we assess the uncertainty induced by potential errors in our photometric redshift estimation, which should present the main systematic effect influencing our results. For the calculation of photometric redshifts all optical (CFHT) and near-IR (VIDEO) bands were used and a nominal accuracy of $\Delta z/(1+z)=0.1$ is achieved for the VIDEO survey as a whole (\citealt{Jarvis2013}).}

%
{However, one of the main uncertainties concerning the use of photometric redshifts pertains to the SED templates to derive them. In particular, it is known that photometric redshift estimation for active galaxies is particularly difficult (e.g., \citealt{Salvato2009} and references therein; \citealt{Salvato2011}; \citealt{Fotopoulou2012}), as the emission coming from the active nucleus can partly or fully mask the host galaxy emission, on which most SED templates rely for redshift estimation. For the photometric redshifts used here we have used the composite templates from \citet{Salvato2009}, which have been explicitly constructed in order to take into account the emission of an AGN (in particular X-ray AGN). These templates are especially relevant for our study here, as we want to characterize the environment of AGN and AGN-composite systems. }

\begin{figure}
\begin{center}
\includegraphics[width=0.45\textwidth,angle=0]{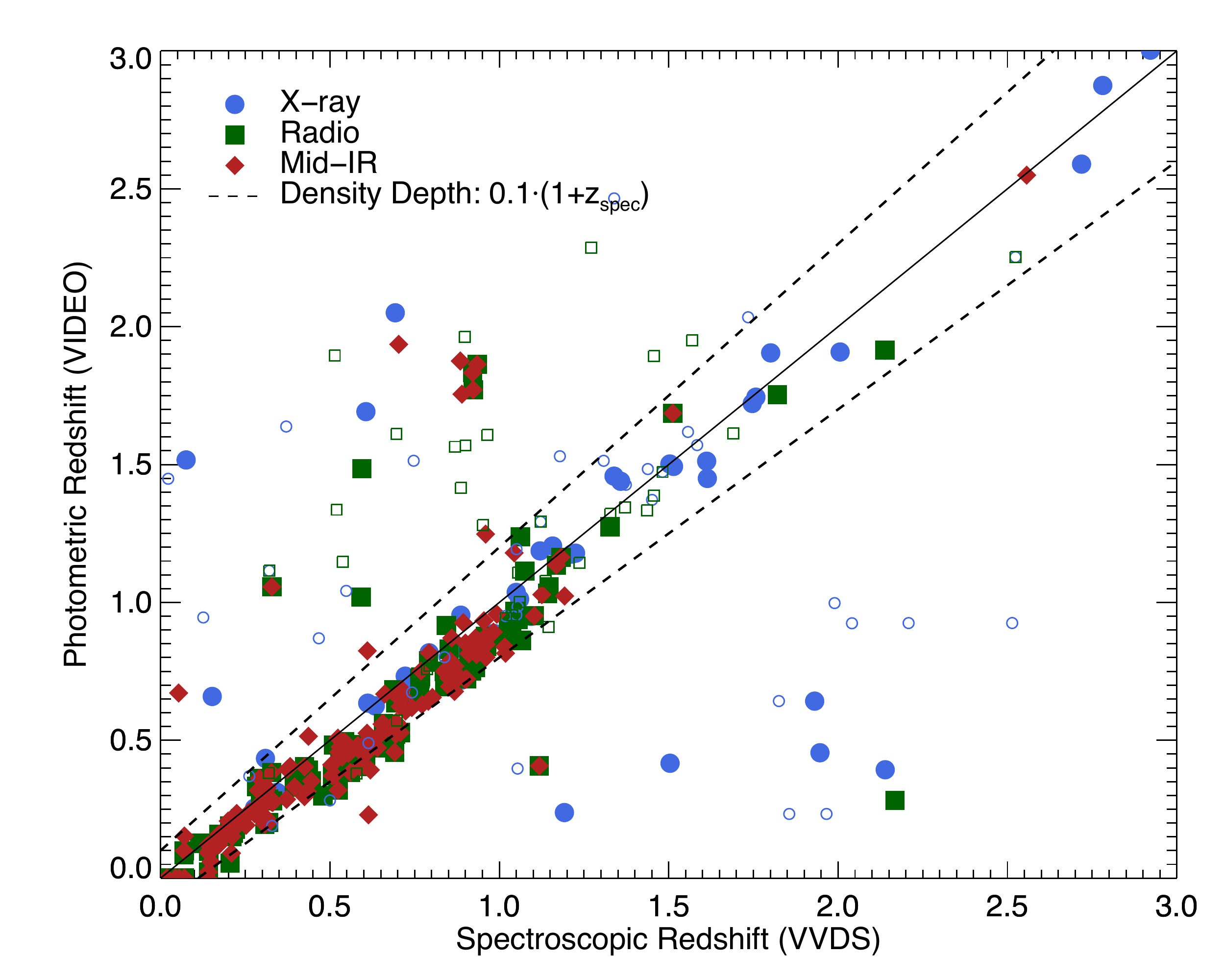}
\caption{{Spectroscopic redshifts from the VVDS survey (\citealt{LeFevre2005}) versus photometric redshifts calculated for the VIDEO sources. Different colours and symbols denote different wavelength selection: X-ray (blue circles), radio (green squares), and mid-IR (red diamonds). Filled symbols denote the best quality spectroscopic redshifts (flags:  4, 13, 14, 23, 24), while open symbols show low quality redshifts (flags: 1, 2, 11, 12, 21, 22). The dashed lines show the depth of the cylinders within which we calculate the ``pseudo-3D'' density, while the solid line show the equality line.}}
\label{fig:z_comp}
\end{center}
\end{figure}

\begin{figure}
\begin{center}
\includegraphics[width=0.45\textwidth,angle=0]{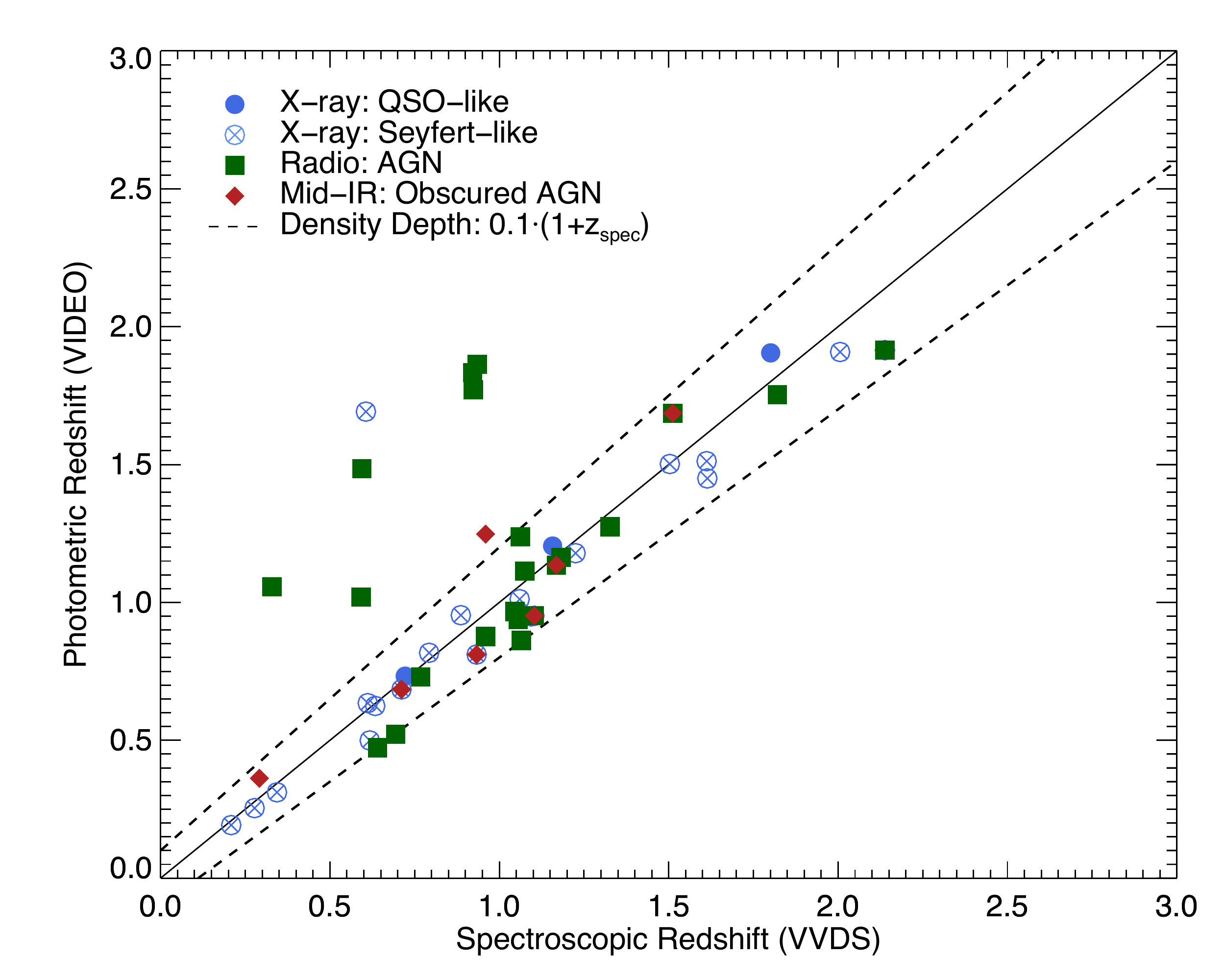}
\caption{{Same as in Fig. \ref{fig:z_comp} but now considering only the AGN-selected sources (following the criteria explained in Section \ref{sec:intro}) with high quality spectroscopic redshifts.}}
\label{fig:z_comp_AGN}
\end{center}
\end{figure}

{In Fig. \ref{fig:z_comp} we use the available spectroscopic redshift from the VIRMOS-VLT Deep Survey (VVDS;\citealt{LeFevre2005}) survey in the XMM-LSS field to assess the accuracy of the photometric redshifts of the multi-wavelength samples. For the comparison we take into account the redshift quality flag from the VVDS and focus on those sources with spectroscopic redshift quality of 3 or 4 (13 or 14 and 23 or 24 for QSO primary targets and serendipitous sources). As can be seen from Fig. \ref{fig:z_comp} there is some scatter in the estimated photometric redshifts, with a fraction of 15 per cent to 9 per cent outliers, for all and only high quality spectroscopic redshifts, respectively\footnote{Outliers are defined here as sources whose photometric redshift deviates more than $0.15\cdot(1+z_{spec})$ from their true redshift.}. The outlier fraction is almost three times that of the total VIDEO sample within the CFHTLS-VIDEO field (\citealt{Jarvis2013}). This is not surprising as the presence of an AGN can severely contaminate the host galaxy emission and therefore make photometric redshift calculation problematic. Let us now consider only those multi-wavelength sources selected as AGN (following the different criteria explain in Section \ref{sec:intro}), the comparison for which is shown in Fig.~\ref{fig:z_comp_AGN}. Out of 54 sources selected as AGN and with available spectroscopic redshifts, 7 are defined as outliers according to the usual criterion.} 

{As was described previously, we use bins of redshifts of width $0.1\cdot(1+z)$ to estimate ``pseudo-3D'' densities around a source. We are interested in quantifying how many sources would not be included in their respective redshift bin due to their miscalculated photometric redshift. The limit of $0.1\cdot(1+z_{spec})$ is shown in Figs. \ref{fig:z_comp} and \ref{fig:z_comp_AGN} with the dashed lines. Focusing on Fig. \ref{fig:z_comp_AGN}, we find that 8 out of the 54 AGN-selected sources with spectroscopic redshifts are found beyond that boundary (a fraction of 15\%). The outliers are dominated by the radio-selected AGN, with most X-ray and mid-IR selected AGN found within the boundaries defined by the dashed lines. Interestingly, all outliers are placed at higher photometric redshifts than their actual spectroscopic redshifts. This implies that in absolute terms, as Fig. \ref{fig:field} shows, these AGN would on average be found in lower number density environments than their control sample (given the flux-limited nature of the sample, the number density decreases with increasing redshift). As a result, we may be under-estimating the density around some AGN, especially the radio-selected ones. Given that especially for radio-AGN we find strong evidence that they inhabit over-dense environments, this effect would make our conclusions even stronger.}

{Finally, given the sensitivity of the density parameters to the photometric redshifts, the fact that our density differences and ratios (i.e., the difference and ratio between the density of AGN to non-AGN sources) converge to unity for the largest distances reassures us that the uncertainties of the photometric redshift do not result in a systematic offset to our environment density calculation. It is also worth noting that it is likely that the photometric redshifts for the less luminous AGN are more robust due to the lower level of AGN contamination. Thus, the comparison with the VVDS spectra, for which spectroscopic redshifts are more easily obtained for brighter sources, should be seen as a conservatively high estimate of the number of photometric-redshift outlier in our AGN sample. In reality we expect a more accurate photometric redshift for those lower-luminosity AGN.}

\subsection{Mergers as triggers for nuclear activity}
We have found that a significant fraction of the AGN sub-samples in our study show a significant excess of nearby galaxies than their respective control samples. We have extended this comparison beyond our carefully selected control sample (which in principal accounts for the influence of stellar mass on the environment of our sources) to  a sample of randomly selected field positions. Most of these comparisons support the view that these AGN  do live in dense environments, albeit not necessarily denser than their control sources. Active galaxies appear to inhabit diverse environments, as is exhibited by the distributions of $\Sigma$ parameters. It should be noted as well that a relation between environment density and AGN luminosity is not found. Individual AGN sub-samples do however show evidence of systematically denser environments than their parent samples. For example, both Seyfert-like hard X-ray sources and flat-spectrum radio-sources appear to exhibit denser surroundings than their control samples. In particular, the radio-AGN samples show a significant preference for group environments (or an excess of close companions) compared to cluster environments.

This situation is in good agreement with the most recent results concerning the morphology of X-ray and radio-selected AGN compared to respective control samples. \citet{Kocevski2012} and \citet{Cisternas2011} both use X-ray selected samples of AGN to study the morphology of active galaxies from high quality HST images and compare them to non-active galaxies. Both studies find no appreciable over-representation of mergers in the AGN samples compared to their control samples. In particular both the X-ray samples employed by \citet{Kocevski2012} and \citet{Cisternas2011} cover a significantly narrower range of X-ray luminosities ($10^{42-44}$ erg/s at 2-10 keV) than our sample. In that range of intermediate X-ray luminosities our results are in agreement (e.g., Figs. \ref{fig:X_dif_lum} and \ref{fig:X_S2_lum}).

Additionally, there has been studies advocating both that X-ray selected AGN lie in under-dense or similar environments compared to non-active galaxies (e.g., \citealt{Waskett2005},\citealt{Montero2009}, \citealt{Tasse2011}) and that X-ray selected AGN avoid under-dense environments or prefer denser ones (e.g., \citealt{Georgakakis2007},\citealt{Coil2009}). It is therefore difficult to compare our results with these studies, especially since here we cover a much wider range of luminosities and redshifts. The most extreme behavior in our X-ray samples is observed for the Seyfert-like sources, which show over-densities at all scales. Given the depth of the VIDEO survey, for the first time we can probe these type of X-ray AGN with a high-energy component relatively weaker than their IR SED. In contrast, traditionally X-ray bright QSO-like objects in our study are found in {slightly over-dense} or similar environments compared to their control sources. None of the previous studies probe the highest or lowest X-ray luminosities as our sample, with the combination of small physical distances, where the effects of AGN over-densities become most apparent.

In comparison, for radio-selected AGN, \citet{Almeida2011}, using ultra-deep imaging of the host galaxies of powerful radio-AGN find up to 93\% of their AGN sample to show morphological disturbances compared to 53\% of their control sample. \citet{Karouzos2010} also find a substantial percentage ($\sim 30\%$ at z$<0.4$) of flat-spectrum radio-AGN showing signs of merger events. Although there is a substantial bias given the high radio-luminosity of the radio-AGN in these studies, environment studies of radio-selected AGN also point to such systems being in over-dense environments (e.g., \citealt{Best2004}, \citealt{Tasse2008}, \citealt{Bradshaw2011}, \citealt{Lietzen2011}). We do recover a weak trend for our most radio-luminous sources being preferentially found in more dense environments. Flat-spectrum sources show on average over-dense environments (e.g., see Figs. \ref{fig:R_surfacedens} and \ref{fig:R_surfacecumdens}) and an excess of close companions (Fig. \ref{fig:R_group_cluster}). This may be a result of the fact that the flat-spectrum sources may be intrinsically less luminous that their radio-flux leads us to believe due to the possibility of Doppler Boosting of the radio flux due to the jet being aligned close to out line of sight (e.g. \citealt{JarvisMcLure2002}). If this is indeed the case, then we may be seeing the enhanced environment density typical of the low-luminosity FRI-type sources, or the low-excitation sources which are believed to be powered by the inefficient accretion of hot gas from the intra-cluster medium (e.g. \citealt{Hardcastle2007}).

Interestingly, not many studies have been undertaken concerning the environment of mid-IR AGN. Assuming that the local environment, studied here, can be a proxy for the probability of merger events occurring, then it would be expected that obscured AGN, as selected through their dust-reddened mid-IR colors, are found in over-dense environments. Our results from the density ratio show over-densities around these objects at close distances (around 200 kpc), while the effect disappears when looking at larger radii. In particular, we see that the total 24$\mu$m source sample appears in denser environments than their control sample at distance scales of $\simeq600$ kpc. Combined with the preference for group environments shown in Fig. \ref{fig:M_group_cluster}, this might indeed provide a hint toward the importance of mergers for this population of objects. It should be noted however that for all our samples, there also exists a component of low-density AGN environments, which persists for all density measures and also scales and is independent on luminosity.

It is obvious that there are conflicting accounts both in the literature and seemingly as well in our results here. Two scenarios can be put forward that can potentially explain our results:
\begin{enumerate}
\item Activity is triggered by either wet major mergers or through secular processes, like cooling gas accretion or bars. In this case, we expect AGN selected at difference wavelengths  to fall in either of the two ``families'' of active galaxies. This scenario is connected to the two modes of accretion in AGN, namely the ``quasar'' and ``radio'' mode, with the former triggered by mergers and the latter linked to secular processes. Moreover, luminosity segregation is expected in such a paradigm, with most luminous AGN usually being a result of mergers, while radiatively less-efficient systems are triggered by less ``dramatic'' processes.
\item Activity is predominantly triggered by mergers both major and minor. In this case  AGN selected at different wavelengths reflect different evolutionary phases of an AGN and therefore phases further ``down the line'' of evolution are expected to exhibit weaker links to the merger event that originally triggered them. In such a paradigm (following for example \citealt{Sanders1988}, \citealt{Hopkins2006b}, \citealt{Hopkins2008}) the AGN spectral energy distribution is expected to peak at different wavelength regimes at different times after the initial merger (or galactic interaction even), with the most luminous, X-ray-bright QSO phase being the last after the merger and the dust-embedded phase being the first. Where exactly a radio-phase would lie in such an evolutionary path is not clear. \citet{Karouzos2010} showed that radio-selected flat-spectrum AGN can be traced in all phases of such an evolutionary track, which implies that other factors should also be at play. An explicit luminosity segregation is not expected here.
\end{enumerate}

For both of these scenarios, some confusion is expected to be introduced to any study of AGN environment. If indeed high-luminosity AGN are triggered by mergers, while fainter ones are not, environment studies would be limited by the averaging out of the environment of these two distinct populations. We have addressed this connection here. However, the limited number of sources per wavelength selection, as well as the implicit correlation between the monochromatic luminosity, the redshift, the bolometric luminosity of a source and its stellar mass (and therefore its environment) does not allow us yet to fully address the luminosity dependance that we have recovered. The way we select our control samples is a step towards addressing this issue. More rigorous stellar mass estimation, although it would probably reduce the number of sources, might provide further insights on this problem. As it stands however, we do not find any significant correlation between the density in which the AGN resides and its luminosity. 

The second scenario is somewhat favoured by our results in the sense of the lack of any correlation between density and luminosities. It should be noted however that the AGN sources in under-dense environments we showed to exist cover the whole AGN luminosity range probed here. This implies that either these systems are exceptional cases or, more convincingly, that there exists a second fundamental parameter that governs the dependence of an AGN to its environment. By comparison to a well selected control sample, we  have shown that the stellar mass, may be the key parameter in dictating whether AGN reside in dense environments.\\
In addition, a scenario where AGN are triggered during or shortly after a merger-event (something that is supported by the duty-cycle timescale of AGN and the relaxation timescale of mergers), can be accommodated by our results. In that case (scenario 2 above) we expect a strong dependence of AGN activity on the merger rate and hence dense environments. This dependence is seen, particularly for the small scales. The fact that we find low-luminosity Seyfert-like sources and 24 $\mu$m selected sources to reside in over-dense environments might be a manifestation of the time lag between the peak of a merger and the onset of the active phase of the galaxy (e.g., \citealt{Wild2010}).

Finally, if we assume that both scenaria above might be relevant to certain AGN hosts, then further confusion is to be expected. The large uncertainties seen in our results can not be purely attributed to small number of sources in the samples studied. Case in point, large uncertainties are equally seen in our mid-IR sample which contains the largest number of sources compared to both the X-ray and radio-selected samples. Assuming that indeed our control samples selection accounts for the implicit link between the stellar mass and luminosity of a galaxy to its environment, then the large scatter observed in our plots might be indicative of the fact that we are mixing two populations of AGN that have been triggered through different means. This is further corroborated by the fact that for most of the $\Sigma_{2}$ and $\Sigma_{5}$ distributions we observe a fairly concentrated Gaussian-like distribution centred around an AGN-to-control density ratio of one, supplemented by a group of AGN in high-density environments (that form the non-Gaussian tail seen in the distributions in Figs. \ref{fig:X_histo_S2S5}, \ref{fig:r_histo_S2S5}, and \ref{fig:m_histo_S2S5}). Therefore, although we can say that for a significant fraction of active galaxies in our samples, over-dense environments point towards a merger-induced triggering, this can by no means be generalised for the whole AGN sample.

It should be noted here that VIDEO has allowed us to  investigate galaxy environments down to a very deep level with enough area to create good control samples for a range of AGN selected at multiple wavelengths. It is therefore conceivable that with the completion of this rich new dataset we are able for the first time to place AGN sources not only in the high-mass end of the galaxy mass function but rather associate nuclear activity with normal, Milky Way-like galaxies out to a redshift of $z \sim 4$. In essence, we can disentangle the study of AGN from the study of massive galaxies and thus over-dense galactic environments.

\section{Conclusions}
\label{sec:conclusion}
We use the first data release of the VISTA-VIDEO near-IR survey of part of the XMM-LSS field to study the close environment of active galaxies selected in different wavelengths, namely in the X-ray, radio, and mid-IR/24$\mu$m. To do this we employ two different environment density measures, (1) counts within a given surface area and redshift slice (pseudo-3D number density) and (2) density defined through the distance to the 2nd and 5th closest neighbour of a source. We select our AGN samples using a variety of criteria, (a) in the X-rays using a hard X-ray to infrared flux ratio (QSO- and Seyfert-like selection), (b) in the radio employing both radio-luminosity cuts and radio spectral index cuts, and (c) in the mid-IR utilising color-cuts to select dust-reddened AGN.

Summarizing, we find the following:
\begin{itemize}
\item individual sub-groups of AGN on average appear to inhabit significantly over-dense environments. In particular, flat-spectrum radio-AGN and Seyfert-like hard X-ray sources are all found in environments denser than their control samples, albeit at different scales.
\item over-density distributions for our AGN samples, although peaking at values implying no significant difference to their control samples, suggest that at least some AGN do indeed reside in some of the densest environments at the given epoch.
\item the above items combined lead us to the conclusion that AGN do live in diverse environments and thus merger-induced activity can only be relevant to a sub-population of AGN. Alternatively, systematic time lags between the different processes may be smearing out the link between mergers and nuclear activity.
\item we find no correlation between AGN luminosity and over-densities within the range of luminosities we probe here. As a result, the traditionally believed scenario of merger-induced activity for the most luminous AGN is not applicable or at least not  directly translated in terms of environment over-density.
\end{itemize}

From our results it becomes apparent that the environment density properties of the different AGN populations (mainly in terms of AGN wavelength selection and to a lesser extend AGN luminosity) are markedly different, and highly dependent on the scales at which these densities are calculated. Although it is not straightforward to disentangle the effects of stellar mass and near-IR luminosity, our results support a scenario where AGN are mostly not or are very loosely connected to their environments. However a significant fraction of AGN do seem to occupy the densest environments in the field.

Be that as it may, this study does showcase the diversity of galaxy environments in which AGN are found and reinforces a rather complicated picture of activity triggering in galaxies. We argue for a scenario where galactic interaction events (potentially minor mergers or even of the ``harassing'' type) should play a significant role in triggering activity, although considerable time lags between a merger event and the emergence of an AGN at a given wavelength regime lead to a dilution of the observational signature of this scenario. We do however find support for mechanisms where a significant fraction of AGN are triggered through secular processes. After the completion of the VIDEO survey, an excellent dataset will be available to re-address the same questions with much higher precision.

\section*{Acknowledgments}
The authors thank the anonymous referee for his suggestions that improved this manuscript. This work was supported by the National Research Foundation of Korea (NRF) grant, No. 2008-0060544, funded by the Korea government (MSIP). MK was partially supported by the EU COST Action ``Black Holes in a Violent Universe''. MK acknowledges the support and hospitality of the Centre for Astrophysics Research (CAR) at the University of Hertfordshire, where part of this work has taken place. MK also wants to warmly thank Mar Mezcua for carefully reading this manuscript and for providing insightful comments and Jae-Woo Kim for interesting discussions that improved this work.

\bibliographystyle{mn2e}
\bibliography{bibtex}

\begin{thebibliography}{76}
\expandafter\ifx\csname natexlab\endcsname\relax\def\natexlab#1{#1}\fi

\bibitem[{{Arnouts} {et~al.}(1999){Arnouts}, {Cristiani}, {Moscardini},
  {Matarrese}, {Lucchin}, {Fontana}, \& {Giallongo}}]{Arnouts1999}
{Arnouts}, S., {Cristiani}, S., {Moscardini}, L., {Matarrese}, S., {Lucchin},
  F., {Fontana}, A., \& {Giallongo}, E. 1999, \mnras, 310, 540

\bibitem[{{Bennert} {et~al.}(2008){Bennert}, {Canalizo}, {Jungwiert},
  {Stockton}, {Schweizer}, {Peng}, \& {Lacy}}]{Bennert2008}
{Bennert}, N., {Canalizo}, G., {Jungwiert}, B., {Stockton}, A., {Schweizer},
  F., {Peng}, C.~Y., \& {Lacy}, M. 2008, \apj, 677, 846

\bibitem[{{Bertin} \& {Arnouts}(1996)}]{Bertin1996}
{Bertin}, E. \& {Arnouts}, S. 1996, \aaps, 117, 393

\bibitem[{{Bertin} {et~al.}(2002){Bertin}, {Mellier}, {Radovich}, {Missonnier},
  {Didelon}, \& {Morin}}]{Bertin2002}
{Bertin}, E., {Mellier}, Y., {Radovich}, M., {Missonnier}, G., {Didelon}, P.,
  \& {Morin}, B. 2002, in Astronomical Society of the Pacific Conference
  Series, Vol. 281, Astronomical Data Analysis Software and Systems XI, ed.
  {D.~A.~Bohlender, D.~Durand, \& T.~H.~Handley}, 228--+

\bibitem[{{Best}(2004)}]{Best2004}
{Best}, P.~N. 2004, \mnras, 351, 70

\bibitem[{{Blandford}(1986)}]{Blandford1986}
{Blandford}, R.~D. 1986, in IAU Symposium, Vol. 119, Quasars, ed. G.~{Swarup}
  \& V.~K. {Kapahi}, 359--368

\bibitem[{{Bondi} {et~al.}(2007){Bondi}, {Ciliegi}, {Venturi}, {Dallacasa},
  {Bardelli}, {Zucca}, {Athreya}, {Gregorini}, {Zanichelli}, {Le F{\`e}vre},
  {Contini}, {Garilli}, {Iovino}, {Temporin}, \& {Vergani}}]{Bondi2007}
{Bondi}, M., {et~al.} 2007, \aap, 463, 519

\bibitem[{{Bondi} {et~al.}(2003){Bondi}, {Ciliegi}, {Zamorani}, {Gregorini},
  {Vettolani}, {Parma}, {de Ruiter}, {Le Fevre}, {Arnaboldi}, {Guzzo}, \&
  {Maccagni}}]{Bondi2003}
{Bondi}, M., {et~al.} 2003, \aap, 403, 857

\bibitem[{{Bradshaw} {et~al.}(2011){Bradshaw}, {Almaini}, {Hartley}, {Chuter},
  {Simpson}, {Conselice}, {Dunlop}, {McLure}, \& {Cirasuolo}}]{Bradshaw2011}
{Bradshaw}, E.~J., {et~al.} 2011, ArXiv e-prints

\bibitem[{{Canalizo} \& {Stockton}(2001)}]{Canalizo2001}
{Canalizo}, G. \& {Stockton}, A. 2001, \apj, 555, 719

\bibitem[{{Cattaneo} {et~al.}(2005){Cattaneo}, {Combes}, {Colombi}, {Bertin},
  \& {Melchior}}]{Cattaneo2005}
{Cattaneo}, A., {Combes}, F., {Colombi}, S., {Bertin}, E., \& {Melchior}, A.
  2005, \mnras, 359, 1237

\bibitem[{{Cisternas} {et~al.}(2011){Cisternas}, {Jahnke}, {Inskip},
  {Kartaltepe}, {Koekemoer}, {Lisker}, {Robaina}, \&
  {Scodeggio}}]{Cisternas2011}
{Cisternas}, M., {Jahnke}, K., {Inskip}, K.~J., {Kartaltepe}, J., {Koekemoer},
  A.~M., {Lisker}, T., {Robaina}, A.~R., \& {Scodeggio}, M. e.~a. 2011, \apj,
  726, 57

\bibitem[{Coil {et~al.}(2009)Coil, Georgakakis, Newman, Cooper, Croton, Davis,
  Koo, Laird, Nandra, Weiner, Willmer, \& Yan}]{Coil2009}
Coil, A.~L., {et~al.} 2009, The Astrophysical Journal, 701, 1484

\bibitem[{{Condon}(1992)}]{Condon1992}
{Condon}, J.~J. 1992, \araa, 30, 575

\bibitem[{{Cooper} {et~al.}(2005){Cooper}, {Newman}, {Madgwick}, {Gerke},
  {Yan}, \& {Davis}}]{Cooper2005}
{Cooper}, M.~C., {Newman}, J.~A., {Madgwick}, D.~S., {Gerke}, B.~F., {Yan}, R.,
  \& {Davis}, M. 2005, \apj, 634, 833

\bibitem[{{Darg} {et~al.}(2010){Darg}, {Kaviraj}, {Lintott}, {Schawinski},
  {Sarzi}, {Bamford}, {Silk}, {Andreescu}, {Murray}, {Nichol}, {Raddick},
  {Slosar}, {Szalay}, {Thomas}, \& {Vandenberg}}]{Darg2010}
{Darg}, D.~W., {et~al.} 2010, \mnras, 401, 1552

\bibitem[{{Deng} {et~al.}(2011){Deng}, {Chen}, \& {Jiang}}]{Deng2011}
{Deng}, X.-F., {Chen}, Y.-Q., \& {Jiang}, P. 2011, \mnras, 417, 453

\bibitem[{{Donley} {et~al.}(2012){Donley}, {Koekemoer}, {Brusa}, {Capak},
  {Cardamone}, {Civano}, {Ilbert}, {Impey}, {Kartaltepe}, {Miyaji}, {Salvato},
  {Sanders}, {Trump}, \& {Zamorani}}]{Donley2012}
{Donley}, J.~L., {et~al.} 2012, \apj, 748, 142

\bibitem[{{Downes} {et~al.}(1986){Downes}, {Peacock}, {Savage}, \&
  {Carrie}}]{Downes1986}
{Downes}, A.~J.~B., {Peacock}, J.~A., {Savage}, A., \& {Carrie}, D.~R. 1986,
  \mnras, 218, 31

\bibitem[{{Dressler}(1980)}]{Dressler1980}
{Dressler}, A. 1980, \apj, 236, 351

\bibitem[{{Eisenhauer} {et~al.}(2005){Eisenhauer}, {Genzel}, {Alexander},
  {Abuter}, {Paumard}, {Ott}, {Gilbert}, {Gillessen}, \&
  {Horrobin}}]{Eisenhauer2005}
{Eisenhauer}, F., {et~al.} 2005, \apj, 628, 246

\bibitem[{{Ellison} {et~al.}(2011){Ellison}, {Patton}, {Mendel}, \&
  {Scudder}}]{Ellison2011}
{Ellison}, S.~L., {Patton}, D.~R., {Mendel}, J.~T., \& {Scudder}, J.~M. 2011,
  \mnras, 418, 2043

\bibitem[{{Emerson} \& {Sutherland}(2010)}]{Emerson2010}
{Emerson}, J.~P. \& {Sutherland}, W.~J. 2010, in Society of Photo-Optical
  Instrumentation Engineers (SPIE) Conference Series, Vol. 7733, Society of
  Photo-Optical Instrumentation Engineers (SPIE) Conference Series

\bibitem[{{Fotopoulou} {et~al.}(2012){Fotopoulou}, {Salvato}, {Hasinger},
  {Rovilos}, {Brusa}, {Egami}, {Lutz}, {Burwitz}, {Henry}, {Huang},
  {Rigopoulou}, \& {Vaccari}}]{Fotopoulou2012}
{Fotopoulou}, S., {et~al.} 2012, \apjs, 198, 1

\bibitem[{{Frenk} {et~al.}(1988){Frenk}, {White}, {Davis}, \&
  {Efstathiou}}]{Frenk1988}
{Frenk}, C.~S., {White}, S.~D.~M., {Davis}, M., \& {Efstathiou}, G. 1988, \apj,
  327, 507

\bibitem[{{Gavazzi} {et~al.}(1996){Gavazzi}, {Pierini}, \&
  {Boselli}}]{Gavazzi1996}
{Gavazzi}, G., {Pierini}, D., \& {Boselli}, A. 1996, \aap, 312, 397

\bibitem[{{Genzel} {et~al.}(2000){Genzel}, {Pichon}, {Eckart}, {Gerhard}, \&
  {Ott}}]{Genzel2000}
{Genzel}, R., {Pichon}, C., {Eckart}, A., {Gerhard}, O.~E., \& {Ott}, T. 2000,
  \mnras, 317, 348

\bibitem[{{Georgakakis} {et~al.}(2007){Georgakakis}, {Nandra}, {Laird},
  {Cooper}, {Gerke}, {Newman}, {Croton}, {Davis}, {Faber}, \&
  {Coil}}]{Georgakakis2007}
{Georgakakis}, A., {et~al.} 2007, \apjl, 660, L15

\bibitem[{{Grogin} {et~al.}(2005){Grogin}, {Conselice}, {Chatzichristou},
  {Alexander}, {Bauer}, {Hornschemeier}, {Jogee}, {Koekemoer}, {Laidler},
  {Livio}, {Lucas}, {Paolillo}, {Ravindranath}, {Schreier}, {Simmons}, \&
  {Urry}}]{Grogin2005}
{Grogin}, N.~A., {et~al.} 2005, \apjl, 627, L97

\bibitem[{{Hardcastle} {et~al.}(2007){Hardcastle}, {Evans}, \&
  {Croston}}]{Hardcastle2007}
{Hardcastle}, M.~J., {Evans}, D.~A., \& {Croston}, J.~H. 2007, \mnras, 376,
  1849

\bibitem[{{Hernquist}(1989)}]{Hernquist1989}
{Hernquist}, L. 1989, \nat, 340, 687

\bibitem[{{Hopkins} \& {Hernquist}(2006)}]{Hopkins2006b}
{Hopkins}, P.~F. \& {Hernquist}, L. 2006, \apjs, 166, 1

\bibitem[{{Hopkins} {et~al.}(2006){Hopkins}, {Hernquist}, {Cox}, {Di Matteo},
  {Robertson}, \& {Springel}}]{Hopkins2006}
{Hopkins}, P.~F., {Hernquist}, L., {Cox}, T.~J., {Di Matteo}, T., {Robertson},
  B., \& {Springel}, V. 2006, \apjs, 163, 1

\bibitem[{{Hopkins} {et~al.}(2008){Hopkins}, {Hernquist}, {Cox}, \& {Kere{\v
  s}}}]{Hopkins2008}
{Hopkins}, P.~F., {Hernquist}, L., {Cox}, T.~J., \& {Kere{\v s}}, D. 2008,
  \apjs, 175, 356

\bibitem[{{Ilbert} {et~al.}(2006){Ilbert}, {Arnouts}, {McCracken},
  {Bolzonella}, {Bertin}, {Le F{\`e}vre}, {Mellier}, {Zamorani}, {Pell{\`o}},
  {Iovino}, {Tresse}, \& {Le Brun}}]{Ilbert2006}
{Ilbert}, O., {et~al.} 2006, \aap, 457, 841

\bibitem[{{Jarvis} {et~al.}(2013){Jarvis}, {Bonfield}, {Bruce}, {Geach},
  {McAlpine}, {McLure}, {Gonz{\'a}lez-Solares}, {Irwin}, {Lewis}, {Yoldas},
  {Andreon}, {Cross}, {Emerson}, {Dalton}, {Dunlop}, {Hodgkin}, {Le}, \&
  {Karouzos}}]{Jarvis2013}
{Jarvis}, M.~J., {et~al.} 2013, \mnras, 428, 1281

\bibitem[{{Jarvis} \& {McLure}(2002)}]{JarvisMcLure2002}
{Jarvis}, M.~J. \& {McLure}, R.~J. 2002, \mnras, 336, L38

\bibitem[{{Karouzos} {et~al.}(2010){Karouzos}, {Britzen}, {Eckart}, {Witzel},
  \& {Zensus}}]{Karouzos2010}
{Karouzos}, M., {Britzen}, S., {Eckart}, A., {Witzel}, A., \& {Zensus}, A.
  2010, ArXiv e-prints

\bibitem[{{Kauffmann} \& {Haehnelt}(2000)}]{Kauffmann2000}
{Kauffmann}, G. \& {Haehnelt}, M. 2000, \mnras, 311, 576

\bibitem[{{Kauffmann} {et~al.}(2004){Kauffmann}, {White}, {Heckman},
  {M{\'e}nard}, {Brinchmann}, {Charlot}, {Tremonti}, \&
  {Brinkmann}}]{Kauffmann2004}
{Kauffmann}, G., {White}, S.~D.~M., {Heckman}, T.~M., {M{\'e}nard}, B.,
  {Brinchmann}, J., {Charlot}, S., {Tremonti}, C., \& {Brinkmann}, J. 2004,
  \mnras, 353, 713

\bibitem[{{Kocevski} {et~al.}(2012){Kocevski}, {Faber}, {Mozena}, {Koekemoer},
  {Nandra}, {Rangel}, {Laird}, {Brusa}, {Wuyts}, {Trump}, {Koo}, {Somerville},
  {Bell}, {Lotz}, {Alexander}, {Bournaud}, \& {Conselice}}]{Kocevski2012}
{Kocevski}, D.~D., {et~al.} 2012, \apj, 744, 148

\bibitem[{{Komatsu} {et~al.}(2011){Komatsu}, {Smith}, {Dunkley}, {Bennett},
  {Gold}, {Hinshaw}, {Jarosik}, {Larson}, {Nolta}, {Page}, \&
  {Spergel}}]{Komatsu2011}
{Komatsu}, E., {et~al.} 2011, \apjs, 192, 18

\bibitem[{{Kormendy} \& {Ho}(2013)}]{Kormendy2013}
{Kormendy}, J. \& {Ho}, L.~C. 2013, \araa, 51, 511

\bibitem[{{Lacy} {et~al.}(2007){Lacy}, {Petric}, {Sajina}, {Canalizo},
  {Storrie-Lombardi}, {Armus}, {Fadda}, \& {Marleau}}]{Lacy2007}
{Lacy}, M., {Petric}, A.~O., {Sajina}, A., {Canalizo}, G., {Storrie-Lombardi},
  L.~J., {Armus}, L., {Fadda}, D., \& {Marleau}, F.~R. 2007, \aj, 133, 186

\bibitem[{{Lacy} {et~al.}(2004){Lacy}, {Storrie-Lombardi}, {Sajina},
  {Appleton}, {Armus}, {Chapman}, {Choi}, {Fadda}, {Fang}, {Frayer}, \&
  {Heinrichsen}}]{Lacy2004}
{Lacy}, M., {et~al.} 2004, \apjs, 154, 166

\bibitem[{{Le F{\`e}vre} {et~al.}(2005){Le F{\`e}vre}, {Vettolani}, {Garilli},
  {Tresse}, {Bottini}, {Le Brun}, {Maccagni}, {Picat}, {Scaramella},
  {Scodeggio}, {Zanichelli}, {Adami}, {Arnaboldi}, {Arnouts}, {Bardelli},
  {Bolzonella}, {Cappi}, {Charlot}, {Ciliegi}, {Contini}, {Foucaud},
  {Franzetti}, {Gavignaud}, {Guzzo}, {Ilbert}, {Iovino}, {McCracken}, {Marano},
  {Marinoni}, {Mathez}, {Mazure}, {Meneux}, {Merighi}, {Paltani}, {Pell{\`o}},
  {Pollo}, {Pozzetti}, {Radovich}, {Zamorani}, {Zucca}, {Bondi}, {Bongiorno},
  {Busarello}, {Lamareille}, {Mellier}, {Merluzzi}, {Ripepi}, \&
  {Rizzo}}]{LeFevre2005}
{Le F{\`e}vre}, O., {et~al.} 2005, \aap, 439, 845

\bibitem[{{Le F{\`e}vre} {et~al.}(2002){Le F{\`e}vre}, {Vettolani}, {Maccagni},
  {Mancini}, {Picat}, \& {Saisse}}]{Lefevre2002}
{Le F{\`e}vre}, O., {Vettolani}, G., {Maccagni}, D., {Mancini}, D., {Picat},
  J.-P., \& {Saisse}, M. 2002, in Astronomical Society of the Pacific
  Conference Series, Vol. 280, Next Generation Wide-Field Multi-Object
  Spectroscopy, ed. {M.~J.~I.~Brown \& A.~Dey}, 117--+

\bibitem[{{Lee} {et~al.}(2010){Lee}, {Lee}, {Park}, \& {Choi}}]{Lee2010}
{Lee}, J.~H., {Lee}, M.~G., {Park}, C., \& {Choi}, Y.-Y. 2010, \mnras, 403,
  1930

\bibitem[{{Li} {et~al.}(2006){Li}, {Kauffmann}, {Jing}, {White}, {B{\"o}rner},
  \& {Cheng}}]{Li2006}
{Li}, C., {Kauffmann}, G., {Jing}, Y.~P., {White}, S.~D.~M., {B{\"o}rner}, G.,
  \& {Cheng}, F.~Z. 2006, \mnras, 368, 21

\bibitem[{{Lietzen} {et~al.}(2011){Lietzen}, {Hein{\"a}m{\"a}ki}, {Nurmi},
  {Liivam{\"a}gi}, {Saar}, {Tago}, {Takalo}, \& {Einasto}}]{Lietzen2011}
{Lietzen}, H., {Hein{\"a}m{\"a}ki}, P., {Nurmi}, P., {Liivam{\"a}gi}, L.~J.,
  {Saar}, E., {Tago}, E., {Takalo}, L.~O., \& {Einasto}, M. 2011, ArXiv
  e-prints

\bibitem[{{Lonsdale} {et~al.}(2003){Lonsdale}, {Smith}, {Rowan-Robinson},
  {Surace}, {Shupe}, {Xu}, {Oliver}, {Padgett}, {Fang}, {Conrow},
  {Franceschini}, {Gautier}, {Griffin}, \& {Hacking}}]{Lonsdale2003}
{Lonsdale}, C.~J., {et~al.} 2003, \pasp, 115, 897

\bibitem[{{Lotz} {et~al.}(2008){Lotz}, {Jonsson}, {Cox}, \&
  {Primack}}]{Lotz2008}
{Lotz}, J.~M., {Jonsson}, P., {Cox}, T.~J., \& {Primack}, J.~R. 2008, ArXiv
  e-prints

\bibitem[{{Mauch} \& {Sadler}(2007)}]{Mauch2007}
{Mauch}, T. \& {Sadler}, E.~M. 2007, \mnras, 375, 931

\bibitem[{{McAlpine} {et~al.}(2012){McAlpine}, {Smith}, {Jarvis}, {Bonfield},
  \& {Fleuren}}]{McAlpine2012}
{McAlpine}, K., {Smith}, D.~J.~B., {Jarvis}, M.~J., {Bonfield}, D.~G., \&
  {Fleuren}, S. 2012, \mnras, 423, 132

\bibitem[{{McLure} \& {Dunlop}(2002)}]{McLure2002}
{McLure}, R.~J. \& {Dunlop}, J.~S. 2002, \mnras, 331, 795

\bibitem[{{Merritt} \& {Ferrarese}(2001)}]{Merritt2001}
{Merritt}, D. \& {Ferrarese}, L. 2001, \apj, 547, 140

\bibitem[{{Miller} {et~al.}(2003){Miller}, {Nichol}, {G{\'o}mez}, {Hopkins}, \&
  {Bernardi}}]{Miller2003}
{Miller}, C.~J., {Nichol}, R.~C., {G{\'o}mez}, P.~L., {Hopkins}, A.~M., \&
  {Bernardi}, M. 2003, \apj, 597, 142

\bibitem[{{Montero-Dorta} {et~al.}(2009){Montero-Dorta}, {Croton}, {Yan},
  {Cooper}, {Newman}, {Georgakakis}, {Prada}, {Davis}, {Nandra}, \&
  {Coil}}]{Montero2009}
{Montero-Dorta}, A.~D., {et~al.} 2009, \mnras, 392, 125

\bibitem[{{Nagar} {et~al.}(2003){Nagar}, {Wilson}, {Falcke}, {Veilleux}, \&
  {Maiolino}}]{Nagar2003}
{Nagar}, N.~M., {Wilson}, A.~S., {Falcke}, H., {Veilleux}, S., \& {Maiolino},
  R. 2003, \aap, 409, 115

\bibitem[{{Pierre} {et~al.}(2007){Pierre}, {Chiappetti}, {Pacaud}, {Gueguen},
  {Libbrecht}, {Altieri}, {Aussel}, {Gandhi}, {Garcet}, {Gosset}, {Paioro},
  {Ponman}, {Read}, {Refregier}, \& {Starck}}]{Pierre2007}
{Pierre}, M., {et~al.} 2007, \mnras, 382, 279

\bibitem[{{Ramos Almeida} {et~al.}(2011){Ramos Almeida}, {Bessiere},
  {Tadhunter}, {P{\'e}rez-Gonz{\'a}lez}, {Barro}, {Inskip}, {Morganti}, {Holt},
  \& {Dicken}}]{Almeida2011}
{Ramos Almeida}, C., {et~al.} 2011, \mnras, 1702

\bibitem[{{Rees}(1978)}]{Rees1978}
{Rees}, M.~J. 1978, \physscr, 17, 193

\bibitem[{{Richstone} {et~al.}(1998){Richstone}, {Ajhar}, {Bender}, {Bower},
  {Dressler}, {Faber}, {Filippenko}, {Gebhardt}, {Green}, {Ho}, {Kormendy},
  {Lauer}, {Magorrian}, \& {Tremaine}}]{Richstone1998}
{Richstone}, D., {et~al.} 1998, \nat, 395, A14+

\bibitem[{{Salvato} {et~al.}(2009){Salvato}, {Hasinger}, {Ilbert}, {Zamorani},
  {Brusa}, {Scoville}, {Rau}, {Capak}, {Arnouts}, {Aussel}, {Bolzonella},
  {Buongiorno}, {Cappelluti}, {Caputi}, \& {Civano}}]{Salvato2009}
{Salvato}, M., {et~al.} 2009, \apj, 690, 1250

\bibitem[{{Salvato} {et~al.}(2011){Salvato}, {Ilbert}, {Hasinger}, {Rau},
  {Civano}, {Zamorani}, {Brusa}, {Elvis}, {Vignali}, {Aussel}, {Comastri},
  {Fiore}, {Le Floc'h}, {Mainieri}, \& {Bardelli}}]{Salvato2011}
{Salvato}, M., {et~al.} 2011, \apj, 742, 61

\bibitem[{{Sanders} {et~al.}(1988){Sanders}, {Soifer}, {Elias}, {Neugebauer},
  \& {Matthews}}]{Sanders1988}
{Sanders}, D.~B., {Soifer}, B.~T., {Elias}, J.~H., {Neugebauer}, G., \&
  {Matthews}, K. 1988, \apjl, 328, L35

\bibitem[{{Serber} {et~al.}(2006){Serber}, {Bahcall}, {M{\'e}nard}, \&
  {Richards}}]{Serber2006}
{Serber}, W., {Bahcall}, N., {M{\'e}nard}, B., \& {Richards}, G. 2006, \apj,
  643, 68

\bibitem[{{Silverman} {et~al.}(2011){Silverman}, {Kampczyk}, {Jahnke},
  {Andrae}, {Lilly}, {Elvis}, {Civano}, {Mainieri}, {Vignali}, {Zamorani},
  {Nair}, {Le F{\`e}vre}, {de Ravel}, {Bardelli}, {Bongiorno}, {Bolzonella},
  {Cappi}, {Caputi}, {Carollo}, {Contini}, {Coppa}, {Cucciati}, {de la Torre},
  {Franzetti}, {Garilli}, {Halliday}, {Hasinger}, {Iovino}, {Knobel},
  {Koekemoer}, {Kova{\v c}}, {Lamareille}, {Le Borgne}, {Le Brun}, {Maier},
  {Mignoli}, {Pello}, {P{\'e}rez-Montero}, {Ricciardelli}, {Peng}, {Scodeggio},
  {Tanaka}, {Tasca}, {Tresse}, {Vergani}, {Zucca}, {Brusa}, {Cappelluti},
  {Comastri}, {Finoguenov}, {Fu}, {Gilli}, {Hao}, {Ho}, \&
  {Salvato}}]{Silverman2011}
{Silverman}, J.~D., {et~al.} 2011, \apj, 743, 2

\bibitem[{{Strand} {et~al.}(2008){Strand}, {Brunner}, \& {Myers}}]{Strand2008}
{Strand}, N.~E., {Brunner}, R.~J., \& {Myers}, A.~D. 2008, \apj, 688, 180

\bibitem[{{Tasse} {et~al.}(2008){Tasse}, {Best}, {R{\"o}ttgering}, \& {Le
  Borgne}}]{Tasse2008}
{Tasse}, C., {Best}, P.~N., {R{\"o}ttgering}, H., \& {Le Borgne}, D. 2008,
  \aap, 490, 893

\bibitem[{{Tasse} {et~al.}(2011){Tasse}, {R{\"o}ttgering}, \&
  {Best}}]{Tasse2011}
{Tasse}, C., {R{\"o}ttgering}, H., \& {Best}, P.~N. 2011, \aap, 525, A127+

\bibitem[{{Toomre} \& {Toomre}(1972)}]{Toomre1972}
{Toomre}, A. \& {Toomre}, J. 1972, \apj, 178, 623

\bibitem[{{Tremaine} {et~al.}(2002){Tremaine}, {Gebhardt}, {Bender}, {Bower},
  {Dressler}, {Faber}, {Filippenko}, {Green}, {Grillmair}, {Ho}, {Kormendy},
  {Lauer}, {Magorrian}, {Pinkney}, \& {Richstone}}]{Tremaine2002}
{Tremaine}, S., {et~al.} 2002, \apj, 574, 740

\bibitem[{{Veilleux} {et~al.}(1995){Veilleux}, {Kim}, {Sanders}, {Mazzarella},
  \& {Soifer}}]{Veilleux1995}
{Veilleux}, S., {Kim}, D., {Sanders}, D.~B., {Mazzarella}, J.~M., \& {Soifer},
  B.~T. 1995, \apjs, 98, 171

\bibitem[{{Waskett} {et~al.}(2005){Waskett}, {Eales}, {Gear}, {McCracken},
  {Lilly}, \& {Brodwin}}]{Waskett2005}
{Waskett}, T.~J., {Eales}, S.~A., {Gear}, W.~K., {McCracken}, H.~J., {Lilly},
  S., \& {Brodwin}, M. 2005, \mnras, 363, 801

\bibitem[{{Wild} {et~al.}(2010){Wild}, {Heckman}, \& {Charlot}}]{Wild2010}
{Wild}, V., {Heckman}, T., \& {Charlot}, S. 2010, \mnras, 405, 933

\end{thebibliography}

\label{lastpage}
\end{document}